%
%
%

\ifx\mnmacrosloaded\undefined 
%
%
%
%

\catcode `\@=11 

\def\@version{1.4}
\def\@verdate{22nd Feb 1994}

%
%
%
%


\newif\ifprod@font

\ifx\@typeface\undefined
  \def\@typeface{Comp. Modern}\prod@fontfalse
\else
  \prod@fonttrue 
\fi

\def\newfam{\alloc@8\fam\chardef\sixt@@n} 

\ifprod@font
\font\fiverm=mtr10 at 5pt
\font\fivebf=mtbx10 at 5pt
\font\fiveit=mtti10 at 5pt
\font\fivesl=mtsl10 at 5pt
\font\fivett=mttt10 at 5pt     \hyphenchar\fivett=-1
\font\fivecsc=mtcsc10 at 5pt
\font\fivesf=mtss10 at 5pt
\font\fivei=mtmi10 at 5pt      \skewchar\fivei='177
\font\fivemib=mtmib10 at 5pt   \skewchar\fivemib='177
\font\fivesy=mtsy10 at 5pt     \skewchar\fivesy='60
\font\fivesyb=mtbsy10 at 5pt   \skewchar\fivesyb='60

\font\sixrm=mtr10 at 6pt
\font\sixbf=mtbx10 at 6pt
\font\sixit=mtti10 at 6pt
\font\sixsl=mtsl10 at 6pt
\font\sixtt=mttt10 at 6pt      \hyphenchar\sixtt=-1
\font\sixcsc=mtcsc10 at 6pt
\font\sixsf=mtss10 at 6pt
\font\sixi=mtmi10 at 6pt       \skewchar\sixi='177
\font\sixmib=mtmib10 at 6pt    \skewchar\sixmib='177
\font\sixsy=mtsy10 at 6pt      \skewchar\sixsy='60
\font\sixsyb=mtbsy10 at 6pt    \skewchar\sixsyb='60

\font\sevenrm=mtr10 at 7pt
\font\sevenbf=mtbx10 at 7pt
\font\sevenit=mtti10 at 7pt
\font\sevensl=mtsl10 at 7pt
\font\seventt=mttt10 at 7pt     \hyphenchar\seventt=-1
\font\sevencsc=mtcsc10 at 7pt
\font\sevensf=mtss10 at 7pt
\font\seveni=mtmi10 at 7pt      \skewchar\seveni='177
\font\sevenmib=mtmib10 at 7pt   \skewchar\sevenmib='177
\font\sevensy=mtsy10 at 7pt     \skewchar\sevensy='60
\font\sevensyb=mtbsy10 at 7pt   \skewchar\sevensyb='60

\font\eightrm=mtr10 at 8pt
\font\eightbf=mtbx10 at 8pt
\font\eightit=mtti10 at 8pt
\font\eighti=mtmi10 at 8pt      \skewchar\eighti='177
\font\eightmib=mtmib10 at 8pt   \skewchar\eightmib='177
\font\eightsy=mtsy10 at 8pt     \skewchar\eightsy='60
\font\eightsyb=mtbsy10 at 8pt   \skewchar\eightsyb='60
\font\eightsl=mtsl10 at 8pt
\font\eighttt=mttt10 at 8pt     \hyphenchar\eighttt=-1
\font\eightcsc=mtcsc10 at 8pt
\font\eightsf=mtss10 at 8pt

\font\ninerm=mtr10 at 9pt
\font\ninebf=mtbx10 at 9pt
\font\nineit=mtti10 at 9pt
\font\ninei=mtmi10 at 9pt      \skewchar\ninei='177
\font\ninemib=mtmib10 at 9pt   \skewchar\ninemib='177
\font\ninesy=mtsy10 at 9pt     \skewchar\ninesy='60
\font\ninesyb=mtbsy10 at 9pt   \skewchar\ninesyb='60
\font\ninesl=mtsl10 at 9pt
\font\ninett=mttt10 at 9pt     \hyphenchar\ninett=-1
\font\ninecsc=mtcsc10 at 9pt
\font\ninesf=mtss10 at 9pt

\font\tenrm=mtr10
\font\tenbf=mtbx10
\font\tenit=mtti10
\font\teni=mtmi10		\skewchar\teni='177
\font\tenmib=mtmib10	\skewchar\tenmib='177
\font\tensy=mtsy10		\skewchar\tensy='60
\font\tensyb=mtbsy10	\skewchar\tensyb='60
\font\tenex=cmex10
\font\tensl=mtsl10
\font\tentt=mttt10		\hyphenchar\tentt=-1
\font\tencsc=mtcsc10
\font\tensf=mtss10

\font\elevenrm=mtr10 at 11pt
\font\elevenbf=mtbx10 at 11pt
\font\elevenit=mtti10 at 11pt
\font\eleveni=mtmi10 at 11pt      \skewchar\eleveni='177
\font\elevenmib=mtmib10 at 11pt   \skewchar\elevenmib='177
\font\elevensy=mtsy10 at 11pt     \skewchar\elevensy='60
\font\elevensyb=mtbsy10 at 11pt   \skewchar\elevensyb='60
\font\elevensl=mtsl10 at 11pt
\font\eleventt=mttt10 at 11pt     \hyphenchar\eleventt=-1
\font\elevencsc=mtcsc10 at 11pt
\font\elevensf=mtss10 at 11pt

\font\twelverm=mtr10 at 12pt
\font\twelvebf=mtbx10 at 12pt
\font\twelveit=mtti10 at 12pt
\font\twelvesl=mtsl10 at 12pt
\font\twelvett=mttt10 at 12pt     \hyphenchar\twelvett=-1
\font\twelvecsc=mtcsc10 at 12pt
\font\twelvesf=mtss10 at 12pt
\font\twelvei=mtmi10 at 12pt      \skewchar\twelvei='177
\font\twelvemib=mtmib10 at 12pt   \skewchar\twelvemib='177
\font\twelvesy=mtsy10 at 12pt     \skewchar\twelvesy='60
\font\twelvesyb=mtbsy10 at 12pt   \skewchar\twelvesyb='60

\font\fourteenrm=mtr10 at 14pt
\font\fourteenbf=mtbx10 at 14pt
\font\fourteenit=mtti10 at 14pt
\font\fourteeni=mtmi10 at 14pt      \skewchar\fourteeni='177
\font\fourteenmib=mtmib10 at 14pt   \skewchar\fourteenmib='177
\font\fourteensy=mtsy10 at 14pt     \skewchar\fourteensy='60
\font\fourteensyb=mtbsy10 at 14pt   \skewchar\fourteensyb='60
\font\fourteensl=mtsl10 at 14pt
\font\fourteentt=mttt10 at 14pt     \hyphenchar\fourteentt=-1
\font\fourteencsc=mtcsc10 at 14pt
\font\fourteensf=mtss10 at 14pt

\font\seventeenrm=mtr10 at 17pt
\font\seventeenbf=mtbx10 at 17pt
\font\seventeenit=mtti10 at 17pt
\font\seventeeni=mtmi10 at 17pt      \skewchar\seventeeni='177
\font\seventeenmib=mtmib10 at 17pt   \skewchar\seventeenmib='177
\font\seventeensy=mtsy10 at 17pt     \skewchar\seventeensy='60
\font\seventeensyb=mtbsy10 at 17pt   \skewchar\seventeensyb='60
\font\seventeensl=mtsl10 at 17pt
\font\seventeentt=mttt10 at 17pt     \hyphenchar\seventeentt=-1
\font\seventeencsc=mtcsc10 at 17pt
\font\seventeensf=mtss10 at 17pt


\newfam\xmfam
\newfam\ymfam

\font\fivexm=mtxm10 at 5pt
\font\sixxm=mtxm10 at 6pt
\font\sevenxm=mtxm10 at 7pt
\font\eightxm=mtxm10 at 8pt
\font\ninexm=mtxm10 at 9pt
\font\tenxm=mtxm10
\font\elevenxm=mtxm10 at 11pt
\font\twelvexm=mtxm10 at 12pt
\font\fourteenxm=mtxm10 at 14pt
\font\seventeenxm=mtxm10 at 17pt

\font\fiveym=mtym10 at 5pt
\font\sixym=mtym10 at 6pt
\font\sevenym=mtym10 at 7pt
\font\eightym=mtym10 at 8pt
\font\nineym=mtym10 at 9pt
\font\tenym=mtym10
\font\elevenym=mtym10 at 11pt
\font\twelveym=mtym10 at 12pt
\font\fourteenym=mtym10 at 14pt
\font\seventeenym=mtym10 at 17pt
\else
\font\fiverm=cmr5
\font\fivei=cmmi5             \skewchar\fivei='177
\font\fivemib=cmmib10 at 5pt  \skewchar\fivemib='177
\font\fivesy=cmsy5            \skewchar\fivesy='60
\font\fivesyb=cmbsy10 at 5pt  \skewchar\fivesyb='60
\font\fivebf=cmbx5

\font\sixrm=cmr6
\font\sixi=cmmi6             \skewchar\sixi='177
\font\sixmib=cmmib10 at 6pt  \skewchar\sixmib='177
\font\sixsy=cmsy6            \skewchar\sixsy='60
\font\sixsyb=cmbsy10 at 6pt  \skewchar\sixsyb='60
\font\sixbf=cmbx6

\font\sevenrm=cmr7
\font\seveni=cmmi7             \skewchar\seveni='177
\font\sevenmib=cmmib10 at 7pt  \skewchar\sevenmib='177
\font\sevensy=cmsy7            \skewchar\sevensy='60
\font\sevensyb=cmbsy10 at 7pt  \skewchar\sevensyb='60
\font\sevenbf=cmbx7

\font\eightrm=cmr8
\font\eightbf=cmbx8
\font\eightit=cmti8
\font\eighti=cmmi8			\skewchar\eighti='177
\font\eightmib=cmmib10 at 8pt	\skewchar\eightmib='177
\font\eightsy=cmsy8			\skewchar\eightsy='60
\font\eightsyb=cmbsy10 at 8pt	\skewchar\eightsyb='60
\font\eightsl=cmsl8
\font\eighttt=cmtt8			\hyphenchar\eighttt=-1
\font\eightcsc=cmcsc10 at 8pt
\font\eightsf=cmss8

\font\ninerm=cmr9
\font\ninebf=cmbx9
\font\nineit=cmti9
\font\ninei=cmmi9			\skewchar\ninei='177
\font\ninemib=cmmib10 at 9pt	\skewchar\ninemib='177
\font\ninesy=cmsy9			\skewchar\ninesy='60
\font\ninesyb=cmbsy10 at 9pt	\skewchar\ninesyb='60
\font\ninesl=cmsl9
\font\ninett=cmtt9			\hyphenchar\ninett=-1
\font\ninecsc=cmcsc10 at 9pt
\font\ninesf=cmss9

\font\tenrm=cmr10
\font\tenbf=cmbx10
\font\tenit=cmti10
\font\teni=cmmi10		\skewchar\teni='177
\font\tenmib=cmmib10	\skewchar\tenmib='177
\font\tensy=cmsy10		\skewchar\tensy='60
\font\tensyb=cmbsy10	\skewchar\tensyb='60
\font\tenex=cmex10
\font\tensl=cmsl10
\font\tentt=cmtt10		\hyphenchar\tentt=-1
\font\tencsc=cmcsc10
\font\tensf=cmss10

\font\elevenrm=cmr10 scaled \magstephalf
\font\elevenbf=cmbx10 scaled \magstephalf
\font\elevenit=cmti10 scaled \magstephalf
\font\eleveni=cmmi10 scaled \magstephalf	\skewchar\eleveni='177
\font\elevenmib=cmmib10 scaled \magstephalf	\skewchar\elevenmib='177
\font\elevensy=cmsy10 scaled \magstephalf	\skewchar\elevensy='60
\font\elevensyb=cmbsy10 scaled \magstephalf	\skewchar\elevensyb='60
\font\elevensl=cmsl10 scaled \magstephalf
\font\eleventt=cmtt10 scaled \magstephalf	\hyphenchar\eleventt=-1
\font\elevencsc=cmcsc10 scaled \magstephalf
\font\elevensf=cmss10 scaled \magstephalf

\font\twelverm=cmr10 scaled \magstep1
\font\twelvebf=cmbx10 scaled \magstep1
\font\twelvei=cmmi10 scaled \magstep1      \skewchar\twelvei='177
\font\twelvemib=cmmib10 scaled \magstep1   \skewchar\twelvemib='177
\font\twelvesy=cmsy10 scaled \magstep1     \skewchar\twelvesy='60
\font\twelvesyb=cmbsy10 scaled \magstep1   \skewchar\twelvesyb='60

\font\fourteenrm=cmr10 scaled \magstep2
\font\fourteenbf=cmbx10 scaled \magstep2
\font\fourteenit=cmti10 scaled \magstep2
\font\fourteeni=cmmi10 scaled \magstep2		\skewchar\fourteeni='177
\font\fourteenmib=cmmib10 scaled \magstep2	\skewchar\fourteenmib='177
\font\fourteensy=cmsy10 scaled \magstep2	\skewchar\fourteensy='60
\font\fourteensyb=cmbsy10 scaled \magstep2	\skewchar\fourteensyb='60
\font\fourteensl=cmsl10 scaled \magstep2
\font\fourteentt=cmtt10 scaled \magstep2	\hyphenchar\fourteentt=-1
\font\fourteencsc=cmcsc10 scaled \magstep2
\font\fourteensf=cmss10 scaled \magstep2

\font\seventeenrm=cmr10 scaled \magstep3
\font\seventeenbf=cmbx10 scaled \magstep3
\font\seventeenit=cmti10 scaled \magstep3
\font\seventeeni=cmmi10 scaled \magstep3	\skewchar\seventeeni='177
\font\seventeenmib=cmmib10 scaled \magstep3	\skewchar\seventeenmib='177
\font\seventeensy=cmsy10 scaled \magstep3	\skewchar\seventeensy='60
\font\seventeensyb=cmbsy10 scaled \magstep3	\skewchar\seventeensyb='60
\font\seventeensl=cmsl10 scaled \magstep3
\font\seventeentt=cmtt10 scaled \magstep3	\hyphenchar\seventeentt=-1
\font\seventeencsc=cmcsc10 scaled \magstep3
\font\seventeensf=cmss10 scaled \magstep3
\fi

\def\hexnumber#1{\ifcase#1 0\or1\or2\or3\or4\or5\or6\or7\or8\or9\or
  A\or B\or C\or D\or E\or F\fi}

\ifprod@font
  \edef\@xm{\hexnumber\xmfam}
  \edef\@ym{\hexnumber\ymfam}
\fi

\def\makestrut{%
  \setbox\strutbox=\hbox{%
    \vrule height.7\baselineskip depth.3\baselineskip width \z@}%
}

\def\baselinestretch{1}
\newskip\tmp@bls

\def\b@ls#1{
  \tmp@bls=#1\relax
  \baselineskip=#1\relax\makestrut
  \normalbaselineskip=\baselinestretch\tmp@bls
  \normalbaselines
}

\def\nostb@ls#1{
  \normalbaselineskip=#1\relax
  \normalbaselines
  \makestrut
}

%

\newfam\mibfam 
\newfam\sybfam 
\newfam\scfam  
\newfam\sffam  

\def\mit{\fam\@ne}

\def\cal{\fam\tw@}

\def\em{\ifdim\fontdimen1\font>\z@ \rm\else\it\fi}

\textfont3=\tenex
\scriptfont3=\tenex
\scriptscriptfont3=\tenex

\setbox0=\hbox{\tenex B} \p@renwd=\wd0 

\def\eightpoint{
  \def\rm{\fam0\eightrm}%
  \textfont0=\eightrm \scriptfont0=\sixrm \scriptscriptfont0=\fiverm%
  \textfont1=\eighti  \scriptfont1=\sixi  \scriptscriptfont1=\fivei%
  \textfont2=\eightsy \scriptfont2=\sixsy \scriptscriptfont2=\fivesy%
  \textfont\itfam=\eightit\def\it{\fam\itfam\eightit}%
  \ifprod@font
    \scriptfont\itfam=\sixit
      \scriptscriptfont\itfam=\fiveit
  \else
    \scriptfont\itfam=\eightit
      \scriptscriptfont\itfam=\eightit
  \fi
  \textfont\bffam=\eightbf%
    \scriptfont\bffam=\sixbf%
      \scriptscriptfont\bffam=\fivebf%
  \def\bf{\fam\bffam\eightbf}%
  \textfont\slfam=\eightsl\def\sl{\fam\slfam\eightsl}%
  \ifprod@font
    \scriptfont\slfam=\sixsl
      \scriptscriptfont\slfam=\fivesl
  \else
    \scriptfont\slfam=\eightsl
      \scriptscriptfont\slfam=\eightsl
  \fi
  \textfont\ttfam=\eighttt\def\tt{\fam\ttfam\eighttt}%
  \ifprod@font
    \scriptfont\ttfam=\sixtt
      \scriptscriptfont\ttfam=\fivett
  \else
    \scriptfont\ttfam=\eighttt
      \scriptscriptfont\ttfam=\eighttt
  \fi
  \textfont\scfam=\eightcsc\def\sc{\fam\scfam\eightcsc}%
  \ifprod@font
    \scriptfont\scfam=\sixcsc
      \scriptscriptfont\scfam=\fivecsc
  \else
    \scriptfont\scfam=\eightcsc
      \scriptscriptfont\scfam=\eightcsc
  \fi
  \textfont\sffam=\eightsf\def\sf{\fam\sffam\eightsf}%
  \ifprod@font
    \scriptfont\sffam=\sixsf
      \scriptscriptfont\sffam=\fivesf
  \else
    \scriptfont\sffam=\eightsf
      \scriptscriptfont\sffam=\eightsf
  \fi
  \textfont\mibfam=\eightmib
    \scriptfont\mibfam=\sixmib
      \scriptscriptfont\mibfam=\fivemib
  \textfont\sybfam=\eightsyb
    \scriptfont\sybfam=\sixsyb
      \scriptscriptfont\sybfam=\fivesyb
  \ifprod@font
    \textfont\xmfam=\eightxm
      \scriptfont\xmfam=\sixxm
        \scriptscriptfont\xmfam=\fivexm
    \textfont\ymfam=\eightym
      \scriptfont\ymfam=\sixym
        \scriptscriptfont\ymfam=\fiveym
  \fi
  \def\oldstyle{\fam\@ne\eighti}%
  \def\boldstyle{\fam\mibfam\eightmib}%
  \b@ls{10pt}\rm%
}

\def\ninepoint{
  \def\rm{\fam0\ninerm}%
  \textfont0=\ninerm \scriptfont0=\sixrm \scriptscriptfont0=\fiverm%
  \textfont1=\ninei  \scriptfont1=\sixi  \scriptscriptfont1=\fivei%
  \textfont2=\ninesy \scriptfont2=\sixsy \scriptscriptfont2=\fivesy%
  \textfont\itfam=\nineit\def\it{\fam\itfam\nineit}%
  \ifprod@font
    \scriptfont\itfam=\sixit
      \scriptscriptfont\itfam=\fiveit
  \else
    \scriptfont\itfam=\nineit
      \scriptscriptfont\itfam=\nineit
  \fi
  \textfont\bffam=\ninebf%
    \scriptfont\bffam=\sixbf%
      \scriptscriptfont\bffam=\fivebf%
  \def\bf{\fam\bffam\ninebf}%
  \textfont\slfam=\ninesl\def\sl{\fam\slfam\ninesl}%
  \ifprod@font
    \scriptfont\slfam=\sixsl
      \scriptscriptfont\slfam=\fivesl
  \else
    \scriptfont\slfam=\ninesl
      \scriptscriptfont\slfam=\ninesl
  \fi
  \textfont\ttfam=\ninett\def\tt{\fam\ttfam\ninett}%
  \ifprod@font
    \scriptfont\ttfam=\sixtt
      \scriptscriptfont\ttfam=\fivett
  \else
    \scriptfont\ttfam=\ninett
      \scriptscriptfont\ttfam=\ninett
  \fi
  \textfont\scfam=\ninecsc\def\sc{\fam\scfam\ninecsc}%
  \ifprod@font
    \scriptfont\scfam=\sixcsc
      \scriptscriptfont\scfam=\fivecsc
  \else
    \scriptfont\scfam=\ninecsc
      \scriptscriptfont\scfam=\ninecsc
  \fi
  \textfont\sffam=\ninesf\def\sf{\fam\sffam\ninesf}%
  \ifprod@font
    \scriptfont\sffam=\sixsf
      \scriptscriptfont\sffam=\fivesf
  \else
    \scriptfont\sffam=\ninesf
      \scriptscriptfont\sffam=\ninesf
  \fi
  \textfont\mibfam=\ninemib
    \scriptfont\mibfam=\sixmib
      \scriptscriptfont\mibfam=\fivemib
  \textfont\sybfam=\ninesyb
    \scriptfont\sybfam=\sixsyb
      \scriptscriptfont\sybfam=\fivesyb
  \ifprod@font
    \textfont\xmfam=\ninexm
      \scriptfont\xmfam=\sixxm
        \scriptscriptfont\xmfam=\fivexm
    \textfont\ymfam=\nineym
      \scriptfont\ymfam=\sixym
        \scriptscriptfont\ymfam=\fiveym
  \fi
  \def\oldstyle{\fam\@ne\ninei}%
  \def\boldstyle{\fam\mibfam\ninemib}%
  \b@ls{\TextLeading plus \Feathering}\rm%
}

\def\tenpoint{
  \def\rm{\fam0\tenrm}%
  \textfont0=\tenrm \scriptfont0=\sevenrm \scriptscriptfont0=\fiverm%
  \textfont1=\teni  \scriptfont1=\seveni  \scriptscriptfont1=\fivei%
  \textfont2=\tensy \scriptfont2=\sevensy \scriptscriptfont2=\fivesy%
  \textfont\itfam=\tenit\def\it{\fam\itfam\tenit}%
  \ifprod@font
    \scriptfont\itfam=\sevenit
      \scriptscriptfont\itfam=\fiveit
  \else
    \scriptfont\itfam=\tenit
      \scriptscriptfont\itfam=\tenit
  \fi
  \textfont\bffam=\tenbf%
    \scriptfont\bffam=\sevenbf%
      \scriptscriptfont\bffam=\fivebf%
  \def\bf{\fam\bffam\tenbf}%
  \textfont\slfam=\tensl\def\sl{\fam\slfam\tensl}%
  \ifprod@font
    \scriptfont\slfam=\sevensl
      \scriptscriptfont\slfam=\fivesl
  \else
    \scriptfont\slfam=\tensl
      \scriptscriptfont\slfam=\tensl
  \fi
  \textfont\ttfam=\tentt\def\tt{\fam\ttfam\tentt}%
  \ifprod@font
    \scriptfont\ttfam=\seventt
      \scriptscriptfont\ttfam=\fivett
  \else
    \scriptfont\ttfam=\tentt
      \scriptscriptfont\ttfam=\tentt
  \fi
  \textfont\scfam=\tencsc\def\sc{\fam\scfam\tencsc}%
  \ifprod@font
    \scriptfont\scfam=\sevencsc
      \scriptscriptfont\scfam=\fivecsc
  \else
    \scriptfont\scfam=\tencsc
      \scriptscriptfont\scfam=\tencsc
  \fi
  \textfont\sffam=\tensf\def\sf{\fam\sffam\tensf}%
  \ifprod@font
    \scriptfont\sffam=\sevensf
      \scriptscriptfont\sffam=\fivesf
  \else
    \scriptfont\sffam=\tensf
      \scriptscriptfont\sffam=\tensf
  \fi
  \textfont\mibfam=\tenmib
    \scriptfont\mibfam=\sevenmib
      \scriptscriptfont\mibfam=\fivemib
  \textfont\sybfam=\tensyb
    \scriptfont\sybfam=\sevensyb
      \scriptscriptfont\sybfam=\fivesyb
  \ifprod@font
    \textfont\xmfam=\tenxm
      \scriptfont\xmfam=\sevenxm
        \scriptscriptfont\xmfam=\fivexm
    \textfont\ymfam=\tenym
      \scriptfont\ymfam=\sevenym
        \scriptscriptfont\ymfam=\fiveym
  \fi
  \def\oldstyle{\fam\@ne\teni}%
  \def\boldstyle{\fam\mibfam\tenmib}%
  \b@ls{11pt}\rm%
}

\def\elevenpoint{
  \def\rm{\fam0\elevenrm}%
  \textfont0=\elevenrm \scriptfont0=\eightrm \scriptscriptfont0=\sixrm%
  \textfont1=\eleveni  \scriptfont1=\eighti  \scriptscriptfont1=\sixi%
  \textfont2=\elevensy \scriptfont2=\eightsy \scriptscriptfont2=\sixsy%
  \textfont\itfam=\elevenit\def\it{\fam\itfam\elevenit}%
  \ifprod@font
    \scriptfont\itfam=\eightit
      \scriptscriptfont\itfam=\sixit
  \else
    \scriptfont\itfam=\elevenit
      \scriptscriptfont\itfam=\elevenit
  \fi
  \textfont\bffam=\elevenbf%
    \scriptfont\bffam=\eightbf%
      \scriptscriptfont\bffam=\sixbf%
  \def\bf{\fam\bffam\elevenbf}%
  \textfont\slfam=\elevensl\def\sl{\fam\slfam\elevensl}%
  \ifprod@font
    \scriptfont\slfam=\eightsl
      \scriptscriptfont\slfam=\sixsl
  \else
    \scriptfont\slfam=\elevensl
      \scriptscriptfont\slfam=\elevensl
  \fi
  \textfont\ttfam=\eleventt\def\tt{\fam\ttfam\eleventt}%
  \ifprod@font
    \scriptfont\ttfam=\eighttt
      \scriptscriptfont\ttfam=\sixtt
  \else
    \scriptfont\ttfam=\eleventt
      \scriptscriptfont\ttfam=\eleventt
  \fi
  \textfont\scfam=\elevencsc\def\sc{\fam\scfam\elevencsc}%
  \ifprod@font
    \scriptfont\scfam=\eightcsc
      \scriptscriptfont\scfam=\sixcsc
  \else
    \scriptfont\scfam=\elevencsc
      \scriptscriptfont\scfam=\elevencsc
  \fi
  \textfont\sffam=\elevensf\def\sf{\fam\sffam\elevensf}%
  \ifprod@font
    \scriptfont\sffam=\eightsf
      \scriptscriptfont\sffam=\sixsf
  \else
    \scriptfont\sffam=\elevensf
      \scriptscriptfont\sffam=\elevensf
  \fi
  \textfont\mibfam=\elevenmib
    \scriptfont\mibfam=\eightmib
      \scriptscriptfont\mibfam=\sixmib
  \textfont\sybfam=\elevensyb
    \scriptfont\sybfam=\eightsyb
      \scriptscriptfont\sybfam=\sixsyb
  \ifprod@font
    \textfont\xmfam=\elevenxm
      \scriptfont\xmfam=\eightxm
       \scriptscriptfont\xmfam=\sixxm
    \textfont\ymfam=\elevenym
      \scriptfont\ymfam=\eightym
        \scriptscriptfont\ymfam=\sixym
   \fi
  \def\oldstyle{\fam\@ne\eleveni}%
  \def\boldstyle{\fam\mibfam\elevenmib}%
  \b@ls{13pt}\rm%
}

\def\fourteenpoint{
  \def\rm{\fam0\fourteenrm}%
  \textfont0\fourteenrm  \scriptfont0\tenrm  \scriptscriptfont0\sevenrm%
  \textfont1\fourteeni   \scriptfont1\teni   \scriptscriptfont1\seveni%
  \textfont2\fourteensy  \scriptfont2\tensy  \scriptscriptfont2\sevensy%
  \textfont\itfam=\fourteenit\def\it{\fam\itfam\fourteenit}%
  \ifprod@font
    \scriptfont\itfam=\tenit
      \scriptscriptfont\itfam=\sevenit
  \else
    \scriptfont\itfam=\fourteenit
      \scriptscriptfont\itfam=\fourteenit
  \fi
  \textfont\bffam=\fourteenbf%
    \scriptfont\bffam=\tenbf%
      \scriptscriptfont\bffam=\sevenbf%
  \def\bf{\fam\bffam\fourteenbf}%
  \textfont\slfam=\fourteensl\def\sl{\fam\slfam\fourteensl}%
  \ifprod@font
    \scriptfont\slfam=\tensl
      \scriptscriptfont\slfam=\sevensl
  \else
    \scriptfont\slfam=\fourteensl
      \scriptscriptfont\slfam=\fourteensl
  \fi
  \textfont\ttfam=\fourteentt\def\tt{\fam\ttfam\fourteentt}%
  \ifprod@font
    \scriptfont\ttfam=\tentt
      \scriptscriptfont\ttfam=\seventt
  \else
    \scriptfont\ttfam=\fourteentt
      \scriptscriptfont\ttfam=\fourteentt
  \fi
  \textfont\scfam=\fourteencsc\def\sc{\fam\scfam\fourteencsc}%
  \ifprod@font
    \scriptfont\scfam=\tencsc
      \scriptscriptfont\scfam=\sevencsc
  \else
    \scriptfont\scfam=\fourteencsc
      \scriptscriptfont\scfam=\fourteencsc
  \fi
  \textfont\sffam=\fourteensf\def\sf{\fam\sffam\fourteensf}%
  \ifprod@font
    \scriptfont\sffam=\tensf
      \scriptscriptfont\sffam=\sevensf
  \else
    \scriptfont\sffam=\fourteensf
      \scriptscriptfont\sffam=\fourteensf
  \fi
  \textfont\mibfam=\fourteenmib
    \scriptfont\mibfam=\tenmib
      \scriptscriptfont\mibfam=\sevenmib
  \textfont\sybfam=\fourteensyb
    \scriptfont\sybfam=\tensyb
      \scriptscriptfont\sybfam=\sevensyb
  \ifprod@font
    \textfont\xmfam=\fourteenxm
      \scriptfont\xmfam=\tenxm
        \scriptscriptfont\xmfam=\sevenxm
   \textfont\ymfam=\fourteenym
      \scriptfont\ymfam=\tenym
        \scriptscriptfont\ymfam=\sevenym
  \fi
  \def\oldstyle{\fam\@ne\fourteeni}%
  \def\boldstyle{\fam\mibfam\fourteenmib}%
  \b@ls{17pt}\rm%
}

\def\seventeenpoint{
  \def\rm{\fam0\seventeenrm}%
  \textfont0\seventeenrm  \scriptfont0\twelverm  \scriptscriptfont0\tenrm%
  \textfont1\seventeeni   \scriptfont1\twelvei   \scriptscriptfont1\teni%
  \textfont2\seventeensy  \scriptfont2\twelvesy  \scriptscriptfont2\tensy%
  \textfont\itfam=\seventeenit\def\it{\fam\itfam\seventeenit}%
  \ifprod@font
    \scriptfont\itfam=\twelveit
      \scriptscriptfont\itfam=\tenit
  \else
    \scriptfont\itfam=\seventeenit
      \scriptscriptfont\itfam=\seventeenit
  \fi
  \textfont\bffam=\seventeenbf%
    \scriptfont\bffam=\twelvebf%
      \scriptscriptfont\bffam=\tenbf%
  \def\bf{\fam\bffam\seventeenbf}%
  \textfont\slfam=\seventeensl\def\sl{\fam\slfam\seventeensl}%
  \ifprod@font
    \scriptfont\slfam=\twelvesl
      \scriptscriptfont\slfam=\tensl
  \else
    \scriptfont\slfam=\seventeensl
      \scriptscriptfont\slfam=\seventeensl
  \fi
  \textfont\ttfam=\seventeentt\def\tt{\fam\ttfam\seventeentt}%
  \ifprod@font
    \scriptfont\ttfam=\twelvett
      \scriptscriptfont\ttfam=\tentt
  \else
    \scriptfont\ttfam=\seventeentt
      \scriptscriptfont\ttfam=\seventeentt
  \fi
  \textfont\scfam=\seventeencsc\def\sc{\fam\scfam\seventeencsc}%
  \ifprod@font
    \scriptfont\scfam=\twelvecsc
      \scriptscriptfont\scfam=\tencsc
  \else
    \scriptfont\scfam=\seventeencsc
      \scriptscriptfont\scfam=\seventeencsc
  \fi
  \textfont\sffam=\seventeensf\def\sf{\fam\sffam\seventeensf}%
  \ifprod@font
    \scriptfont\sffam=\twelvesf
      \scriptscriptfont\sffam=\tensf
  \else
    \scriptfont\sffam=\seventeensf
      \scriptscriptfont\sffam=\seventeensf
  \fi
  \textfont\mibfam=\seventeenmib
    \scriptfont\mibfam=\twelvemib
      \scriptscriptfont\mibfam=\tenmib
  \textfont\sybfam=\seventeensyb
    \scriptfont\sybfam=\twelvesyb
      \scriptscriptfont\sybfam=\tensyb
  \ifprod@font
    \textfont\xmfam=\seventeenxm
      \scriptfont\xmfam=\twelvexm
        \scriptscriptfont\xmfam=\tenxm
    \textfont\ymfam=\seventeenym
      \scriptfont\ymfam=\twelveym
        \scriptscriptfont\ymfam=\tenym
  \fi
  \def\oldstyle{\fam\@ne\seventeeni}%
  \def\boldstyle{\fam\mibfam\seventeenmib}%
  \b@ls{20pt}\rm%
}

\lineskip=1pt      \normallineskip=\lineskip
\lineskiplimit=\z@ \normallineskiplimit=\lineskiplimit



\def\la{\mathrel{\mathchoice {\vcenter{\offinterlineskip\halign{\hfil
$\displaystyle##$\hfil\cr<\cr\sim\cr}}}
{\vcenter{\offinterlineskip\halign{\hfil$\textstyle##$\hfil\cr
<\cr\sim\cr}}}
{\vcenter{\offinterlineskip\halign{\hfil$\scriptstyle##$\hfil\cr
<\cr\sim\cr}}}
{\vcenter{\offinterlineskip\halign{\hfil$\scriptscriptstyle##$\hfil\cr
<\cr\sim\cr}}}}}

\def\ga{\mathrel{\mathchoice {\vcenter{\offinterlineskip\halign{\hfil
$\displaystyle##$\hfil\cr>\cr\sim\cr}}}
{\vcenter{\offinterlineskip\halign{\hfil$\textstyle##$\hfil\cr
>\cr\sim\cr}}}
{\vcenter{\offinterlineskip\halign{\hfil$\scriptstyle##$\hfil\cr
>\cr\sim\cr}}}
{\vcenter{\offinterlineskip\halign{\hfil$\scriptscriptstyle##$\hfil\cr
>\cr\sim\cr}}}}}

\def\getsto{\mathrel{\mathchoice {\vcenter{\offinterlineskip
\halign{\hfil
$\displaystyle##$\hfil\cr\gets\cr\to\cr}}}
{\vcenter{\offinterlineskip\halign{\hfil$\textstyle##$\hfil\cr\gets
\cr\to\cr}}}
{\vcenter{\offinterlineskip\halign{\hfil$\scriptstyle##$\hfil\cr\gets
\cr\to\cr}}}
{\vcenter{\offinterlineskip\halign{\hfil$\scriptscriptstyle##$\hfil\cr
\gets\cr\to\cr}}}}}

\def\lid{\mathrel{\mathchoice {\vcenter{\offinterlineskip\halign{\hfil
$\displaystyle##$\hfil\cr<\cr\noalign{\vskip1.2pt}=\cr}}}
{\vcenter{\offinterlineskip\halign{\hfil$\textstyle##$\hfil\cr<\cr
\noalign{\vskip1.2pt}=\cr}}}
{\vcenter{\offinterlineskip\halign{\hfil$\scriptstyle##$\hfil\cr<\cr
\noalign{\vskip1pt}=\cr}}}
{\vcenter{\offinterlineskip\halign{\hfil$\scriptscriptstyle##$\hfil\cr
<\cr
\noalign{\vskip0.9pt}=\cr}}}}}

\def\gid{\mathrel{\mathchoice {\vcenter{\offinterlineskip\halign{\hfil
$\displaystyle##$\hfil\cr>\cr\noalign{\vskip1.2pt}=\cr}}}
{\vcenter{\offinterlineskip\halign{\hfil$\textstyle##$\hfil\cr>\cr
\noalign{\vskip1.2pt}=\cr}}}
{\vcenter{\offinterlineskip\halign{\hfil$\scriptstyle##$\hfil\cr>\cr
\noalign{\vskip1pt}=\cr}}}
{\vcenter{\offinterlineskip\halign{\hfil$\scriptscriptstyle##$\hfil\cr
>\cr
\noalign{\vskip0.9pt}=\cr}}}}}

\def\grole{\mathrel{\mathchoice {\vcenter{\offinterlineskip\halign{\hfil
$\displaystyle##$\hfil\cr>\cr\noalign{\vskip-1.5pt}<\cr}}}
{\vcenter{\offinterlineskip\halign{\hfil$\textstyle##$\hfil\cr
>\cr\noalign{\vskip-1.5pt}<\cr}}}
{\vcenter{\offinterlineskip\halign{\hfil$\scriptstyle##$\hfil\cr
>\cr\noalign{\vskip-1pt}<\cr}}}
{\vcenter{\offinterlineskip\halign{\hfil$\scriptscriptstyle##$\hfil\cr
>\cr\noalign{\vskip-0.5pt}<\cr}}}}}

\def\leogr{\mathrel{\mathchoice {\vcenter{\offinterlineskip\halign{\hfil
$\displaystyle##$\hfil\cr<\cr\noalign{\vskip-1.5pt}>\cr}}}
{\vcenter{\offinterlineskip\halign{\hfil$\textstyle##$\hfil\cr
<\cr\noalign{\vskip-1.5pt}>\cr}}}
{\vcenter{\offinterlineskip\halign{\hfil$\scriptstyle##$\hfil\cr
<\cr\noalign{\vskip-1pt}>\cr}}}
{\vcenter{\offinterlineskip\halign{\hfil$\scriptscriptstyle##$\hfil\cr
<\cr\noalign{\vskip-0.5pt}>\cr}}}}}

\def\loa{\mathrel{\mathchoice {\vcenter{\offinterlineskip\halign{\hfil
$\displaystyle##$\hfil\cr<\cr\approx\cr}}}
{\vcenter{\offinterlineskip\halign{\hfil$\textstyle##$\hfil\cr
<\cr\approx\cr}}}
{\vcenter{\offinterlineskip\halign{\hfil$\scriptstyle##$\hfil\cr
<\cr\approx\cr}}}
{\vcenter{\offinterlineskip\halign{\hfil$\scriptscriptstyle##$\hfil\cr
<\cr\approx\cr}}}}}

\def\goa{\mathrel{\mathchoice {\vcenter{\offinterlineskip\halign{\hfil
$\displaystyle##$\hfil\cr>\cr\approx\cr}}}
{\vcenter{\offinterlineskip\halign{\hfil$\textstyle##$\hfil\cr
>\cr\approx\cr}}}
{\vcenter{\offinterlineskip\halign{\hfil$\scriptstyle##$\hfil\cr
>\cr\approx\cr}}}
{\vcenter{\offinterlineskip\halign{\hfil$\scriptscriptstyle##$\hfil\cr
>\cr\approx\cr}}}}}

\def\diameter{{\ifmmode\mathchoice
{\ooalign{\hfil\hbox{$\displaystyle/$}\hfil\crcr
{\hbox{$\displaystyle\mathchar"20D$}}}}
{\ooalign{\hfil\hbox{$\textstyle/$}\hfil\crcr
{\hbox{$\textstyle\mathchar"20D$}}}}
{\ooalign{\hfil\hbox{$\scriptstyle/$}\hfil\crcr
{\hbox{$\scriptstyle\mathchar"20D$}}}}
{\ooalign{\hfil\hbox{$\scriptscriptstyle/$}\hfil\crcr
{\hbox{$\scriptscriptstyle\mathchar"20D$}}}}
\else{\ooalign{\hfil/\hfil\crcr\mathhexbox20D}}%
\fi}}

\def\sq{\ifmmode\squareforqed\else{\unskip\nobreak\hfil
\penalty50\hskip1em\null\nobreak\hfil\squareforqed
\parfillskip=0pt\finalhyphendemerits=0\endgraf}\fi}
\def\squareforqed{\hbox{\rlap{$\sqcap$}$\sqcup$}}


\def\bbbc{{\mathchoice {\setbox0=\hbox{$\displaystyle\rm C$}\hbox{\hbox
to0pt{\kern0.4\wd0\vrule height0.9\ht0\hss}\box0}}
{\setbox0=\hbox{$\textstyle\rm C$}\hbox{\hbox
to0pt{\kern0.4\wd0\vrule height0.9\ht0\hss}\box0}}
{\setbox0=\hbox{$\scriptstyle\rm C$}\hbox{\hbox
to0pt{\kern0.4\wd0\vrule height0.9\ht0\hss}\box0}}
{\setbox0=\hbox{$\scriptscriptstyle\rm C$}\hbox{\hbox
to0pt{\kern0.4\wd0\vrule height0.9\ht0\hss}\box0}}}}
\def\bbbq{{\mathchoice {\setbox0=\hbox{$\displaystyle\rm
Q$}\hbox{\raise
0.15\ht0\hbox to0pt{\kern0.4\wd0\vrule height0.8\ht0\hss}\box0}}
{\setbox0=\hbox{$\textstyle\rm Q$}\hbox{\raise
0.15\ht0\hbox to0pt{\kern0.4\wd0\vrule height0.8\ht0\hss}\box0}}
{\setbox0=\hbox{$\scriptstyle\rm Q$}\hbox{\raise
0.15\ht0\hbox to0pt{\kern0.4\wd0\vrule height0.7\ht0\hss}\box0}}
{\setbox0=\hbox{$\scriptscriptstyle\rm Q$}\hbox{\raise
0.15\ht0\hbox to0pt{\kern0.4\wd0\vrule height0.7\ht0\hss}\box0}}}}
\def\bbbt{{\mathchoice {\setbox0=\hbox{$\displaystyle\rm
T$}\hbox{\hbox to0pt{\kern0.3\wd0\vrule height0.9\ht0\hss}\box0}}
{\setbox0=\hbox{$\textstyle\rm T$}\hbox{\hbox
to0pt{\kern0.3\wd0\vrule height0.9\ht0\hss}\box0}}
{\setbox0=\hbox{$\scriptstyle\rm T$}\hbox{\hbox
to0pt{\kern0.3\wd0\vrule height0.9\ht0\hss}\box0}}
{\setbox0=\hbox{$\scriptscriptstyle\rm T$}\hbox{\hbox
to0pt{\kern0.3\wd0\vrule height0.9\ht0\hss}\box0}}}}
\def\bbbs{{\mathchoice
{\setbox0=\hbox{$\displaystyle     \rm S$}\hbox{\raise0.5\ht0\hbox
to0pt{\kern0.35\wd0\vrule height0.45\ht0\hss}\hbox
to0pt{\kern0.55\wd0\vrule height0.5\ht0\hss}\box0}}
{\setbox0=\hbox{$\textstyle        \rm S$}\hbox{\raise0.5\ht0\hbox
to0pt{\kern0.35\wd0\vrule height0.45\ht0\hss}\hbox
to0pt{\kern0.55\wd0\vrule height0.5\ht0\hss}\box0}}
{\setbox0=\hbox{$\scriptstyle      \rm S$}\hbox{\raise0.5\ht0\hbox
to0pt{\kern0.35\wd0\vrule height0.45\ht0\hss}\raise0.05\ht0\hbox
to0pt{\kern0.5\wd0\vrule height0.45\ht0\hss}\box0}}
{\setbox0=\hbox{$\scriptscriptstyle\rm S$}\hbox{\raise0.5\ht0\hbox
to0pt{\kern0.4\wd0\vrule height0.45\ht0\hss}\raise0.05\ht0\hbox
to0pt{\kern0.55\wd0\vrule height0.45\ht0\hss}\box0}}}}
\def\bbbz{{\mathchoice {\hbox{$\sf\textstyle Z\kern-0.4em Z$}}
{\hbox{$\sf\textstyle Z\kern-0.4em Z$}}
{\hbox{$\sf\scriptstyle Z\kern-0.3em Z$}}
{\hbox{$\sf\scriptscriptstyle Z\kern-0.2em Z$}}}}


\ifprod@font
  \mathchardef\la="3\@xm2E
  \mathchardef\getsto="3\@xm1C
  \mathchardef\lid="3\@xm35
  \mathchardef\grole="3\@xm3F
  \mathchardef\loa="3\@xm2F
  \mathchardef\ga="3\@xm26
  \mathchardef\gid="3\@xm3D
  \mathchardef\leogr="3\@xm37
  \mathchardef\goa="3\@xm27
  \mathchardef\sq="0\@xm03
%
%
\def\diameter{{%
  \ifmmode
    \mathchoice
    {\ooalign{\hfil\hbox{$\displaystyle/$}\hfil\crcr
    {\lower.2ex\hbox{$\displaystyle\mathchar"20D$}}}}%
    {\ooalign{\hfil\hbox{$\textstyle/$}\hfil\crcr
    {\lower.2ex\hbox{$\textstyle\mathchar"20D$}}}}%
    {\ooalign{\hfil\hbox{$\scriptstyle/$}\hfil\crcr
    {\lower.1ex\hbox{$\scriptstyle\mathchar"20D$}}}}%
    {\ooalign{\hfil\hbox{$\scriptscriptstyle/$}\hfil\crcr
    {\lower.1ex\hbox{$\scriptscriptstyle\mathchar"20D$}}}}%
  \else
    {\ooalign{\hfil/\hfil\crcr\lower.2ex\hbox{\mathhexbox20D}}}%
  \fi
}}
%
%

\def\bbbc{{\Bbb{C}}}
\def\bbbq{{\Bbb{Q}}}
\def\bbbt{{\Bbb{T}}}
\def\bbbs{{\Bbb{S}}}
\def\bbbz{{\Bbb{Z}}}
\fi


\ifprod@font
\mathchardef\boxdot="2\@xm00
\mathchardef\boxplus="2\@xm01
\mathchardef\boxtimes="2\@xm02
\mathchardef\square="0\@xm03
\mathchardef\blacksquare="0\@xm04
\mathchardef\centerdot="2\@xm05
\mathchardef\lozenge="0\@xm06
\mathchardef\blacklozenge="0\@xm07
\mathchardef\circlearrowright="3\@xm08
\mathchardef\circlearrowleft="3\@xm09
\mathchardef\rightleftharpoons="3\@xm0A
\mathchardef\leftrightharpoons="3\@xm0B
\mathchardef\boxminus="2\@xm0C
\mathchardef\Vdash="3\@xm0D
\mathchardef\Vvdash="3\@xm0E
\mathchardef\vDash="3\@xm0F
\mathchardef\twoheadrightarrow="3\@xm10
\mathchardef\twoheadleftarrow="3\@xm11
\mathchardef\leftleftarrows="3\@xm12
\mathchardef\rightrightarrows="3\@xm13
\mathchardef\upuparrows="3\@xm14
\mathchardef\downdownarrows="3\@xm15
\mathchardef\upharpoonright="3\@xm16

\mathchardef\downharpoonright="3\@xm17
\mathchardef\upharpoonleft="3\@xm18
\mathchardef\downharpoonleft="3\@xm19
\mathchardef\rightarrowtail="3\@xm1A
\mathchardef\leftarrowtail="3\@xm1B
\mathchardef\leftrightarrows="3\@xm1C
\mathchardef\rightleftarrows="3\@xm1D
\mathchardef\Lsh="3\@xm1E
\mathchardef\Rsh="3\@xm1F
\mathchardef\rightsquigarrow="3\@xm20
\mathchardef\leftrightsquigarrow="3\@xm21
\mathchardef\looparrowleft="3\@xm22
\mathchardef\looparrowright="3\@xm23
\mathchardef\circeq="3\@xm24
\mathchardef\succsim="3\@xm25
\mathchardef\gtrsim="3\@xm26
\mathchardef\gtrapprox="3\@xm27
\mathchardef\multimap="3\@xm28
\mathchardef\therefore="3\@xm29
\mathchardef\because="3\@xm2A
\mathchardef\doteqdot="3\@xm2B

\mathchardef\triangleq="3\@xm2C
\mathchardef\precsim="3\@xm2D
\mathchardef\lesssim="3\@xm2E
\mathchardef\lessapprox="3\@xm2F
\mathchardef\eqslantless="3\@xm30
\mathchardef\eqslantgtr="3\@xm31
\mathchardef\curlyeqprec="3\@xm32
\mathchardef\curlyeqsucc="3\@xm33
\mathchardef\preccurlyeq="3\@xm34
\mathchardef\leqq="3\@xm35
\mathchardef\leqslant="3\@xm36
\mathchardef\lessgtr="3\@xm37
\mathchardef\backprime="0\@xm38
\mathchardef\risingdotseq="3\@xm3A
\mathchardef\fallingdotseq="3\@xm3B
\mathchardef\succcurlyeq="3\@xm3C
\mathchardef\geqq="3\@xm3D
\mathchardef\geqslant="3\@xm3E
\mathchardef\gtrless="3\@xm3F
\mathchardef\sqsubset="3\@xm40
\mathchardef\sqsupset="3\@xm41
\mathchardef\vartriangleright="3\@xm42
\mathchardef\vartriangleleft="3\@xm43
\mathchardef\trianglerighteq="3\@xm44
\mathchardef\trianglelefteq="3\@xm45
\mathchardef\bigstar="0\@xm46
\mathchardef\between="3\@xm47
\mathchardef\blacktriangledown="0\@xm48
\mathchardef\blacktriangleright="3\@xm49
\mathchardef\blacktriangleleft="3\@xm4A
\mathchardef\vartriangle="0\@xm4D
\mathchardef\blacktriangle="0\@xm4E
\mathchardef\triangledown="0\@xm4F
\mathchardef\eqcirc="3\@xm50
\mathchardef\lesseqgtr="3\@xm51
\mathchardef\gtreqless="3\@xm52
\mathchardef\lesseqqgtr="3\@xm53
\mathchardef\gtreqqless="3\@xm54
\mathchardef\Rrightarrow="3\@xm56
\mathchardef\Lleftarrow="3\@xm57
\mathchardef\veebar="2\@xm59
\mathchardef\barwedge="2\@xm5A
\mathchardef\doublebarwedge="2\@xm5B
\mathchardef\angle="0\@xm5C
\mathchardef\measuredangle="0\@xm5D
\mathchardef\sphericalangle="0\@xm5E
\mathchardef\varpropto="3\@xm5F
\mathchardef\smallsmile="3\@xm60
\mathchardef\smallfrown="3\@xm61
\mathchardef\Subset="3\@xm62
\mathchardef\Supset="3\@xm63
\mathchardef\Cup="2\@xm64

\mathchardef\Cap="2\@xm65

\mathchardef\curlywedge="2\@xm66
\mathchardef\curlyvee="2\@xm67
\mathchardef\leftthreetimes="2\@xm68
\mathchardef\rightthreetimes="2\@xm69
\mathchardef\subseteqq="3\@xm6A
\mathchardef\supseteqq="3\@xm6B
\mathchardef\bumpeq="3\@xm6C
\mathchardef\Bumpeq="3\@xm6D
\mathchardef\lll="3\@xm6E

\mathchardef\ggg="3\@xm6F

\mathchardef\circledS="0\@xm73
\mathchardef\pitchfork="3\@xm74
\mathchardef\dotplus="2\@xm75
\mathchardef\backsim="3\@xm76
\mathchardef\backsimeq="3\@xm77
\mathchardef\complement="0\@xm7B
\mathchardef\intercal="2\@xm7C
\mathchardef\circledcirc="2\@xm7D
\mathchardef\circledast="2\@xm7E
\mathchardef\circleddash="2\@xm7F
\def\ulcorner{\delimiter"4\@xm70\@xm70 }
\def\urcorner{\delimiter"5\@xm71\@xm71 }
\def\llcorner{\delimiter"4\@xm78\@xm78 }
\def\lrcorner{\delimiter"5\@xm79\@xm79 }
\def\yen{\mathhexbox\@xm55 }
\def\checkmark{\mathhexbox\@xm58 }
\def\circledR{\mathhexbox\@xm72 }
\def\maltese{\mathhexbox\@xm7A }
\mathchardef\lvertneqq="3\@ym00
\mathchardef\gvertneqq="3\@ym01
\mathchardef\nleq="3\@ym02
\mathchardef\ngeq="3\@ym03
\mathchardef\nless="3\@ym04
\mathchardef\ngtr="3\@ym05
\mathchardef\nprec="3\@ym06
\mathchardef\nsucc="3\@ym07
\mathchardef\lneqq="3\@ym08
\mathchardef\gneqq="3\@ym09
\mathchardef\nleqslant="3\@ym0A
\mathchardef\ngeqslant="3\@ym0B
\mathchardef\lneq="3\@ym0C
\mathchardef\gneq="3\@ym0D
\mathchardef\npreceq="3\@ym0E
\mathchardef\nsucceq="3\@ym0F
\mathchardef\precnsim="3\@ym10
\mathchardef\succnsim="3\@ym11
\mathchardef\lnsim="3\@ym12
\mathchardef\gnsim="3\@ym13
\mathchardef\nleqq="3\@ym14
\mathchardef\ngeqq="3\@ym15
\mathchardef\precneqq="3\@ym16
\mathchardef\succneqq="3\@ym17
\mathchardef\precnapprox="3\@ym18
\mathchardef\succnapprox="3\@ym19
\mathchardef\lnapprox="3\@ym1A
\mathchardef\gnapprox="3\@ym1B
\mathchardef\nsim="3\@ym1C
\mathchardef\ncong="3\@ym1D

\mathchardef\varsubsetneq="3\@ym20
\mathchardef\varsupsetneq="3\@ym21
\mathchardef\nsubseteqq="3\@ym22
\mathchardef\nsupseteqq="3\@ym23
\mathchardef\subsetneqq="3\@ym24
\mathchardef\supsetneqq="3\@ym25
\mathchardef\varsubsetneqq="3\@ym26
\mathchardef\varsupsetneqq="3\@ym27
\mathchardef\subsetneq="3\@ym28
\mathchardef\supsetneq="3\@ym29
\mathchardef\nsubseteq="3\@ym2A
\mathchardef\nsupseteq="3\@ym2B
\mathchardef\nparallel="3\@ym2C
\mathchardef\nmid="3\@ym2D
\mathchardef\nshortmid="3\@ym2E
\mathchardef\nshortparallel="3\@ym2F
\mathchardef\nvdash="3\@ym30
\mathchardef\nVdash="3\@ym31
\mathchardef\nvDash="3\@ym32
\mathchardef\nVDash="3\@ym33
\mathchardef\ntrianglerighteq="3\@ym34
\mathchardef\ntrianglelefteq="3\@ym35
\mathchardef\ntriangleleft="3\@ym36
\mathchardef\ntriangleright="3\@ym37
\mathchardef\nleftarrow="3\@ym38
\mathchardef\nrightarrow="3\@ym39
\mathchardef\nLeftarrow="3\@ym3A
\mathchardef\nRightarrow="3\@ym3B
\mathchardef\nLeftrightarrow="3\@ym3C
\mathchardef\nleftrightarrow="3\@ym3D
\mathchardef\divideontimes="2\@ym3E
\mathchardef\varnothing="0\@ym3F
\mathchardef\nexists="0\@ym40
\mathchardef\mho="0\@ym66
\mathchardef\eth="0\@ym67
\mathchardef\eqsim="3\@ym68
\mathchardef\beth="0\@ym69
\mathchardef\gimel="0\@ym6A
\mathchardef\daleth="0\@ym6B
\mathchardef\lessdot="3\@ym6C
\mathchardef\gtrdot="3\@ym6D
\mathchardef\ltimes="2\@ym6E
\mathchardef\rtimes="2\@ym6F
\mathchardef\shortmid="3\@ym70
\mathchardef\shortparallel="3\@ym71
\mathchardef\smallsetminus="2\@ym72
\mathchardef\thicksim="3\@ym73
\mathchardef\thickapprox="3\@ym74
\mathchardef\approxeq="3\@ym75
\mathchardef\succapprox="3\@ym76
\mathchardef\precapprox="3\@ym77
\mathchardef\curvearrowleft="3\@ym78
\mathchardef\curvearrowright="3\@ym79
\mathchardef\digamma="0\@ym7A
\mathchardef\varkappa="0\@ym7B
\mathchardef\hslash="0\@ym7D
\mathchardef\hbar="0\@ym7E
\mathchardef\backepsilon="3\@ym7F


\def\Bbb{\ifmmode\let\next\Bbb@\else
\def\next{\errmessage{Use \string\Bbb\space only in math mode}}\fi\next}
\def\Bbb@#1{{\Bbb@@{#1}}}
\def\Bbb@@#1{\fam\ymfam#1}
\fi


\def\Nulle{0} 
\def\Afe{1}   
\def\Hae{2}   
\def\Hbe{3}   
\def\Hce{4}   
\def\Hde{5}   


\newcount\LastMac       \LastMac=\Nulle

\newskip\half      \half=5.5pt plus 1.5pt minus 2.25pt
\newskip\one       \one=11pt plus 3pt minus 5.5pt
\newskip\onehalf   \onehalf=16.5pt plus 5.5pt minus 8.25pt
\newskip\two       \two=22pt plus 5.5pt minus 11pt

\def\Half{\addvspace{\half}}
\def\One{\addvspace{\one}}
\def\OneHalf{\addvspace{\onehalf}}
\def\Two{\addvspace{\two}}


\def\Raggedright{
  \rightskip=\z@ plus \hsize\relax
}

\def\Fullout{
  \rightskip=\z@\relax
}

\def\Hang#1#2{
  \hangindent=#1%
  \hangafter=#2\relax
}


\newif\ifsp@page
\def\pagestyle#1{\csname ps@#1\endcsname}
\def\thispagestyle#1{\global\sp@pagetrue\gdef\sp@type{#1}}

\def\ps@titlepage{%
  \def\@oddhead{\eightpoint\noindent \the\CatchLine
    \ifprod@font\else\qquad Printed\ \today\fi \hfil}%
  \let\@evenhead=\@oddhead
}

\def\ps@headings{%
  \def\@oddhead{\elevenpoint\it\noindent
    \hfill\the\RightHeader\hskip1.5em\rm\folio}%
  \def\@evenhead{\elevenpoint\noindent
    \folio\hskip1.5em\it\the\LeftHeader\hfill}%
}

\def\ps@plate{%
  \def\@oddhead{\eightpoint\noindent\plt@cap\hfil}%
  \def\@evenhead{\eightpoint\noindent\plt@cap\hfil}%
}



\def\title#1{
  \bgroup
    \vbox to 8pt{\vss}%
    \seventeenpoint
    \Raggedright
    \noindent \strut{\bf #1}\par
  \egroup
}

\def\author#1{
  \bgroup
    \ifnum\LastMac=\Afe \OneHalf\else \vskip 21pt\fi
    \fourteenpoint
    \Raggedright
    \noindent \strut #1\par
    \vskip 3pt%
  \egroup
}

\def\affiliation#1{
  \bgroup
    \vskip -4pt%
    \eightpoint
    \Raggedright
    \noindent \strut {\it #1}\par
  \egroup
  \LastMac=\Afe\relax
}

\def\acceptedline#1{
  \bgroup
    \Two
    \eightpoint
    \Raggedright
    \noindent \strut #1\par
  \egroup
}

\long\def\abstract#1{%
  \bgroup
    \vskip 20pt%
    \everypar{\Hang{11pc}{0}}%
    \noindent{\ninebf ABSTRACT}\par
    \tenpoint
    \Fullout
    \noindent #1\par
  \egroup
}

\long\def\keywords#1{
  \bgroup
    \Half
    \everypar{\Hang{11pc}{0}}%
    \tenpoint
    \Fullout
    \noindent\hbox{\bf Key words:}\ #1\par
  \egroup
}


\def\maketitle{%
  \EndOpening
  \ifsinglecol \else \MakePage\fi
}


\def\pageoffset#1#2{\hoffset=#1\relax\voffset=#2\relax}


\def\Autonumber{
  \global\AutoNumbertrue  
}

\newif\ifAutoNumber \AutoNumberfalse
\newcount\Sec        
\newcount\SecSec
\newcount\SecSecSec

\Sec=\z@

\def\:{\let\@sptoken= } \:  
\def\:{\@xifnch} \expandafter\def\: {\futurelet\@tempc\@ifnch}

\def\@ifnextchar#1#2#3{%
  \let\@tempMACe #1%
  \def\@tempMACa{#2}%
  \def\@tempMACb{#3}%
  \futurelet \@tempMACc\@ifnch%
}

\def\@ifnch{%
\ifx \@tempMACc \@sptoken%
  \let\@tempMACd\@xifnch%
\else%
  \ifx \@tempMACc \@tempMACe%
    \let\@tempMACd\@tempMACa%
  \else%
    \let\@tempMACd\@tempMACb%
  \fi%
\fi%
\@tempMACd%
}

\def\@ifstar#1#2{\@ifnextchar *{\def\@tempMACa*{#1}\@tempMACa}{#2}}

\newskip\@tempskipb

\def\addvspace#1{%
  \ifvmode\else \endgraf\fi%
  \ifdim\lastskip=\z@%
    \vskip #1\relax%
  \else%
    \@tempskipb#1\relax\@xaddvskip%
  \fi%
}

\def\@xaddvskip{%
  \ifdim\lastskip<\@tempskipb%
    \vskip-\lastskip%
    \vskip\@tempskipb\relax%
  \else%
    \ifdim\@tempskipb<\z@%
      \ifdim\lastskip<\z@ \else%
        \advance\@tempskipb\lastskip%
        \vskip-\lastskip\vskip\@tempskipb%
      \fi%
    \fi%
  \fi%
}

\newskip\@tmpSKIP

\def\addpen#1{%
  \ifvmode
    \if@nobreak
    \else
      \ifdim\lastskip=\z@
        \penalty#1\relax
      \else
        \@tmpSKIP=\lastskip
        \vskip -\lastskip
        \penalty#1\vskip\@tmpSKIP
      \fi
    \fi
  \fi
}

\newcount\@clubpen   \@clubpen=\clubpenalty
\newif\if@nobreak    \@nobreakfalse

\def\@noafterindent{%
  \global\@nobreaktrue
  \everypar{\if@nobreak
              \global\@nobreakfalse
              \clubpenalty \@M
              {\setbox\z@\lastbox}%
              \LastMac=\Nulle\relax%
            \else
              \clubpenalty \@clubpen
              \everypar{}%
            \fi}
}

\newcount\gds@cbrk   \gds@cbrk=-300

\def\@nohdbrk{\interlinepenalty \@M\relax}

\let\@par=\par
\def\@restorepar{\def\par{\@par}}

\newif\if@endpe   \@endpefalse

\def\@doendpe{\@endpetrue \@nobreakfalse \LastMac=\Nulle\relax%
     \def\par{\@restorepar\everypar{}\par\@endpefalse}%
              \everypar{\setbox\z@\lastbox\everypar{}\@endpefalse}%
}

\def\section{\@ifstar{\@ssection}{\@section}}

\def\@section#1{
  \if@nobreak
    \everypar{}%
    \ifnum\LastMac=\Hae \addvspace{\half}\fi
  \else
    \addpen{\gds@cbrk}%
    \addvspace{\two}%
  \fi
  \bgroup
    \ninepoint\bf
    \Raggedright
    \ifAutoNumber
      \global\advance\Sec \@ne
      \noindent\@nohdbrk\number\Sec\hskip 1pc \uppercase{#1}\par
      \global\SecSec=\z@
    \else
      \noindent\@nohdbrk\uppercase{#1}\par
    \fi
  \egroup
  \nobreak
  \vskip\half
  \nobreak
  \@noafterindent
  \LastMac=\Hae\relax
}

\def\@ssection#1{
  \if@nobreak
    \everypar{}%
    \ifnum\LastMac=\Hae \addvspace{\half}\fi
  \else
    \addpen{\gds@cbrk}%
    \addvspace{\two}%
  \fi
  \bgroup
    \ninepoint\bf
    \Raggedright
    \noindent\@nohdbrk\uppercase{#1}\par
  \egroup
  \nobreak
  \vskip\half
  \nobreak
  \@noafterindent
  \LastMac=\Hae\relax
}

\def\subsection#1{
  \if@nobreak
    \everypar{}%
    \ifnum\LastMac=\Hae \addvspace{1pt plus 1pt minus .5pt}\fi
  \else
    \addpen{\gds@cbrk}%
    \addvspace{\onehalf}%
  \fi
  \bgroup
    \ninepoint\bf
    \Raggedright
    \ifAutoNumber
      \global\advance\SecSec \@ne
      \noindent\@nohdbrk\number\Sec.\number\SecSec \hskip 1pc\relax #1\par
      \global\SecSecSec=\z@
    \else
      \noindent\@nohdbrk #1\par
    \fi
  \egroup
  \nobreak
  \vskip\half
  \nobreak
  \@noafterindent
  \LastMac=\Hbe\relax
}

\def\subsubsection#1{
  \if@nobreak
    \everypar{}%
    \ifnum\LastMac=\Hbe \addvspace{1pt plus 1pt minus .5pt}\fi
  \else
    \addpen{\gds@cbrk}%
    \addvspace{\onehalf}%
  \fi
  \bgroup
    \ninepoint\it
    \Raggedright
    \ifAutoNumber
      \global\advance\SecSecSec \@ne
      \noindent\@nohdbrk\number\Sec.\number\SecSec.\number\SecSecSec
        \hskip 1pc\relax #1\par
    \else
      \noindent\@nohdbrk #1\par
    \fi
  \egroup
  \nobreak
  \vskip\half
  \nobreak
  \@noafterindent
  \LastMac=\Hce\relax
}

\def\paragraph#1{
  \if@nobreak
    \everypar{}%
  \else
    \addpen{\gds@cbrk}%
    \addvspace{\one}%
  \fi%
  \bgroup%
    \ninepoint\it
    \noindent #1\ \nobreak%
  \egroup
  \LastMac=\Hde\relax
  \ignorespaces
}




\def\beginlist{%
  \par\if@nobreak \else\addvspace{\half}\fi%
  \bgroup%
    \ninepoint
    \let\item=\list@item%
}

\def\list@item{%
  \par\noindent\hskip 1em\relax%
  \ignorespaces%
}

\def\endlist{\par\egroup\addvspace{\half}\@doendpe}


\def\beginrefs{%
  \par
  \bgroup
    \eightpoint
    \Raggedright
    \let\bibitem=\bib@item
}

\def\bib@item{%
  \par\parindent=1.5em\Hang{1.5em}{1}%
  \everypar={\Hang{1.5em}{1}\ignorespaces}%
  \noindent\ignorespaces
}

\def\endrefs{\par\egroup\@doendpe}


\newtoks\CatchLine

\def\@journal{Mon.\ Not.\ R.\ Astron.\ Soc.\ }  
\def\@pubyear{1994}        
\def\@pagerange{000--000}  
\def\@volume{000}          
\def\@microfiche{}         %

\def\pubyear#1{\gdef\@pubyear{#1}\@makecatchline}
\def\pagerange#1{\gdef\@pagerange{#1}\@makecatchline}
\def\volume#1{\gdef\@volume{#1}\@makecatchline}
\def\microfiche#1{\gdef\@microfiche{and Microfiche\ #1}\@makecatchline}

\def\@makecatchline{%
  \global\CatchLine{%
    {\rm \@journal {\bf \@volume},\ \@pagerange\ (\@pubyear)\ \@microfiche}}%
}

\@makecatchline 

\newtoks\LeftHeader
\def\shortauthor#1{
  \global\LeftHeader{#1}%
}

\newtoks\RightHeader
\def\shorttitle#1{
  \global\RightHeader{#1}%
}

\def\PageHead{
  \begingroup
    \ifsp@page
      \csname ps@\sp@type\endcsname
      \global\sp@pagefalse
    \fi
    \ifodd\pageno
      \let\the@head=\@oddhead
    \else
      \let\the@head=\@evenhead
    \fi
    \vbox to \z@{\vskip-22.5\p@%
      \hbox to \PageWidth{\vbox to8.5\p@{}%
        \the@head
      }%
    \vss}%
  \endgroup
  \nointerlineskip
}

\def\today{%
  \number\day\space
  \ifcase\month\or January\or February\or March\or April\or May\or June\or
    July\or August\or September\or October\or November\or December\fi
  \space\number\year%
}

\def\PageFoot{} 

\def\authorcomment#1{%
  \gdef\PageFoot{%
    \nointerlineskip%
    \vbox to 22pt{\vfil%
      \hbox to \PageWidth{\elevenpoint\noindent \hfil #1 \hfil}}%
  }%
}


\newif\ifplate@page
\newbox\plt@box

\def\beginplatepage{%
  \let\plate=\plate@head
  \let\caption=\fig@caption
  \global\setbox\plt@box=\vbox\bgroup
  \TEMPDIMEN=\PageWidth 
  \hsize=\PageWidth\relax
}

\def\endplatepage{\par\egroup\global\plate@pagetrue}
\def\plate@head#1{\gdef\plt@cap{#1}}


\def\letters{%
  \gdef\folio{\ifnum\pageno<\z@ L\romannumeral-\pageno
    \else L\number\pageno \fi}%
}


\everydisplay{\displaysetup}

\newif\ifeqno
\newif\ifleqno

\def\displaysetup#1$${%
 \displaytest#1\eqno\eqno\displaytest
}

\def\displaytest#1\eqno#2\eqno#3\displaytest{%
 \if!#3!\ldisplaytest#1\leqno\leqno\ldisplaytest
 \else\eqnotrue\leqnofalse\def\eqn{#2}\def\eq{#1}\fi
 \generaldisplay$$}

\def\ldisplaytest#1\leqno#2\leqno#3\ldisplaytest{%
 \def\eq{#1}%
 \if!#3!\eqnofalse\else\eqnotrue\leqnotrue
  \def\eqn{#2}\fi}

\def\generaldisplay{%
\ifeqno \ifleqno
   \hbox to \hsize{\noindent
     $\displaystyle\eq$\hfil$\displaystyle\eqn$}
  \else
    \hbox to \hsize{\noindent
     $\displaystyle\eq$\hfil$\displaystyle\eqn$}
  \fi
 \else
 \hbox to \hsize{\vbox{\noindent
  $\displaystyle\eq$\hfil}}
 \fi
}


\def\@notice{%
  \par\Two%
  \noindent{\b@ls{11pt}\ninerm This paper has been produced using the
    Blackwell Scientific Publications \TeX\ macros.\par}%
}

\outer\def\bye{\@notice\par\vfill\supereject\end}


\def\start@mess{%
  Monthly notices of the RAS journal style (\@typeface)\space
    v\@version,\space \@verdate.%
}

\everyjob{\Warn{\start@mess}}



\newif\if@debug \@debugfalse  

\def\Print#1{\if@debug\immediate\write16{#1}\else \fi}
\def\Warn#1{\immediate\write16{#1}}
\def\wlog#1{}

\newcount\Iteration 

\def\Single{0} \def\Double{1}                 
\def\Figure{0} \def\Table{1}                  

\def\InStack{0}  
\def\InZoneA{1}
\def\InZoneB{2}
\def\InZoneC{3}

\newcount\TEMPCOUNT 
\newdimen\TEMPDIMEN 
\newbox\TEMPBOX     
\newbox\VOIDBOX     

\newcount\LengthOfStack 
\newcount\MaxItems      
\newcount\StackPointer
\newcount\Point         
\newcount\NextFigure    
\newcount\NextTable     
\newcount\NextItem      

\newcount\StatusStack   
\newcount\NumStack      
\newcount\TypeStack     
\newcount\SpanStack     
\newcount\BoxStack      

\newcount\ItemSTATUS    
\newcount\ItemNUMBER    
\newcount\ItemTYPE      
\newcount\ItemSPAN      
\newbox\ItemBOX         
\newdimen\ItemSIZE      

\newdimen\PageHeight    
\newdimen\TextLeading   
\newdimen\Feathering    
\newcount\LinesPerPage  
\newdimen\ColumnWidth   
\newdimen\ColumnGap     
\newdimen\PageWidth     
\newdimen\BodgeHeight   
\newcount\Leading       

\newdimen\ZoneBSize  
\newdimen\TextSize   
\newbox\ZoneABOX     
\newbox\ZoneBBOX     
\newbox\ZoneCBOX     

\newif\ifFirstSingleItem
\newif\ifFirstZoneA
\newif\ifMakePageInComplete
\newif\ifMoreFigures \MoreFiguresfalse 
\newif\ifMoreTables  \MoreTablesfalse  

\newif\ifFigInZoneB 
\newif\ifFigInZoneC 
\newif\ifTabInZoneB 
\newif\ifTabInZoneC

\newif\ifZoneAFullPage

\newbox\MidBOX    
\newbox\LeftBOX
\newbox\RightBOX
\newbox\PageBOX   

\newif\ifLeftCOL  
\LeftCOLtrue

\newdimen\ZoneBAdjust

\newcount\ItemFits
\def\Yes{1}
\def\No{2}


\MaxItems=15
\NextFigure=\z@        
\NextTable=\@ne

\BodgeHeight=6pt
\TextLeading=11pt    
\Leading=11
\Feathering=\z@      
\LinesPerPage=61     
\topskip=\TextLeading
\ColumnWidth=20pc    
\ColumnGap=2pc       

\newskip\ItemSepamount  
\ItemSepamount=\TextLeading plus \TextLeading minus 4pt

\parskip=\z@ plus .1pt
\parindent=18pt
\widowpenalty=\z@
\clubpenalty=10000
\tolerance=1500
\hbadness=1500
\abovedisplayskip=6pt plus 2pt minus 2pt
\belowdisplayskip=6pt plus 2pt minus 2pt
\abovedisplayshortskip=6pt plus 2pt minus 2pt
\belowdisplayshortskip=6pt plus 2pt minus 2pt

\ninepoint 


\PageHeight=682pt

\PageWidth=2\ColumnWidth
\advance\PageWidth by \ColumnGap

\pagestyle{headings}




\newcount\DUMMY \StatusStack=\allocationnumber
\newcount\DUMMY \newcount\DUMMY \newcount\DUMMY
\newcount\DUMMY \newcount\DUMMY \newcount\DUMMY
\newcount\DUMMY \newcount\DUMMY \newcount\DUMMY
\newcount\DUMMY \newcount\DUMMY \newcount\DUMMY
\newcount\DUMMY \newcount\DUMMY \newcount\DUMMY

\newcount\DUMMY \NumStack=\allocationnumber
\newcount\DUMMY \newcount\DUMMY \newcount\DUMMY
\newcount\DUMMY \newcount\DUMMY \newcount\DUMMY
\newcount\DUMMY \newcount\DUMMY \newcount\DUMMY
\newcount\DUMMY \newcount\DUMMY \newcount\DUMMY
\newcount\DUMMY \newcount\DUMMY \newcount\DUMMY

\newcount\DUMMY \TypeStack=\allocationnumber
\newcount\DUMMY \newcount\DUMMY \newcount\DUMMY
\newcount\DUMMY \newcount\DUMMY \newcount\DUMMY
\newcount\DUMMY \newcount\DUMMY \newcount\DUMMY
\newcount\DUMMY \newcount\DUMMY \newcount\DUMMY
\newcount\DUMMY \newcount\DUMMY \newcount\DUMMY

\newcount\DUMMY \SpanStack=\allocationnumber
\newcount\DUMMY \newcount\DUMMY \newcount\DUMMY
\newcount\DUMMY \newcount\DUMMY \newcount\DUMMY
\newcount\DUMMY \newcount\DUMMY \newcount\DUMMY
\newcount\DUMMY \newcount\DUMMY \newcount\DUMMY
\newcount\DUMMY \newcount\DUMMY \newcount\DUMMY

\newbox\DUMMY   \BoxStack=\allocationnumber
\newbox\DUMMY   \newbox\DUMMY \newbox\DUMMY
\newbox\DUMMY   \newbox\DUMMY \newbox\DUMMY
\newbox\DUMMY   \newbox\DUMMY \newbox\DUMMY
\newbox\DUMMY   \newbox\DUMMY \newbox\DUMMY
\newbox\DUMMY   \newbox\DUMMY \newbox\DUMMY

\def\wlog{\immediate\write\m@ne}


\def\GetItemAll#1{%
 \GetItemSTATUS{#1}
 \GetItemNUMBER{#1}
 \GetItemTYPE{#1}
 \GetItemSPAN{#1}
 \GetItemBOX{#1}
}

\def\GetItemSTATUS#1{%
 \Point=\StatusStack
 \advance\Point by #1
 \global\ItemSTATUS=\count\Point
}

\def\GetItemNUMBER#1{%
 \Point=\NumStack
 \advance\Point by #1
 \global\ItemNUMBER=\count\Point
}

\def\GetItemTYPE#1{%
 \Point=\TypeStack
 \advance\Point by #1
 \global\ItemTYPE=\count\Point
}

\def\GetItemSPAN#1{%
 \Point\SpanStack
 \advance\Point by #1
 \global\ItemSPAN=\count\Point
}

\def\GetItemBOX#1{%
 \Point=\BoxStack
 \advance\Point by #1
 \global\setbox\ItemBOX=\vbox{\copy\Point}
 \global\ItemSIZE=\ht\ItemBOX
 \global\advance\ItemSIZE by \dp\ItemBOX
 \TEMPCOUNT=\ItemSIZE
 \divide\TEMPCOUNT by \Leading
 \divide\TEMPCOUNT by 65536
 \advance\TEMPCOUNT \@ne
 \ItemSIZE=\TEMPCOUNT pt
 \global\multiply\ItemSIZE by \Leading
}


\def\JoinStack{%
 \ifnum\LengthOfStack=\MaxItems 
  \Warn{WARNING: Stack is full...some items will be lost!}
 \else
  \Point=\StatusStack
  \advance\Point by \LengthOfStack
  \global\count\Point=\ItemSTATUS
  \Point=\NumStack
  \advance\Point by \LengthOfStack
  \global\count\Point=\ItemNUMBER
  \Point=\TypeStack
  \advance\Point by \LengthOfStack
  \global\count\Point=\ItemTYPE
  \Point\SpanStack
  \advance\Point by \LengthOfStack
  \global\count\Point=\ItemSPAN
  \Point=\BoxStack
  \advance\Point by \LengthOfStack
  \global\setbox\Point=\vbox{\copy\ItemBOX}
  \global\advance\LengthOfStack \@ne
  \ifnum\ItemTYPE=\Figure 
   \global\MoreFigurestrue
  \else
   \global\MoreTablestrue
  \fi
 \fi
}


\def\LeaveStack#1{%
 {\Iteration=#1
 \loop
 \ifnum\Iteration<\LengthOfStack
  \advance\Iteration \@ne
  \GetItemSTATUS{\Iteration}
   \advance\Point by \m@ne
   \global\count\Point=\ItemSTATUS
  \GetItemNUMBER{\Iteration}
   \advance\Point by \m@ne
   \global\count\Point=\ItemNUMBER
  \GetItemTYPE{\Iteration}
   \advance\Point by \m@ne
   \global\count\Point=\ItemTYPE
  \GetItemSPAN{\Iteration}
   \advance\Point by \m@ne
   \global\count\Point=\ItemSPAN
  \GetItemBOX{\Iteration}
   \advance\Point by \m@ne
   \global\setbox\Point=\vbox{\copy\ItemBOX}
 \repeat}
 \global\advance\LengthOfStack by \m@ne
}


\newif\ifStackNotClean

\def\CleanStack{%
 \StackNotCleantrue
 {\Iteration=\z@
  \loop
   \ifStackNotClean
    \GetItemSTATUS{\Iteration}
    \ifnum\ItemSTATUS=\InStack
     \advance\Iteration \@ne
     \else
      \LeaveStack{\Iteration}
    \fi
   \ifnum\LengthOfStack<\Iteration
    \StackNotCleanfalse
   \fi
 \repeat}
}


\def\FindItem#1#2{%
 \global\StackPointer=\m@ne 
 {\Iteration=\z@
  \loop
  \ifnum\Iteration<\LengthOfStack
   \GetItemSTATUS{\Iteration}
   \ifnum\ItemSTATUS=\InStack
    \GetItemTYPE{\Iteration}
    \ifnum\ItemTYPE=#1
     \GetItemNUMBER{\Iteration}
     \ifnum\ItemNUMBER=#2
      \global\StackPointer=\Iteration
      \Iteration=\LengthOfStack 
     \fi
    \fi
   \fi
  \advance\Iteration \@ne
 \repeat}
}


\def\FindNext{%
 \global\StackPointer=\m@ne 
 {\Iteration=\z@
  \loop
  \ifnum\Iteration<\LengthOfStack
   \GetItemSTATUS{\Iteration}
   \ifnum\ItemSTATUS=\InStack
    \GetItemTYPE{\Iteration}
   \ifnum\ItemTYPE=\Figure
    \ifMoreFigures
      \global\NextItem=\Figure
      \global\StackPointer=\Iteration
      \Iteration=\LengthOfStack 
    \fi
   \fi
   \ifnum\ItemTYPE=\Table
    \ifMoreTables
      \global\NextItem=\Table
      \global\StackPointer=\Iteration
      \Iteration=\LengthOfStack 
    \fi
   \fi
  \fi
  \advance\Iteration \@ne
 \repeat}
}


\def\ChangeStatus#1#2{%
 \Point=\StatusStack
 \advance\Point by #1
 \global\count\Point=#2
}



\def\Zone{\InZoneA}

\ZoneBAdjust=\z@

\def\MakePage{
 \global\ZoneBSize=\PageHeight
 \global\TextSize=\ZoneBSize
 \global\ZoneAFullPagefalse
 \global\topskip=\TextLeading
 \MakePageInCompletetrue
 \MoreFigurestrue
 \MoreTablestrue
 \FigInZoneBfalse
 \FigInZoneCfalse
 \TabInZoneBfalse
 \TabInZoneCfalse
 \global\FirstSingleItemtrue
 \global\FirstZoneAtrue
 \global\setbox\ZoneABOX=\box\VOIDBOX
 \global\setbox\ZoneBBOX=\box\VOIDBOX
 \global\setbox\ZoneCBOX=\box\VOIDBOX
 \loop
  \ifMakePageInComplete
 \FindNext
 \ifnum\StackPointer=\m@ne
  \NextItem=\m@ne
  \MoreFiguresfalse
  \MoreTablesfalse
 \fi
 \ifnum\NextItem=\Figure
   \FindItem{\Figure}{\NextFigure}
   \ifnum\StackPointer=\m@ne \global\MoreFiguresfalse
   \else
    \GetItemSPAN{\StackPointer}
    \ifnum\ItemSPAN=\Single \def\Zone{\InZoneB}\relax
     \ifFigInZoneC \global\MoreFiguresfalse\fi
    \else
     \def\Zone{\InZoneA}
     \ifFigInZoneB \def\Zone{\InZoneC}\fi
    \fi
   \fi
   \ifMoreFigures\Print{}\FigureItems\fi
 \fi
\ifnum\NextItem=\Table
   \FindItem{\Table}{\NextTable}
   \ifnum\StackPointer=\m@ne \global\MoreTablesfalse
   \else
    \GetItemSPAN{\StackPointer}
    \ifnum\ItemSPAN=\Single\relax
     \ifTabInZoneC \global\MoreTablesfalse\fi
    \else
     \def\Zone{\InZoneA}
     \ifTabInZoneB \def\Zone{\InZoneC}\fi
    \fi
   \fi
   \ifMoreTables\Print{}\TableItems\fi
 \fi
   \MakePageInCompletefalse 
   \ifMoreFigures\MakePageInCompletetrue\fi
   \ifMoreTables\MakePageInCompletetrue\fi
 \repeat
 \ifZoneAFullPage
  \global\TextSize=\z@
  \global\ZoneBSize=\z@
  \global\vsize=\z@\relax
  \global\topskip=\z@\relax
  \vbox to \z@{\vss}
  \eject
 \else
 \global\advance\ZoneBSize by -\ZoneBAdjust
 \global\vsize=\ZoneBSize
 \global\hsize=\ColumnWidth
 \global\ZoneBAdjust=\z@
 \ifdim\TextSize<23pt
 \Warn{}
 \Warn{* Making column fall short: TextSize=\the\TextSize *}
 \vskip-\lastskip\eject\fi
 \fi
}

\def\MakeRightCol{
 \global\TextSize=\ZoneBSize
 \MakePageInCompletetrue
 \MoreFigurestrue
 \MoreTablestrue
 \global\FirstSingleItemtrue
 \global\setbox\ZoneBBOX=\box\VOIDBOX
 \def\Zone{\InZoneB}
 \loop
  \ifMakePageInComplete
 \FindNext
 \ifnum\StackPointer=\m@ne
  \NextItem=\m@ne
  \MoreFiguresfalse
  \MoreTablesfalse
 \fi
 \ifnum\NextItem=\Figure
   \FindItem{\Figure}{\NextFigure}
   \ifnum\StackPointer=\m@ne \MoreFiguresfalse
   \else
    \GetItemSPAN{\StackPointer}
    \ifnum\ItemSPAN=\Double\relax
     \MoreFiguresfalse\fi
   \fi
   \ifMoreFigures\Print{}\FigureItems\fi
 \fi
 \ifnum\NextItem=\Table
   \FindItem{\Table}{\NextTable}
   \ifnum\StackPointer=\m@ne \MoreTablesfalse
   \else
    \GetItemSPAN{\StackPointer}
    \ifnum\ItemSPAN=\Double\relax
     \MoreTablesfalse\fi
   \fi
   \ifMoreTables\Print{}\TableItems\fi
 \fi
   \MakePageInCompletefalse 
   \ifMoreFigures\MakePageInCompletetrue\fi
   \ifMoreTables\MakePageInCompletetrue\fi
 \repeat
 \ifZoneAFullPage
  \global\TextSize=\z@
  \global\ZoneBSize=\z@
  \global\vsize=\z@\relax
  \global\topskip=\z@\relax
  \vbox to \z@{\vss}
  \eject
 \else
 \global\vsize=\ZoneBSize
 \global\hsize=\ColumnWidth
 \ifdim\TextSize<23pt
 \Warn{}
 \Warn{* Making column fall short: TextSize=\the\TextSize *}
 \vskip-\lastskip\eject\fi
\fi
}

\def\FigureItems{
 \Print{Considering...}
 \ShowItem{\StackPointer}
 \GetItemBOX{\StackPointer} 
 \GetItemSPAN{\StackPointer}
  \CheckFitInZone 
  \ifnum\ItemFits=\Yes
   \ifnum\ItemSPAN=\Single
     \ChangeStatus{\StackPointer}{\InZoneB} 
     \global\FigInZoneBtrue
     \ifFirstSingleItem
      \hbox{}\vskip-\BodgeHeight
     \global\advance\ItemSIZE by \TextLeading
     \fi
     \unvbox\ItemBOX\ItemSep
     \global\FirstSingleItemfalse
     \global\advance\TextSize by -\ItemSIZE
     \global\advance\TextSize by -\TextLeading
   \else
    \ifFirstZoneA
     \global\advance\ItemSIZE by \TextLeading
     \global\FirstZoneAfalse\fi
    \global\advance\TextSize by -\ItemSIZE
    \global\advance\TextSize by -\TextLeading
    \global\advance\ZoneBSize by -\ItemSIZE
    \global\advance\ZoneBSize by -\TextLeading
    \ifFigInZoneB\relax
     \else
     \ifdim\TextSize<3\TextLeading
     \global\ZoneAFullPagetrue
     \fi
    \fi
    \ChangeStatus{\StackPointer}{\Zone}
    \ifnum\Zone=\InZoneC \global\FigInZoneCtrue\fi
  \fi
   \Print{TextSize=\the\TextSize}
   \Print{ZoneBSize=\the\ZoneBSize}
  \global\advance\NextFigure \@ne
   \Print{This figure has been placed.}
  \else
   \Print{No space available for this figure...holding over.}
   \Print{}
   \global\MoreFiguresfalse
  \fi
}

\def\TableItems{
 \Print{Considering...}
 \ShowItem{\StackPointer}
 \GetItemBOX{\StackPointer} 
 \GetItemSPAN{\StackPointer}
  \CheckFitInZone 
  \ifnum\ItemFits=\Yes
   \ifnum\ItemSPAN=\Single
    \ChangeStatus{\StackPointer}{\InZoneB}
     \global\TabInZoneBtrue
     \ifFirstSingleItem
      \hbox{}\vskip-\BodgeHeight
     \global\advance\ItemSIZE by \TextLeading
     \fi
     \unvbox\ItemBOX\ItemSep
     \global\FirstSingleItemfalse
     \global\advance\TextSize by -\ItemSIZE
     \global\advance\TextSize by -\TextLeading
   \else
    \ifFirstZoneA
    \global\advance\ItemSIZE by \TextLeading
    \global\FirstZoneAfalse\fi
    \global\advance\TextSize by -\ItemSIZE
    \global\advance\TextSize by -\TextLeading
    \global\advance\ZoneBSize by -\ItemSIZE
    \global\advance\ZoneBSize by -\TextLeading
    \ifFigInZoneB\relax
     \else
     \ifdim\TextSize<3\TextLeading
     \global\ZoneAFullPagetrue
     \fi
    \fi
    \ChangeStatus{\StackPointer}{\Zone}
    \ifnum\Zone=\InZoneC \global\TabInZoneCtrue\fi
   \fi
  \global\advance\NextTable \@ne
   \Print{This table has been placed.}
  \else
  \Print{No space available for this table...holding over.}
   \Print{}
   \global\MoreTablesfalse
  \fi
}


\def\CheckFitInZone{%
{\advance\TextSize by -\ItemSIZE
 \advance\TextSize by -\TextLeading
 \ifFirstSingleItem
  \advance\TextSize by \TextLeading
 \fi
 \ifnum\Zone=\InZoneA\relax
  \else \advance\TextSize by -\ZoneBAdjust
 \fi
 \ifdim\TextSize<3\TextLeading \global\ItemFits=\No
 \else \global\ItemFits=\Yes\fi}
}

\def\BeginOpening{%
  \thispagestyle{titlepage}%
  \global\setbox\ItemBOX=\vbox\bgroup%
    \hsize=\PageWidth%
    \hrule height \z@
    \ifsinglecol\vskip 6pt\fi 
}

\let\begintopmatter=\BeginOpening  

\def\EndOpening{%
  \One
  \egroup
  \ifsinglecol
    \box\ItemBOX%
    \vskip\TextLeading plus 2\TextLeading
    \@noafterindent
  \else
    \ItemNUMBER=\z@%
    \ItemTYPE=\Figure
    \ItemSPAN=\Double
    \ItemSTATUS=\InStack
    \JoinStack
  \fi
}


\newif\if@here  \@herefalse

\def\no@float{\global\@heretrue}
\let\nofloat=\relax 

\def\beginfigure{%
  \@ifstar{\global\@dfloattrue \@bfigure}{\global\@dfloatfalse \@bfigure}%
}

\def\@bfigure#1{%
  \par
  \if@dfloat
    \ItemSPAN=\Double
    \TEMPDIMEN=\PageWidth
  \else
    \ItemSPAN=\Single
    \TEMPDIMEN=\ColumnWidth
  \fi
  \ifsinglecol
    \TEMPDIMEN=\PageWidth
  \else
    \ItemSTATUS=\InStack
    \ItemNUMBER=#1%
    \ItemTYPE=\Figure
  \fi
  \bgroup
    \hsize=\TEMPDIMEN
    \global\setbox\ItemBOX=\vbox\bgroup
      \eightpoint\nostb@ls{10pt}%
      \let\caption=\fig@caption
      \ifsinglecol \let\nofloat=\no@float\fi
}

\def\fig@caption#1{%
  \vskip 5.5pt plus 6pt%
  \bgroup 
    \eightpoint\nostb@ls{10pt}%
    \setbox\TEMPBOX=\hbox{#1}%
    \ifdim\wd\TEMPBOX>\TEMPDIMEN
      \noindent \unhbox\TEMPBOX\par
    \else
      \hbox to \hsize{\hfil\unhbox\TEMPBOX\hfil}%
    \fi
  \egroup
}

\def\endfigure{%
  \par\egroup 
  \egroup
  \ifsinglecol
    \if@here \midinsert\global\@herefalse\else \topinsert\fi
      \unvbox\ItemBOX
    \endinsert
  \else
    \JoinStack
    \Print{Processing source for figure \the\ItemNUMBER}%
  \fi
}


\newbox\tab@cap@box
\def\tab@caption#1{\global\setbox\tab@cap@box=\hbox{#1\par}}

\newtoks\tab@txt@toks
\long\def\tab@txt#1{\global\tab@txt@toks={#1}\global\table@txttrue}

\newif\iftable@txt  \table@txtfalse
\newif\if@dfloat    \@dfloatfalse

\def\begintable{%
  \@ifstar{\global\@dfloattrue \@btable}{\global\@dfloatfalse \@btable}%
}

\def\@btable#1{%
  \par
  \if@dfloat
    \ItemSPAN=\Double
    \TEMPDIMEN=\PageWidth
  \else
    \ItemSPAN=\Single
    \TEMPDIMEN=\ColumnWidth
  \fi
  \ifsinglecol
    \TEMPDIMEN=\PageWidth
  \else
    \ItemSTATUS=\InStack
    \ItemNUMBER=#1%
    \ItemTYPE=\Table
  \fi
  \bgroup
    \eightpoint\nostb@ls{10pt}%
    \global\setbox\ItemBOX=\vbox\bgroup
      \let\caption=\tab@caption
      \let\tabletext=\tab@txt
      \ifsinglecol \let\nofloat=\no@float\fi
}

\def\endtable{%
  \par\egroup 
  \egroup
  \setbox\TEMPBOX=\hbox to \TEMPDIMEN{%
    \hss
    \vbox{%
      \hsize=\wd\ItemBOX
      \ifvoid\tab@cap@box
      \else
        \noindent\unhbox\tab@cap@box
        \vskip 5.5pt plus 6pt%
      \fi
      \box\ItemBOX
      \iftable@txt
        \vskip 10pt%
        \eightpoint\nostb@ls{10pt}%
        \noindent\the\tab@txt@toks
        \global\table@txtfalse
      \fi
    }%
    \hss
  }%
  \ifsinglecol
    \if@here \midinsert\global\@herefalse\else \topinsert\fi
      \box\TEMPBOX
    \endinsert
  \else
    \global\setbox\ItemBOX=\box\TEMPBOX
    \JoinStack
    \Print{Processing source for table \the\ItemNUMBER}%
  \fi
}

\def\UnloadZoneA{%
\FirstZoneAtrue
 \Iteration=\z@
  \loop
   \ifnum\Iteration<\LengthOfStack
    \GetItemSTATUS{\Iteration}
    \ifnum\ItemSTATUS=\InZoneA
     \GetItemBOX{\Iteration}
     \ifFirstZoneA \vbox to \BodgeHeight{\vfil}%
     \FirstZoneAfalse\fi
     \unvbox\ItemBOX\ItemSep
     \LeaveStack{\Iteration}
     \else
     \advance\Iteration \@ne
   \fi
 \repeat
}

\def\UnloadZoneC{%
\Iteration=\z@
  \loop
   \ifnum\Iteration<\LengthOfStack
    \GetItemSTATUS{\Iteration}
    \ifnum\ItemSTATUS=\InZoneC
     \GetItemBOX{\Iteration}
     \ItemSep\unvbox\ItemBOX
     \LeaveStack{\Iteration}
     \else
     \advance\Iteration \@ne
   \fi
 \repeat
}


\def\ShowItem#1{
  {\GetItemAll{#1}
  \Print{\the#1:
  {TYPE=\ifnum\ItemTYPE=\Figure Figure\else Table\fi}
  {NUMBER=\the\ItemNUMBER}
  {SPAN=\ifnum\ItemSPAN=\Single Single\else Double\fi}
  {SIZE=\the\ItemSIZE}}}
}

\def\ShowStack{%
 \Print{}
 \Print{LengthOfStack = \the\LengthOfStack}
 \ifnum\LengthOfStack=\z@ \Print{Stack is empty}\fi
 \Iteration=\z@
 \loop
 \ifnum\Iteration<\LengthOfStack
  \ShowItem{\Iteration}
  \advance\Iteration \@ne
 \repeat
}

\def\B#1#2{%
\hbox{\vrule\kern-0.4pt\vbox to #2{%
\hrule width #1\vfill\hrule}\kern-0.4pt\vrule}
}


\newif\ifsinglecol   \singlecolfalse

\def\onecolumn{%
  \global\output={\singlecoloutput}%
  \global\hsize=\PageWidth
  \global\vsize=\PageHeight
  \global\ColumnWidth=\hsize
  \global\TextLeading=12pt
  \global\Leading=12
  \global\singlecoltrue
  \global\let\onecolumn=\relax
  \global\let\footnote=\sing@footnote
  \global\let\vfootnote=\sing@vfootnote
  \ninepoint 
  \message{(Single column)}%
}

\def\singlecoloutput{%
  \shipout\vbox{\PageHead\pagebody\PageFoot}%
  \advancepageno
  \ifplate@page
    \shipout\vbox{%
      \sp@pagetrue
      \def\sp@type{plate}%
      \global\plate@pagefalse
      \PageHead\vbox to \PageHeight{\unvbox\plt@box\vfil}\PageFoot%
    }%
    \message{[plate]}%
    \advancepageno
  \fi
  \ifnum\outputpenalty>-\@MM \else\dosupereject\fi%
}

\def\ItemSep{\vskip\ItemSepamount\relax}

\def\ItemSepbreak{\par\ifdim\lastskip<\ItemSepamount
  \removelastskip\penalty-200\ItemSep\fi%
}


\let\@@endinsert=\endinsert 

\def\endinsert{\egroup 
  \if@mid \dimen@\ht\z@ \advance\dimen@\dp\z@ \advance\dimen@12\p@
    \advance\dimen@\pagetotal \advance\dimen@-\pageshrink
    \ifdim\dimen@>\pagegoal\@midfalse\p@gefalse\fi\fi
  \if@mid \ItemSep\box\z@\ItemSepbreak
  \else\insert\topins{\penalty100 
    \splittopskip\z@skip
    \splitmaxdepth\maxdimen \floatingpenalty\z@
    \ifp@ge \dimen@\dp\z@
    \vbox to\vsize{\unvbox\z@\kern-\dimen@}
    \else \box\z@\nobreak\ItemSep\fi}\fi\endgroup%
}


\def\gobbleone#1{}
\def\gobbletwo#1#2{}
\let\footnote=\gobbletwo 
\let\vfootnote=\gobbleone

\def\sing@footnote#1{\let\@sf\empty 
  \ifhmode\edef\@sf{\spacefactor\the\spacefactor}\/\fi
  \hbox{$^{\hbox{\eightpoint #1}}$}\@sf\sing@vfootnote{#1}%
}

\def\sing@vfootnote#1{\insert\footins\bgroup\eightpoint\b@ls{9pt}%
  \interlinepenalty\interfootnotelinepenalty
  \splittopskip\ht\strutbox 
  \splitmaxdepth\dp\strutbox \floatingpenalty\@MM
  \leftskip\z@skip \rightskip\z@skip \spaceskip\z@skip \xspaceskip\z@skip
  \noindent $^{\scriptstyle\hbox{#1}}$\hskip 4pt%
    \footstrut\futurelet\next\fo@t%
}

\def\footnoterule{\kern-3\p@ \hrule height \z@ \kern 3\p@}

\skip\footins=19.5pt plus 12pt minus 1pt
\count\footins=1000
\dimen\footins=\maxdimen


\def\landscape{%
  \global\TEMPDIMEN=\PageWidth
  \global\PageWidth=\PageHeight
  \global\PageHeight=\TEMPDIMEN
  \global\let\landscape=\relax
  \onecolumn
  \message{(landscape)}%
  \raggedbottom
}


\output{%
  \ifLeftCOL
    \global\setbox\LeftBOX=\vbox to \ZoneBSize{\box255\unvbox\ZoneBBOX}%
    \global\LeftCOLfalse
    \MakeRightCol
  \else
    \setbox\RightBOX=\vbox to \ZoneBSize{\box255\unvbox\ZoneBBOX}%
    \setbox\MidBOX=\hbox{\box\LeftBOX\hskip\ColumnGap\box\RightBOX}%
    \setbox\PageBOX=\vbox to \PageHeight{%
      \UnloadZoneA\box\MidBOX\UnloadZoneC}%
    \shipout\vbox{\PageHead\box\PageBOX\PageFoot}%
    \advancepageno
    \ifplate@page
      \shipout\vbox{%
        \sp@pagetrue
        \def\sp@type{plate}%
        \global\plate@pagefalse
        \PageHead\vbox to \PageHeight{\unvbox\plt@box\vfil}\PageFoot%
      }%
      \message{[plate]}%
      \advancepageno
    \fi
    \global\LeftCOLtrue
    \CleanStack
    \MakePage
  \fi
}


\Warn{\start@mess}

\def\mnmacrosloaded{} 

\catcode `\@=12 


\fi


\newcount\fignumber\fignumber=0\relax
\def\fign{\advance\fignumber by 1}

\def\figmod#1
{
	\ifnum#1>0

\input psfig

\def\putfig#1#2#3{\setbox20=\hbox
	{
	\psfig{file=#1,height=#2cm,clip=,angle=#3}
	}
	\centerline{$\vcenter{\box20}$}}

\def\putfigl#1#2#3{\setbox20=\hbox
	{
	\psfig{file=#1,height=#2cm,angle=#3}
	}
	\centerline{$\vcenter{\box20}$}}

	\else

		\input fignot

	\fi
}

\def\em{\ifdim\fontdimen1\font>\z@ \rm\else\it\fi}

\figmod{1}			


\pageoffset{-2.5pc}{0pc}

%
  
%
%
%

\Autonumber  


\pagerange{0-0}    
\pubyear{0000}
\volume{000}

\begintopmatter  

\title{Pure Luminosity Evolution models for faint field galaxy samples}
\author{L. Pozzetti$^{1,2}$, G. Bruzual A.$^{3,4}$, and G. Zamorani$^{2,5}$}
\affiliation{$^1$Dipartimento di Astronomia, Universit\`a di Bologna,
via Zamboni 33, I-40126 Bologna, Italy (lucia@astbo3.bo.astro.it)}
\smallskip
\affiliation{$^2$Osservatorio Astronomico di Bologna,
via Zamboni 33, I-40126 Bologna, Italy (zamorani@astbo3.bo.astro.it)}
\smallskip
\affiliation{$^3$Centro de Investigaciones de Astronom{\'\i}a, A.P. 264,
M\'erida 5101-A, Venezuela (bruzual@cida.ve)}
\smallskip
\affiliation{$^4$Landessternwarte Heidelberg K\"onigstuhl,
D-69117 Heidelberg, Germany}
\smallskip
\affiliation{$^5$Istituto di Radioastronomia del CNR,
via Gobetti 101, I-40129 Bologna, Italy}

\shortauthor{Pozzetti, Bruzual and Zamorani}
\shorttitle{PLE faint galaxy models}


\acceptedline{Accepted April 1996}

\abstract {
We have examined a set of pure luminosity evolution (PLE) models in order to
explore up to what extent the rapidly increasing observational constraints
from faint galaxy samples can be understood in this simple framework.
We find that a PLE model, in which galaxies evolve mildly in time
even in the rest frame UV, can reproduce most of the observed
properties of faint galaxies assuming an open ($\Omega\sim0$) universe.
In particular, such a model is able to fit reasonably well
the number counts in the $U,~b_j,~r_f,~I$, and $K$ bands, as
well as the colour and redshift distributions derived from most of the
existing samples. The most significant discrepancy between the predictions
of this model and the data is the $z$ distribution of faint $K$-selected
galaxies. Significantly worse fits are obtained with PLE models
for the theoretically attractive value of $\Omega = 1$, although a
simple number luminosity evolution model with a significant amount of
merger events fits the data also in this cosmology.
}

\keywords {
galaxies: evolution - galaxies: photometry - galaxies: redshifts - 
cosmology: miscellaneous.
}

\maketitle 

\section{Introduction}

An impressively large amount of observations of faint field galaxies has been
collected over the last fifteen years.
The determination of galaxy number counts, colour and redshift $(z)$
distributions
of increasingly deep samples is now part of the observational routine of
several established groups.
Since the review paper by Koo \& Kron (1992), galaxy photometric surveys have
been extended to $b_j = 27.5$ (Metcalfe et al. 1995a) and
to $K < 24 $ (Gardner, Cowie, \& Wainscoat 1993; Soifer et al. 1994;
Djorgovski et al. 1995), and spectroscopic surveys
to $K < 20 $ (Songaila et al. 1994), $I<22$
(Crampton et al. 1995) and to $B<24$ (Glazebrook et al. 1995a).
These data provide useful constraints for both cosmological models and
galaxy evolutionary models.

As soon as deep optical counts of galaxies became available, it was realized
that no-evolution (nE) models,
i.e. models in which the absolute brightness and the spectra of galaxies
do not change in time, predict a surface density of galaxies
at faint magnitude significantly lower than the observations
(see the review by Koo \& Kron 1992).
The excess in the observed number counts with respect to the nE model
predictions is of the order of $\sim$ 4 to 5 at $b_j \sim$ 24 and $\sim$
5 to 10 at $b_j \sim$ 26 (Maddox et al. 1990;
Guiderdoni \& Rocca-Volmerange 1991, hereafter GRV91).

Simple pure luminosity evolution (PLE) models, which allow for brightness
and spectral evolution in the galaxy population, were shown to provide a
better fit to the faint galaxy number
counts (Tinsley 1980; Bruzual \& Kron 1980 (hereafter BK80); Koo 1981,1985).
However, these early PLE models were ruled out
on the basis of comparisons with the results of $z$ surveys of faint galaxies
(Broadhurst, Ellis, \& Shanks 1988; Colless et al. 1990; Koo \& Kron 1992 and
references therein), which failed to reveal the large number of high $z$
galaxies predicted by the models.
The $z$ distributions in these surveys appeared to be in agreement
with the predictions of nE models.

In an attempt to explain all aspects of the data with a single model,
this apparent paradox (counts require PLE model but $z$ distribution only admits
nE model) prompted the development of a number of less straightforward models.
These models relax some of the assumptions adopted in nE and PLE models,
such as, the constancy of the volume number density of galaxies,
and/or the standard cosmology.
The excess observed in the faint counts can be explained, for example,
by either non-conservation of the number of galaxies due to
merger events (GRV91;
Broadhurst, Ellis, \& Glazebrook 1992; Carlberg \& Charlot 1992) or
dwarf galaxies which have faded and/or disappeared in recent epochs
(Broadhurst et al. 1988; Cowie, Songaila \& Hu 1991; Babul \& Rees 1992).
Fukugita et al. (1990) have proposed to revise the cosmological model
introducing a non-zero cosmological constant.

More recent and refined versions of PLE models have shown that
at least some of the discrepancies with existing data
can be substantially reduced also in the framework of these models
(Guiderdoni \& Rocca-Volmerange 1990 (hereafter GRV90);
Gronwall \& Koo 1995 (hereafter GK95);
Metcalfe et al. 1991,1995a).

The main goal of this paper is to explore up to what extent the properties
of the faint galaxy samples can be understood in the framework of PLE models,
without invoking the more exotic scenarios proposed in the literature.
We examine a new set of PLE models and compare their predictions
with the most recent data, including counts in all available photometric
bands, and colours and $z$ distributions as a function of magnitude for various
samples, selected in different bands.
A global comparison of this kind is essential because each different
data set provides constraints on different PLE model parameters.

We find that standard PLE models built along the lines described by BK80,
but with mild galaxy evolution, similar to that {\it assumed} by Metcalfe
et al. (1991, 1995a), but {\it derived} here consistently from the 
Bruzual \& Charlot (1993, hereafter BC93) spectral evolution models,
provide satisfactory fits to most of the observational data
in an $\Omega \sim 0$ universe.
The only data set which our models do not reproduce well is the $z$
distribution of the $K$-selected sample for $K > 17$ by Songaila et al. (1994).
Despite this failure, and before new samples either
confirm or modify the Songaila et al. results, we think that it is useful
to compare the faint galaxy surveys with the predictions of simple PLE models
in order to constrain the range of evolutionary and cosmological parameters
used in these models.

In \S 2 we describe the main ingredients of our PLE model,
i.e. the shape and normalization of the luminosity function of galaxies
of various morphological types and the adopted models for their
spectral evolution.
In \S 3 we compare the predictions of our models with the observational data.
The conclusions of our work are presented in \S 4.

\section{Standard PLE models}

Standard PLE models (BK80; Metcalfe et al. 1991) assume that galaxies
maintain at every $z$ the same proportion of the various morphological
types as observed locally ($z = 0)$,
and that the spectral evolution of these galaxies is well described by models
which at low $z$ reproduce the colours and $k$-corrections determined from
observed galaxy spectra.
Assuming a geometry for the universe, and scaling the galaxy luminosity
function (LF) to match the number counts at a given magnitude in a specific
band, one can compute the expected galaxy number counts in various bands,
as well as the colour and $z$ distributions as a function of magnitude.

\subsection{Modelling faint galaxy counts and colour and $z$ distributions}

In a homogeneous and isotropic universe, the number of galaxies of each type
brighter than a given magnitude $m$ can be calculated from the integral
$$N_i(<m)=\int_o^{z_{max}}\int_{M_{min}}^{M_{max}(m,z)}\phi_i(M) dM
{dV(z,\Omega)\over{dz}} dz,\eqno(1)$$
where $\phi_i(M)$ and $dV(z,\Omega)$ are, respectively, the local LF
for the type and the comoving volume element.
In this paper we will consider only the standard ($\Lambda=0$) Friedmann
cosmology defined by $H_0$ and $\Omega$.
We use $H_0 = 50$ km s$^{-1}$ Mpc$^{-1}$ throughout this paper.

The total number of galaxies brighter than $m$, $N(<m)$, is obtained by 
adding $N_i(<m)$ over all types considered. From (1) we obtain 
the $z$-distribution
$N_i(<m;z,z+dz)$ by integrating over the specific $z$ range ($z,z+dz$).
In (1) $z_{max}=min~(z_f,z_L)$, where $z_f$ is the assumed $z$ of galaxy
formation and $z_L$ is the value of $z$ at which the Lyman continuum break
is shifted to the effective wavelength of the filter being considered, and
the galaxy presumably becomes dark (Madau 1995).
We use the following filters: 
$U$ (Koo 1981), $B$ and $V$ (Buser 1978), $b_j$ and $r_f$
(Couch \& Newell 1980), $I$ and $K$ (Wainscoat \& Cowie 1992).
For the $U,~b_j,~r_f,~I$, and $K$ bands, $z_L \simeq $ 3, 4, 6, 8, and 23,
respectively.

The integration over $dM$ in (1) extends up to
$$M_{max}(m,z)=m-5 log {d_L(z;H_0,\Omega)\over{10}}-corr(z;H_0,\Omega,z_f),
\eqno(2)$$
where $d_L(z;H_0,\Omega)$ is the luminosity distance measured in pc,
and $corr(z;H_0,\Omega,z_f)$ is the {\it correction} needed to obtain the 
galaxy rest frame magnitude from its observer frame magnitude.
In the nE model, it is just the $k$-correction.
In the PLE model, it is given by the ($e+k$)-correction,
which also includes the effects due to the intrinsic galaxy luminosity
evolution.
We computed this correction up to $z=z_{max}$ from the synthetic spectral
energy distribution (SED) of the various galaxy types for each photometric
band.

Finally, the colour distribution $N_i(<m;c,c+dc)$ can be derived from the
$z$ distribution $N_i(<m;z,z+dz)$ for each type of galaxy by using the
cosmology dependent relation between colour and $z$, 
$c_i(z;H_0,\Omega,z_f)$, given by the adopted spectral evolution model.
The colour distributions for the different types are then added up.

\subsection{Luminosity function}

There is increasing evidence that the galaxy LF varies with galaxy type.
Bingelli, Sandage \& Tamman (1988) have derived morphology dependent LFs
for galaxies in the Virgo cluster.
Efstathiou, Ellis \& Peterson (1988), Shanks (1990) and Loveday et al. (1992)
have shown that the LF in the field is well described by the analytical
expression of Schechter (1976), with values of $\alpha$ and $M^*$ which depend
on the galaxy morphological type.
In general, blue galaxies show a steeper slope than red galaxies.

%
\begintable*{1}
\nofloat
\caption{{\bf Table 1.} Local galaxy mix, LF parameters$^a$, and
local colors$^b$}
\halign{%
#\hfil & \hfil#\hfil & \hfil#\hfil & ~~\hfil#\hfil
& ~~\hfil#\hfil & \hfil#\hfil & \hfil#\hfil & \hfil#\hfil & \hfil#\hfil
& \hfil#\hfil \cr
\noalign{\vskip 3pt}\noalign{\hrule}\cr\noalign{\vskip 3pt}
Type~~~ & ~~Fraction$^c$~~ & ~~$\alpha$~~ & ~~$M^*_{b_j}$~~ & ~~$\Phi_i^*$~~
& ~$B-V$~ & ~$U-b_j$~ & ~$b_j-r_f$~ & ~$b_j-I$~ & ~$b_j-K$~ \cr
\noalign{\vskip 3pt}\noalign{\hrule}\cr\noalign{\vskip 3pt}
E/S0     & 0.28 & -0.48 & -20.87 & 0.95 & 0.95 & 0.75  & 1.55 & 2.32 & 4.13 \cr
Sab-Sbc  & 0.47 & -1.24 & -21.14 & 1.15 & 0.68 & 0.32  & 1.21 & 1.83 & 3.47 \cr
Scd-Sdm  & 0.22 & -1.24 & -21.14 & 0.54 & 0.43 & -0.05 & 0.88 & 1.39 & 2.96 \cr
very Blue (vB)  & 0.03 & -1.24 & -21.14 & 0.12 & 0.07 & -0.49 & 0.32 & 0.63 & 2.06 \cr
\noalign{\vskip 3pt}\noalign{\hrule}\cr
}
\tabletext{\noindent
$^a$ $\alpha$, $M^*_{b_j}$ from Efstathiou et al. (1988);
$\Phi_i^*$ in units of $10^{-3}$ Mpc$^{-3}$ (\S 2.3).
$H_0=50$ km s$^{-1}$ Mpc$^{-1}$ throughout this paper;
\par\noindent $^b$ From BC93 ($t_g=16$ Gyr, $z_f = 4.5$, $\Omega = 0$);
\par\noindent $^c$ Mix from Ellis (1983).}
\endtable

We adopt the values of $\alpha$ and $M^*_{b_j}$ derived separately
for early and late type galaxies by Efstathiou et al. (1988) from
the Anglo-Australian Redshift Survey data, listed in Table 1.
We assume that these parameters are valid from
$M_{b_j}=-24$ to $M_{b_j}=-15.5$ (cf. Loveday et al. 1992). To obtain the
local LF in the $U,~r_f,~I$, and $K$ bands we shift $M^*_{b_j}$ according
to the colours at $z=0$ for each galaxy type listed in Table 1.
The $K$ band LF thus obtained is consistent with the
observed one (Mobasher et al. 1995; Glazebrook et al. 1995b).

We have not used the more recent determination of $\alpha$ and $M^*_{b_j}$
by Loveday et al. (1992), because of the bias mentioned by these authors
against identifying early-type galaxies in their sample.
As shown by Zucca, Pozzetti \& Zamorani (1994), the correction for this bias
moves the Loveday et al. values of ($\alpha$, $M^*$) towards those of
Efstathiou et al. (1988).

\subsection{Count normalization}

The uncertainties in the normalization of the local galaxy LF translate
into a major source of uncertainty in the models for faint galaxy counts.
Moreover, different selection effects in photographic and CCD data
may lead to difficulties in comparing bright and deep surveys (McGaugh 1994).
The reduced size of well calibrated samples at bright magnitude and the local
fluctuations caused by galaxy clustering lead to large variations in the
bright counts.
In principle, this problem would be avoided by using the APM galaxy counts
in the range $15 < b_j < 20.5$ derived from 4300 deg$^2$ of the sky
(Maddox et al. 1990). However, while these counts are close to the
mean of previously published data at $b_j \sim 19-20$, they
show a steeper slope for $b_j<19$ than previous determinations.
Thus, the APM counts at $b_j < 17$ are below models normalized to
match the counts at $b_j = 19-20$.
Metcalfe, Fong \& Shanks (1995b) show evidence of a possible magnitude
scale error in the APM galaxy survey, whose size could be
large enough to cause the apparent disagreement between the APM galaxy counts
and the predictions from standard models in the range $17.5 < b_j < 20$.

Because of this uncertainty in the counts at bright magnitude,
we have chosen to scale our model predictions
to the observed number counts in the range $19<b_j<19.5$,
namely $log~N_0=1.98$ gal deg$^{-2}$ (0.5 mag)$^{-1}$ (see also
GRV90). This normalization is consistent with that obtained recently 
for the $K$ band LF (Glazebrook et al. 1995b), as well as for the $b_j$ 
band (Ellis et al. 1995).

Finally, in order to compute $\Phi^*_i$, i.e. the normalization
of the LF for each morphological type,
we have adopted the local galaxy mix derived by Ellis (1983) for
$b_j<16.5$ from the DARS data (see column 2 in Table 1). This galaxy mix
is in good agreement with the percentage of elliptical and spiral
galaxies derived recently by Efstathiou et al. (1988).
The same values of $\Phi^*_i$ are used to model different photometric bands.
The $b_j$ LF summed over all types derived from our models is well described
by the parameters $\alpha= -1.2,~M^*_{b_j}=-21.1$,
$\Phi^*_{ALL}=2.5 \times 10^{-3}$ Mpc$^{-3}$.

\subsection {Spectral energy distributions}

Our galaxy spectral energy distributions (SEDs)
are based on the BC93 galaxy spectral evolution library.
The BC93 models are built from a library of stellar tracks
which includes all evolutionary stages for stars of solar metallicity.
Empirical near-UV to near-IR spectra of galactic stars, extended to the far-UV
by means of model atmospheres, are used in the synthesis.
The inclusion in these models of the Post-AGB stellar evolutionary phase,
which contributes significantly in the UV spectral range in old
stellar populations (Magris \& Bruzual 1993),
represents an improvement over previous spectral evolution codes.

We used the following procedure to select a subset of galaxy models from
the BC93 library.
First, we selected models on the basis of their ability to reproduce the
shape of the continuum and spectral features in the SEDs of local galaxies
of different morphological types, and verified that the $k$-corrections
computed from the models were in agreement with the empirical ones.
Then, among these models we selected by trial and error those which
reproduced better the empirical constraints on galaxy evolutionary 
properties, namely, galaxy number counts and colour and $z$ distributions.

Thus, we have to compute the number count model as outlined in \S2.1
in order to find which galaxy spectral evolution models produce the
best fit to the counts and colour and $z$ distributions.
The feedback in this procedure is unavoidable due to the lack of independent
constraints on galaxy evolution (BK80, GK95).

%
\begintable{2}
\caption{{\bf Table 2.} Initial mass function$^a$}
\halign{%
#\hfil & \hfil# & \hfil# & \hfil# & # \cr
\noalign{\vskip 3pt}\noalign{\hrule}\cr\noalign{\vskip 3pt}
~IMF & $x_i$ & $m_1$ & $m_2$ & $~~c_i$  \cr
\noalign{\vskip 3pt}\noalign{\hrule}\cr\noalign{\vskip 3pt}
\noalign{\vskip 4pt}
Scalo (1986)    & -2.60  &  0.10 &   0.18 & 100.530 \cr
                &  0.01  &  0.18 &   0.42 & 1.14430 \cr
                &  1.75  &  0.42 &   0.62 & 0.25293 \cr
                &  1.08  &  0.62 &   1.18 & 0.34842 \cr
                &  2.50  &  1.18 &   3.50 & 0.44073 \cr
                &  1.63  &  3.50 &    125 & 0.14819 \cr
\noalign{\vskip 4pt}
Salpeter (1955) &  1.35  &  0.10 &    125 & 0.17038 \cr
\noalign{\vskip 3pt}\noalign{\hrule}\cr
}
\tabletext{\noindent $^a f_i(m) = c_i m^{-(1+x_i)}$ for $m_1 \le m \le m_2$,
$m_1$ and $m_2$ in M$_\odot$ units.}
\endtable

\subsubsection {Evolutionary constraints on the SEDs}

The following aspects of the data guided our choice of the SEDs.
Strong evolution for $z<1$ is ruled out for all galaxy types
by the observed $z$ distributions
of Broadhurst et al. (1988) and Colless et al. (1990, 1993), which are close
to the nE prediction up to $b_j=22.5$.
Colless et al. (1993) estimate an upper limit to
the amount of luminosity evolution of $\Delta M_B \sim -1.2 $ for
$z<1$, providing a strong constraint to the evolutionary corrections.
The observed number counts, particularly in the $b_j$ band, require
optically mildly-evolving SEDs, over the entire $z$ range,
in agreement with Koo, Gronwall, \& Bruzual (1993, hereafter KGB93).
Additional evidence in favor of mild evolution derives from recent work on the
$z$ evolution of the LF in $B$-, $I$-, and $K$-selected samples by
Colless (1995), Lilly et al. (1995b), and Glazebrook et al. (1995b),
respectively.

We introduce the spectral class of very Blue (vB) galaxies in order to
explain the bluest colours, $b_j-r_f \la 0.3$,
observed in the deepest $b_j$-selected surveys (Glazebrook et al. 1995a).
vB galaxies are meant to reproduce a population of starburst galaxies
present at each $z$, whose evolution does not follow the pure luminosity
prescription.
We assume that star formation in these galaxies keeps their SED constant 
in time.
For vB galaxies we adopt the LF of spiral galaxies.
The observed fraction of galaxies bluer than $B-V=0.6$ at the faint magnitude
limit ($\sim10\%$), is reproduced assuming that vB galaxies represent
$\sim 3\%$ of the total local mix (Table 1).
Thus, at the few \% level, our PLE models include a fraction
of non-passively evolving galaxies.

%
\begintable{3}
\nofloat
\caption{{\bf Table 3.} Galaxy SED model parameters}
\halign{%
\hfil#\hfil & ~~#\hfil & ~~#\hfil & ~~#\hfil & ~~#\hfil \cr
\noalign{\vskip 3pt}\noalign{\hrule}\cr\noalign{\vskip 3pt}
$\Omega$ & Type & SFR & IMF & $t_g$ (Gyr) \cr
\noalign{\vskip 3pt}\noalign{\hrule}\cr\noalign{\vskip 3pt}
0 & E/S0     & $\tau_1,\tau_2$ & Scalo    & 16 \cr
  & Sab-Sbc  & $\tau_{10}$     & Scalo    & 16 \cr
  & Scd-Sdm  & cons            & Salpeter & 16 \cr
  & vB       & cons            & Salpeter & 0.1 \cr
\noalign{\vskip 3pt}\cr
1 & E/S0     & $B_1,\tau_1$    & Scalo    & 12.7 \cr
  & Sab-Sbc  & $\tau_8$        & Scalo    & 12.7 \cr
  & Scd-Sdm  & cons            & Salpeter & 12.7 \cr
  & vB       & cons            & Salpeter & 0.1 \cr
\noalign{\vskip 3pt}\noalign{\hrule}\cr
}
\endtable

%
\fign\beginfigure{\fignumber}
\putfigl{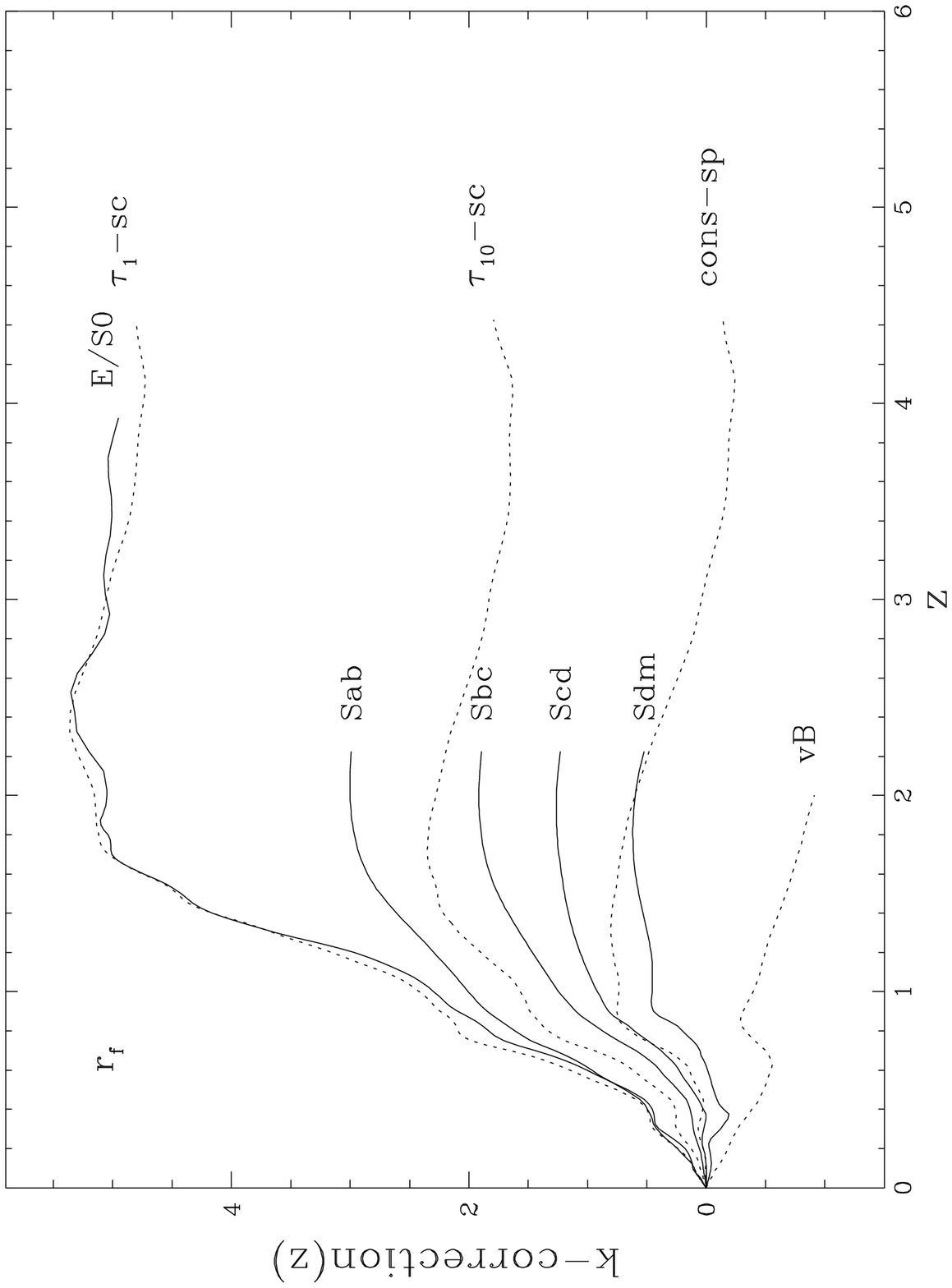}{6}{270}
\caption
{
{\bf Figure 1.}
$k$-correction in the $r_f$ band for galaxies of different morphological
types from BC93 models (dotted lines, {\it sc:} Scalo IMF; {\it sp:} Salpeter
IMF) and from the observed SEDs described in the text (solid lines).
}
\endfigure

\section{Results}

\subsection{Reference models}

A BC93 galaxy evolution model is specified by the star formation rate (SFR)
$\psi(t)$ and the initial mass function (IMF).
Once $H_0,~\Omega$, and $z_f$ are specified, the age $t_g$ of the model SEDs
and their observer frame properties are fixed.
We considered models with $t_g = 16$ and 12.7 Gyr, corresponding
to $z_f = 4.5$ and 10 for $\Omega = 0$ and 1, respectively.
We tested models computed for the Salpeter (1955) and the Scalo (1986) IMFs
(Table 2).
Table 3 defines the BC93 model SEDs selected as described in \S2.4 that
will be used to build our reference models for the counts as indicated in \S2.1.
A brief comment on these SEDs follows.

Models for E/S0 and Sab-Sbc galaxies are characterized by the SFR
$\psi(t) \propto \tau^{-1} ~ \exp(-t/\tau)$,
where $\tau$ is the e-folding time characterizing this form of $\psi(t)$.
The spectral properties of nearby E/S0 galaxies
are reproduced well by both the $\tau=1$ Gyr model (hereafter $\tau_1$ model)
and a model in which star formation takes
place at a constant rate during the first Gyr in the life of the galaxy
(hereafter $B_1$ model) for either the Salpeter or
the Scalo IMF, and by the
$\tau=2$ Gyr model (or $\tau_2$ model) for the Scalo IMF.
The observed $b_j-r_f$ and $B-K$ colour distributions are reproduced more
closely by our $\Omega = 0$ reference PLE model if we represent the E/S0
galaxy SEDs by the $\tau_1$ and $\tau_2$ models, rather than by the $B_1$ model
(Table 3).

The local properties of Sab-Sbc galaxies are well described by the
$\tau=4$ Gyr (or $\tau_4$) Salpeter IMF model, or the $\tau=10$ Gyr
(or $\tau_{10}$) Scalo IMF model. The Scd-Sdm galaxies are described
satisfactorily by a model in which stars form at a constant rate following
the Salpeter IMF (hereafter $cons$ model).
The SED of the vB galaxies can be approximated by a model with constant
SFR seen at an age of 0.1 Gyr after the last major event of star formation.
This SED is even bluer than that of NGC 4449 (BC93).

\subsubsection{Scalo IMF vs. Salpeter IMF}

The constraints mentioned in \S2.4 led us to adopt Scalo IMF models for
early-type galaxies (Table 3).
The Scalo IMF is less rich in massive stars than the Salpeter IMF
because of the steeper slope of the former at the high mass end.
The high number of massive stars in the Salpeter IMF models produces
a large amount of UV flux at early times, making high $z$ E galaxies detectable
in current deep surveys and producing an excess in the faint $b_j$ counts
with respect to the observations.
An alternative way to reduce the UV flux is to assume a significant
amount of dust extinction inside the galaxies (GK95).

There is some ambiguity in the literature on the functional form of the
Scalo IMF. Following a suggestion by D.C. Koo, BC93 have divided the IMF given
by Scalo in graphical form into the 6 different segments listed in Table 2.
For the $i^{th}$-segment the IMF is given by $f_i(m)=c_i m^{-(1+x_i)}$.
The constants $c_i$ are computed from the requirement of continuity of the
IMF (cf. Guiderdoni \& Rocca-Volmerange 1987, hereafter GRV87).

%
\begintable{4}
\nofloat
\caption{{\bf Table 4.} Intrinsic luminosity evolution}
\halign{%
#\hfil & \hfil#\hfil & \hfil#\hfil & \hfil#\hfil & \hfil#\hfil \cr
\noalign{\vskip 3pt}\noalign{\hrule}\cr\noalign{\vskip 3pt}
$\lambda_0^a$ & 2300\AA & 1150\AA & 11000\AA & 5500\AA \cr
Type~~~ & ~$\Delta B_{z=1}$~ & ~$\Delta B_{z=3}$~ & ~$\Delta K_{z=1}$~ &
 ~$\Delta K_{z=3}$~ \cr
\noalign{\vskip 3pt}\noalign{\hrule}\cr\noalign{\vskip 3pt}
E/S0     & -1.6     & -6.0     & -0.6     & -2.3     \cr
Sab-Sbc  & -1.0     & -1.6     & $\sim 0$ & $\sim 0$ \cr
Scd-Sdm  & $\sim 0$ & $\sim 0$ & +0.4     & +0.6     \cr
\noalign{\vskip 3pt}\noalign{\hrule}\cr
}
\tabletext{\noindent $^a$ Rest frame wavelength}
\endtable

\subsection {$k$ and $(e+k)$-corrections}

Fig 1 shows the model and the empirical $k$-corrections in the $r_f$ band.
These models, listed under $\Omega = 0$ in Table 3, reproduce the flattening at
high $z$ shown by the empirical $k$-corrections of ellipticals and spirals
(cf. Cowie et al. 1994).
In contrast, the $k$-corrections of ellipticals modeled by
Rocca-Volmerange \& Guiderdoni (1988) continue to increase with $z$.
This difference is mainly due to the lower UV flux in the GRV87
model for E galaxies, resulting from the lack of
Post-AGB stars in their population synthesis.

The $k$-corrections derived from the synthetic spectra are in agreement with
the $k$-corrections in the $(U,b_j,r_f,I)$ bands computed from the observed
spectra of Pence (1976) up to $z=(0.6,0.9,2.2,2.5)$,
with the compilation in $b_j$ by King \& Ellis (1984) up to $z=1.5$, and with
the $K$ and $B$ band estimates by Cowie et al. (1994) up to $z < 3$.
Our E/S0 models reproduce also the $k$-correction at higher $z$
computed from the average observed E galaxy SED of BC93.

%
\fign\beginfigure{\fignumber}
\putfigl{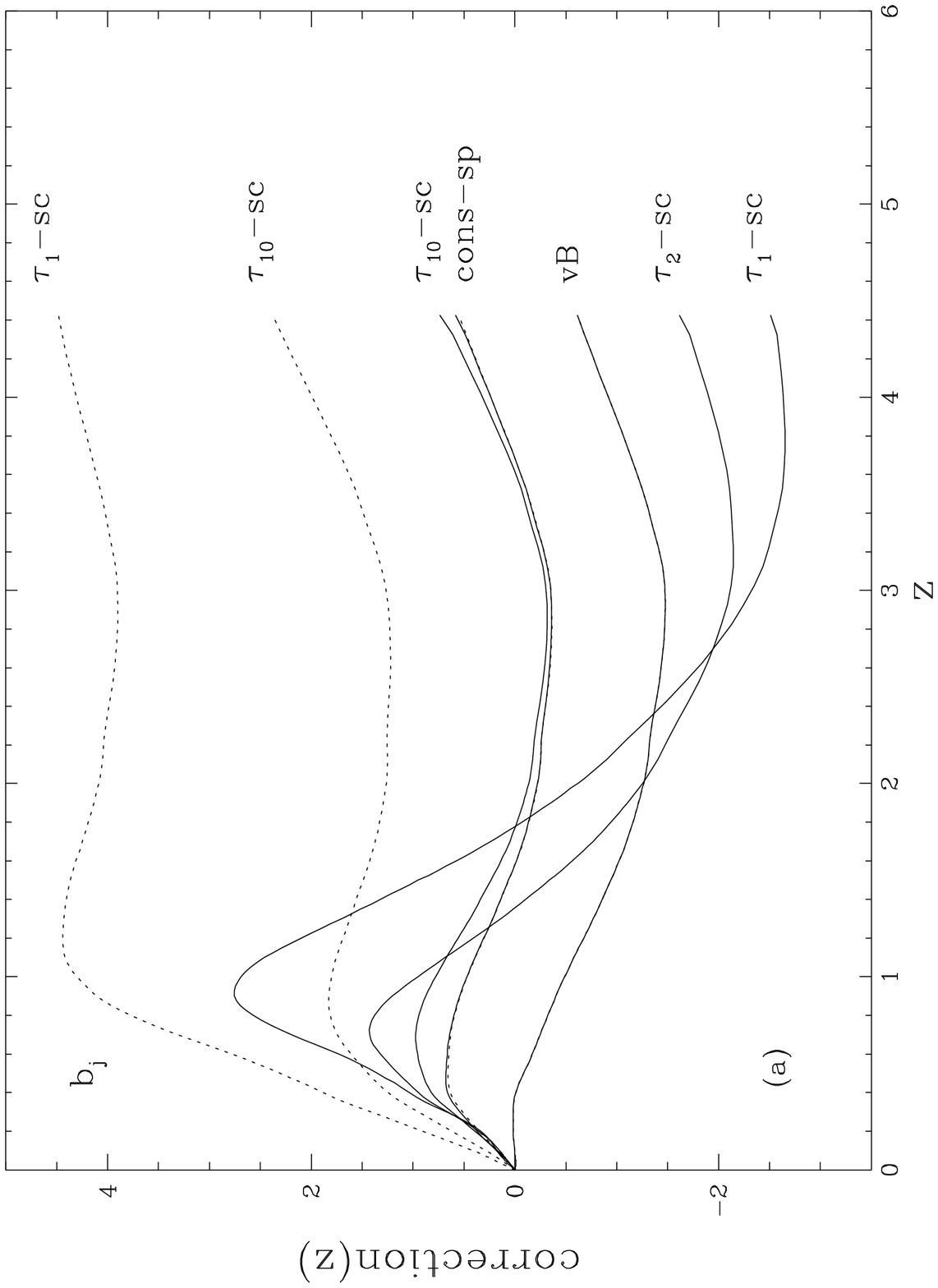}{6}{270}
\putfigl{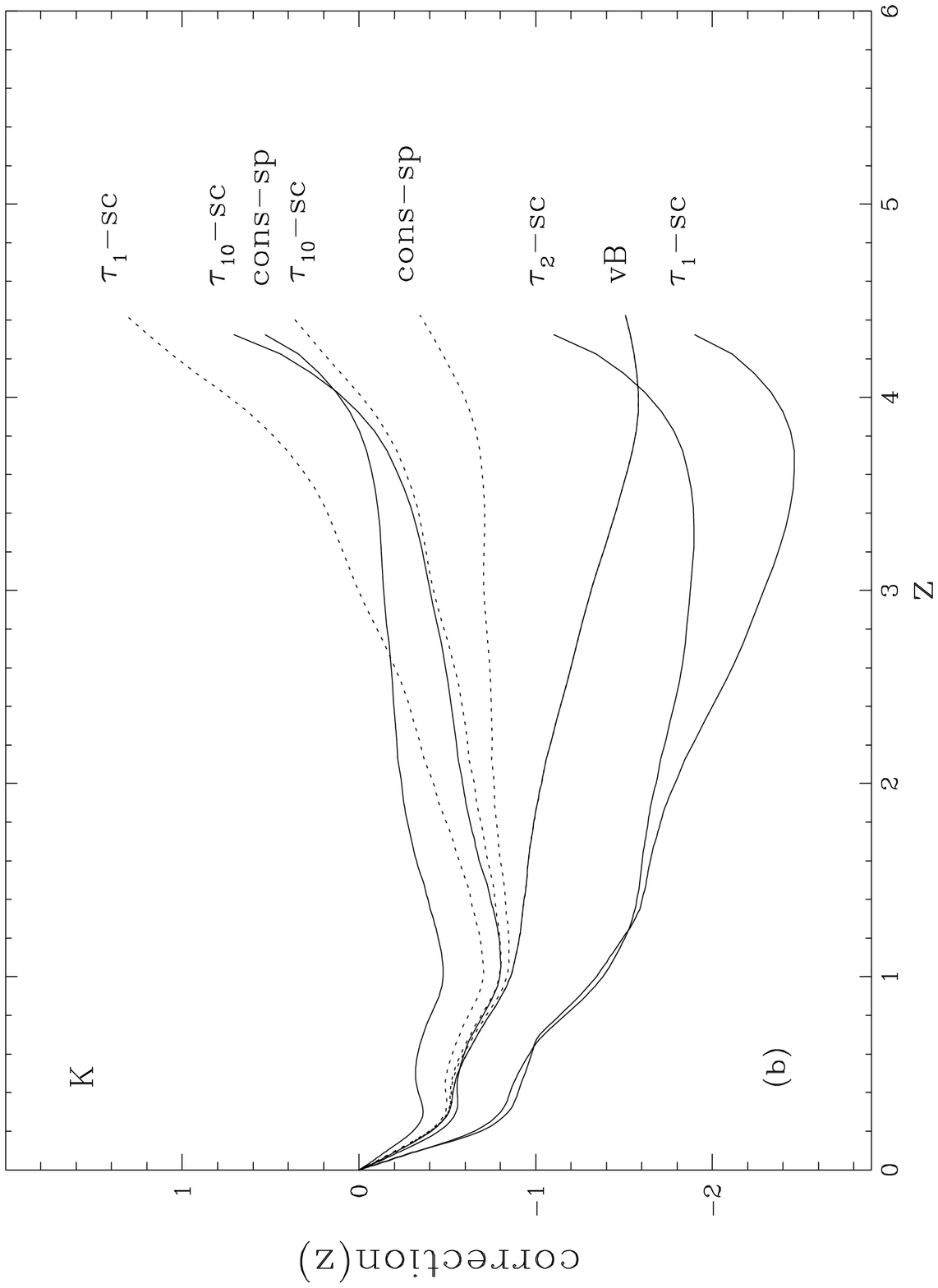}{6}{270}
\caption
{
{\bf Figure 2.}
{\it (a)} $b_j$ band $(e+k)$-corrections (solid lines) and $k$-corrections
(dotted lines) for galaxies of different morphological types derived from the
BC93 models listed under $\Omega = 0$ in Table 3.
{\it (b)} Same as {\it (a)} but for the $K$ band.
}
\endfigure

%
\begintable{5}
\nofloat
\caption{{\bf Table 5.} Slope $\gamma$ at faint magnitudes}
\halign{%
#\hfil & \hfil#\hfil & \hfil#\hfil & \hfil#\hfil & \hfil#\hfil \cr
\noalign{\vskip 3pt}\noalign{\hrule}\cr\noalign{\vskip 3pt}
Band~~~ & ~~mag. range ~~ & ~~$\gamma_{obs}$~~ & ~~$\gamma_{\rm PLE}^a$~~
& ~~$\gamma_f^b$~~ \cr
\noalign{\vskip 3pt}\noalign{\hrule}\cr\noalign{\vskip 3pt}
$U$     & 20--25 & 0.49 & 0.49 & 0.22 \cr
$b_j$   & 20--25 & 0.45 & 0.46 & 0.30 \cr
$r_f$   & 20--25 & 0.37 & 0.36 & 0.21 \cr
$I$     & 19--23 & 0.34 & 0.35 & 0.17 \cr
$K$     & 18--23 & 0.26-0.30 & 0.29 & 0.10 \cr
\noalign{\vskip 3pt}\noalign{\hrule}\cr
}
\tabletext{\noindent $^a$ From the $\Omega=0$ reference PLE model
\par\noindent $^b$ Predicted slope at faint limits: $24 < m < 27$}
\endtable

Figures 2a and 2b show the $k$ and $(e+k)$ corrections in the $b_j$ and $K$
bands, respectively, for the SEDs selected in the $\Omega = 0$ case (Table 3).
We see here the flattening of the ($e+k$)-corrections
in $b_j$ and $r_f$ at high $z$, which was assumed entirely {\it ad hoc} by
Metcalfe et al. (1991, 1995a).
If we use Salpeter IMF models for the Ellipticals, the ($e+k$)-correction
in $b_j$ continues to decrease with $z$, up to $\sim-3(-4)$ at $z=2(3)$.

The difference between the $(e+k)$ and $k$-corrections gives the intrinsic
galaxy luminosity evolution, or $\Delta M_z = e$-correction$(z)$.
This quantity represents the luminosity brightening of a galaxy at
$\lambda_0 = \lambda_{obs} \times (1+z)^{-1}$
with respect to its $z = 0$ luminosity at the same wavelength.
The $e$-corrections at $z=1$ and 3, as observed in the 
$B$ and $K$ bands are given in Table 4.
The amount of brightening up to $z=1$ is relatively small,
as required by the observed $z$ distributions.
For $z \sim 3$, $\Delta B_z \sim -6$ for E/S0 galaxies and
$\Delta B_z \sim -1.6$ for Sab-Sbc galaxies.
In the $K$ band the luminosity evolution is significantly lower than in the
other bands ($\Delta K_{z=1} \sim -0.6$ and $\Delta K_{z=3} \sim -2.3$
for E/S0 galaxies), showing how the concept of mild luminosity evolution
depends on the wavelength range being considered.

\subsection{Number counts}

Optical counts have reached very deep levels thanks to CCD cameras.
IR array detectors have made possible to extend the faint
galaxy work to the $K$ band.
Observations in the $K$ band are particularly important because
light at $2 \mu m$ is relatively insensitive to dust extinction.
At high $z$ the $K$ band samples the well understood rest
frame optical range, in which spectral evolution is less significant than
in the UV region sampled at high $z$ by the optical bands (Table 4).
$K$ band counts are thus less sensitive than optical counts to the
evolution of the stellar population and to the details of the SFR and IMF.
$K$ counts can therefore be used, at least in principle, as a direct test of
cosmological models.
Because of these reasons in this paper we pay special attention to fitting
at the same time all known properties of faint galaxy samples,
from the $U$ to the $K$ band.

%
\fign\beginfigure*{\fignumber}
\psfig{file=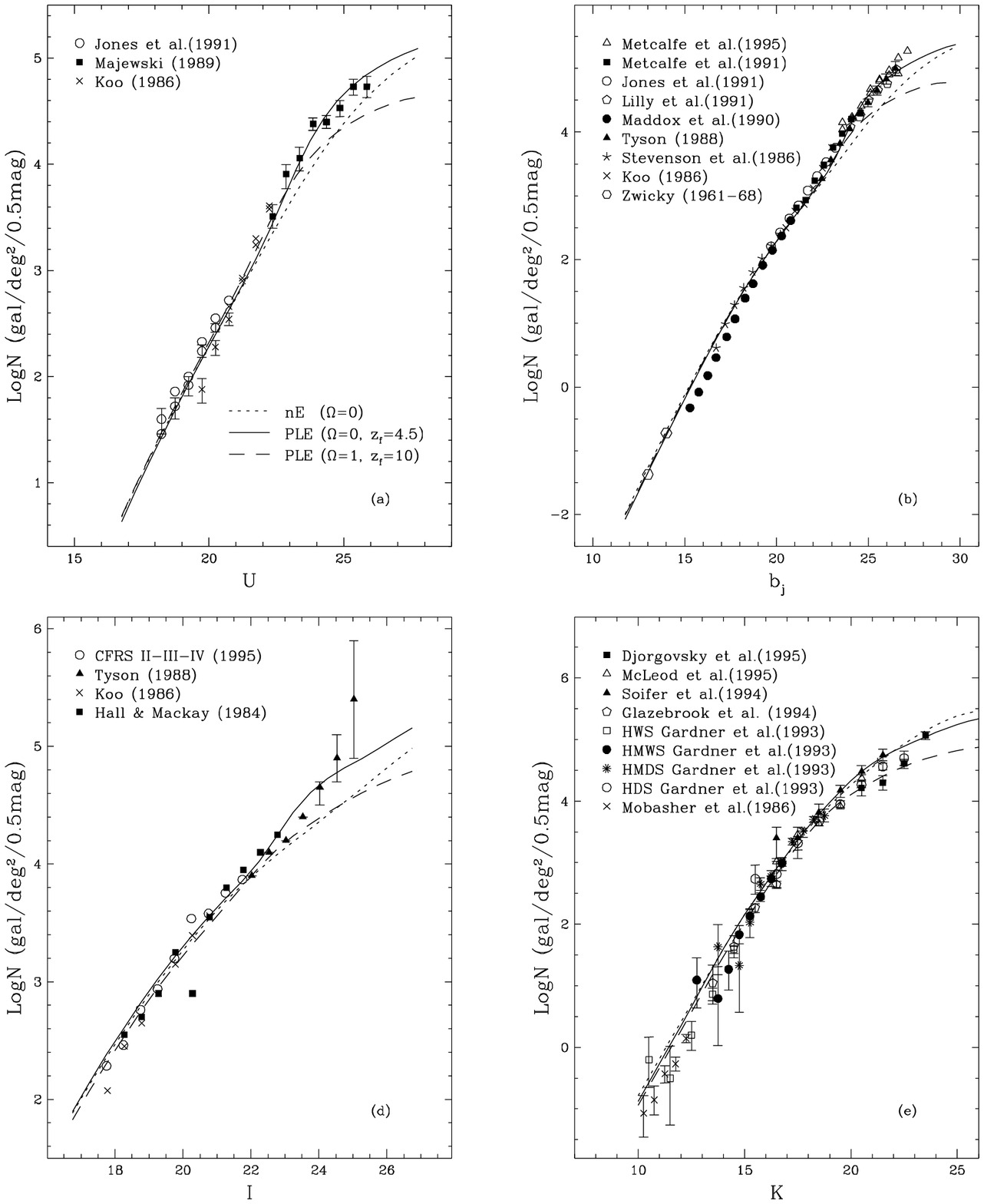,height=16cm,width=13cm,clip=,angle=0}
\caption
{
{\bf Figure 3.}
Differential galaxy number counts per square degree per half
magnitude interval as a function of apparent magnitude.
The sources of the observed data points are indicated in each panel.
The lines show the predicted counts for different models:
dotted line: nE model ($\Omega = 0,\ z_f = 4.5$);
solid line: PLE model ($\Omega = 0,\ z_f = 4.5$);
dashed line: PLE model ($\Omega = 1,\ z_f = 10$).
{\it (a)} $U$ band.
{\it (b)} $b_j$ band.
{\it (c)} $r_f$ band.
{\it (d)} $I$ band.
{\it (e)} $K$ band.
}
\endfigure

Figures 3a-e show the differential number counts in the $(U,b_j,r_f,I,K)$ 
bands. As in most observational papers, from which the data have been taken,
$N(m)$ is plotted per half magnitude bin.
The sources of the data points are indicated in the figures.
Even though in a few cases the counts from different groups show a
relatively large scatter, the slope $\gamma$ of the $log~N - m$
relation is reasonably well defined in all bands.
At bright magnitude $\gamma\sim 0.6$ is close to the Euclidean value.
Table 5 shows that at fainter magnitude $\gamma$ decreases with increasing
filter effective wavelength (Jones et al. 1991; Gardner et al. 1993;
Djorgovski et al. 1995).
Despite the large difference in $\gamma$ between $b_j$ and $K$,
the two bands are sampling the same galaxy population.
We verified that the observed $b_j$ counts can be reproduced (in number
and $\gamma$) by just shifting the observed $K$ band counts according to
the $(B-K)_{med}$ at a given $K$ given by Gardner et al. (1993).
The lines in these figures represent the predictions of the models that are
described in detail below.

\subsubsection{nE model}

The well known excess of faint galaxies above the nE model prediction
is confirmed (Fig 3).
In nE models the SEDs of distant galaxies are represented by SEDs that match
those of nearby galaxies (Table 3),
which correspond, on average, to old stellar populations.
The amount of UV flux produced by Post-AGB stars in E/S0 galaxies
is not as large as that attained at early ages.
The large $k$-correction moves these galaxies towards fainter magnitude
and, hence, they do not contribute to the predicted counts
even at the faint limit reached by current observations.

%
\fign\beginfigure{\fignumber}
\putfig{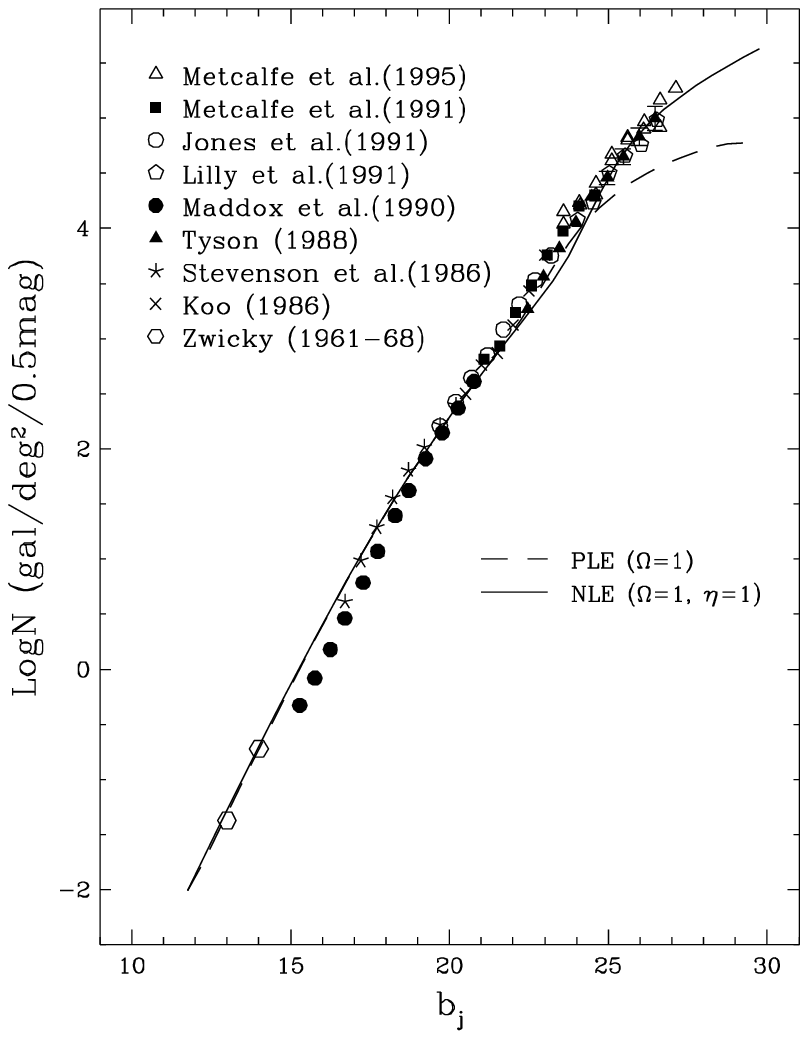}{8}{0}
\caption
{
{\bf Figure 4.}
Differential galaxy number counts as a function of $b_j$ apparent magnitude.
The lines show the predicted counts for two different models:
solid line: NLE model ($\Omega = 1,\ z_f = 10,\ \eta = 1$);
dashed line: PLE model ($\Omega = 1,\ z_f = 10$).
}
\endfigure
%
\fign\beginfigure*{\fignumber}
\nofloat
\putfig{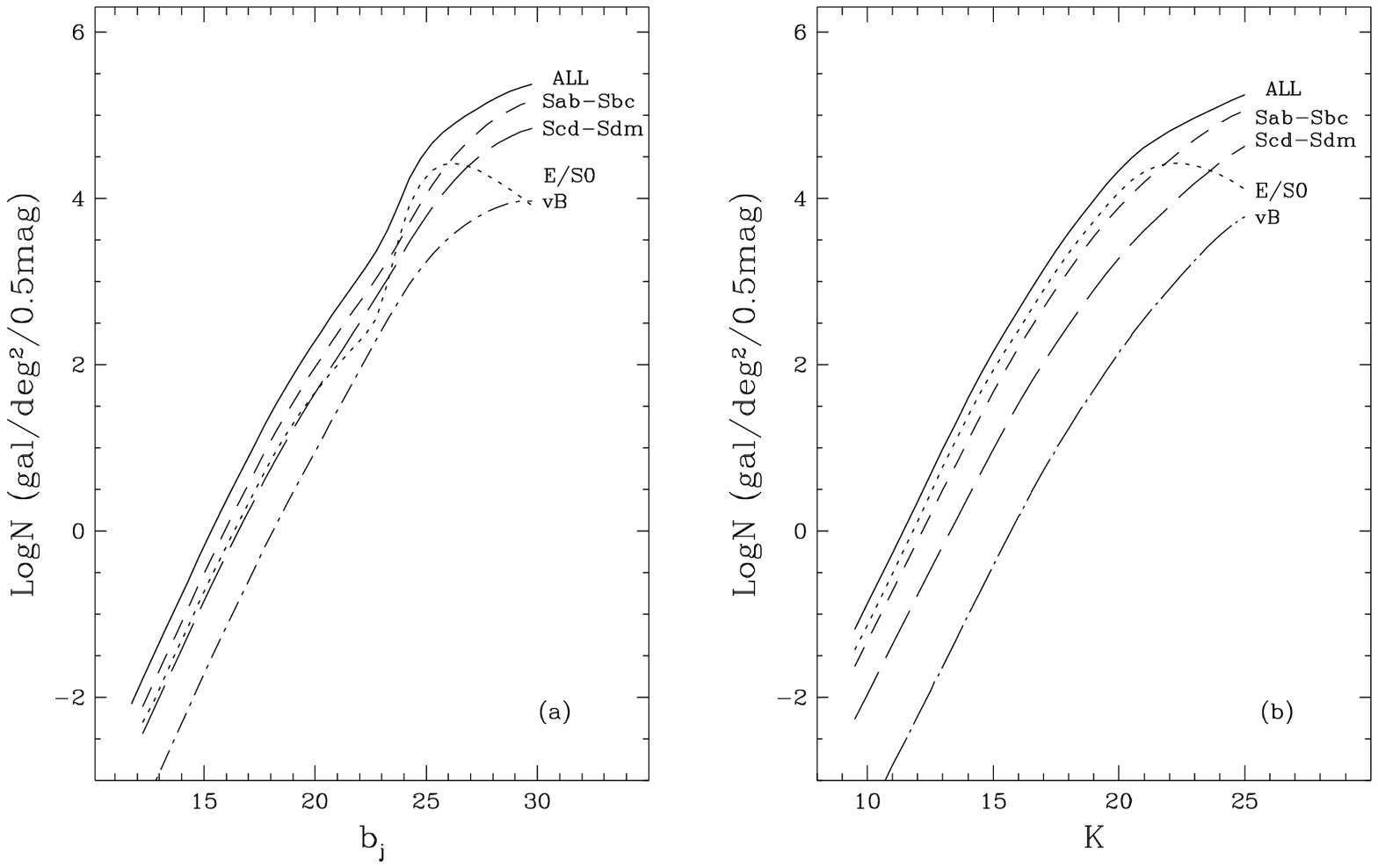}{8.0}{0}
\caption
{
{\bf Figure 5.}
Contribution of each morphological type to the total
differential galaxy number counts as a function of apparent magnitude.
The lines show the predicted counts for the reference $\Omega = 0$ PLE model.
Total counts: solid line;
E/S0: dotted line;
Sab-Sbc: short-dashed line;
Scd-Sdm: long-dashed line;
vB: dot-dashed line.
{\it (a)} $b_j$ band.
{\it (b)} $K$ band.
}
\endfigure

The excess of galaxies with respect to our nE model (with $\Omega=0$)
is about a factor of
2 at $m\sim24$ in all bands, except for the $K$ band in which the nE model is
consistent with the data. Even at fainter magnitude the observed data are
never more than a factor of three higher than the nE model.
The excess reported by other authors (Broadhurst et al. 1988; Maddox et al.
1990; GRV90) is larger by a factor $\sim$ 2 than our result.
This difference is mainly due to our choice of normalizing the model counts
to the observed number in the range $19.0 < b_j < 19.5$ (\S 2.3),
and to our selection of realistic SEDs for early-type galaxies.
Our SEDs reproduce reasonably well the UV flux of local galaxies, which
translates into significant differences between our $k$-corrections at high
$z$ and those used in the quoted papers.
On the other hand, our results are in reasonable agreement with the nE model
of KGB93, supporting their conclusion that only a moderate amount of spectral
evolution is required by the data.

In an $\Omega=1$ universe the difference between data and model (not shown)
at faint magnitude is substantially higher, being a factor of $\sim 3(4)$ at
$b_j\sim 24(26)$.

\subsubsection{PLE models}

{From} Figures 3a-e we see that the $\Omega = 0$ reference PLE model reproduces
reasonably well the observed galaxy counts in the five bands over a wide
magnitude range.
The most significant discrepancies are with respect to the bright $b_j$ counts
of Maddox et al. (1990, \S2.3), and in $K$ with respect to the Mobasher et al.
(1986) counts and the Gardner et al. (1993) HWS counts.
At $K > 20$ the model agrees well with the data from Soifer et al. (1994) and
McLeod et al. (1995), but overestimates the Gardner et al. (1993) HDS data
and the Djorgovski et al. (1995) counts for $K < 23$.
Table 5 shows the excellent agreement between the observed and predicted values
of the slope $\gamma$ in the five bands.
At fainter magnitude than the current limits the model predicts a significant
flattening of $N(m)$ in all bands (last column of Table 5).
The faintest $r_f$ and $I$ band counts of Tyson (1988) do not show this
flattening but show an excess (with large error bars) with respect to the
model predictions.
If faint counts keep increasing with a steep slope, then higher values of
$z_f$ are likely needed to fit the data.
For increasing $z_f$, the flattening predicted at faint magnitude shifts
towards even fainter magnitude.
Deeper data are necessary to test these predictions in detail.

%
\fign\beginfigure*{\fignumber}
\putfigl{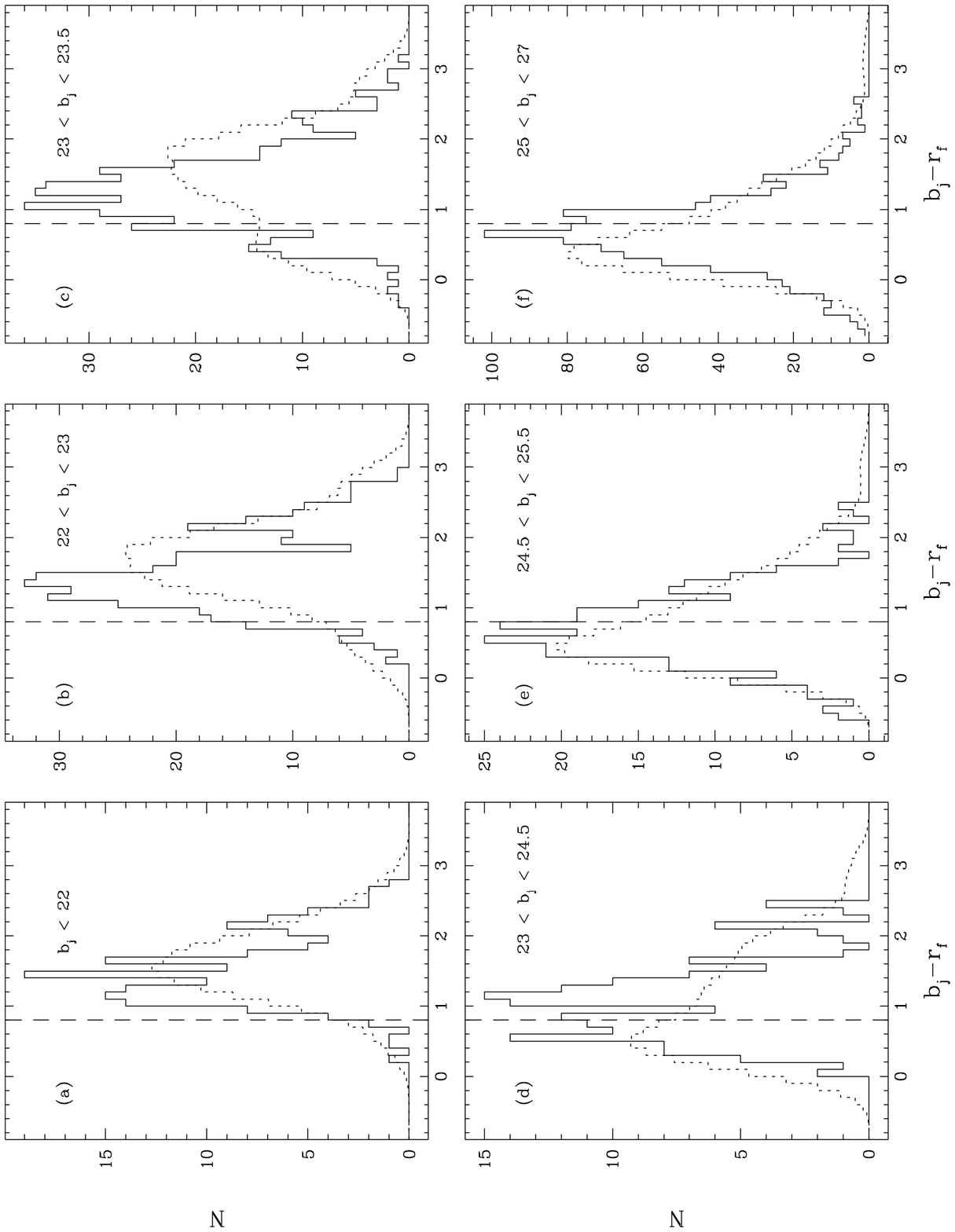}{12.0}{270}
\caption
{
{\bf Figure 6.}
$b_j-r_f$ colour distribution for different $b_j$ bins.
The solid histograms show the observed distributions from
Metcalfe et al. (1991, 1995a) and, for the faintest bin, from
Koo \& Kron (1992; data kindly provided to us by C. Gronwall).
When needed, the data have been converted to the $b_j$ and $r_f$ passbands of
Couch \& Newell (1980).
The dotted histograms show the predictions of the $\Omega=0$ reference PLE
model. To aid the eye a dashed line is drawn at $b_j-r_f = 0.8$.
}
\endfigure

In Figures 3a-e we see that the $\Omega = 1$ model
significantly underestimates the faint counts in all bands.
Although to a lesser extent, this is even true in the $K$ band,
where the influence of evolution is less important.
The deficiency in the counts for the $\Omega = 1$ model is due to
the decreasing comoving volume available for increasing $\Omega$.
The differences between the predictions of this model and the data
are minimized assuming a high $z_f$.
Even for $z_f = 10$ the excess in the observed counts is $\sim 3$ at $b_j = 26$.

\subsubsection{Number luminosity evolution (NLE) model}
\def\sbl{\char91 } 
\def\sbr{\char93 } 

Models that obey the $\Omega = 1$ constraint of the inflation scenario
require physical effects not considered in simple PLE models.
Among several possibilities, strong galaxy merger events at early
ages have been suggested
(Rocca-Volmerange \& Guiderdoni 1990 (hereafter RVG90); Broadhurst et al. 1992).
To test qualitatively this possibility, we have constructed a NLE model
in which the LF is allowed to evolve with $z$ as proposed by RVG90
$$\phi(L,z) = (1+z)^{2\eta} \phi\sbl L(1+z)^{\eta},z=0\sbr.\eqno(3)$$
This function simulates the merging of faint high $z$ galaxies to form
bright local galaxies, while conserving the total comoving mass density.
We have used the same galaxy evolution models as in the PLE models (Table 3).
This simple and crude NLE model fits the galaxy counts quite well in all
bands. A good fit is obtained for $\eta=1$ in (3).
Fig 4 shows the resulting $b_j$ counts.
More realistic models that simulate the change of the photometric
properties of the merging galaxies have to be computed
(cf. Fritze-v.Alvensleben \& Gerhard 1994).

\subsubsection{Contribution of each galaxy type to the total counts}

%
\fign\beginfigure*{\fignumber}
\putfigl{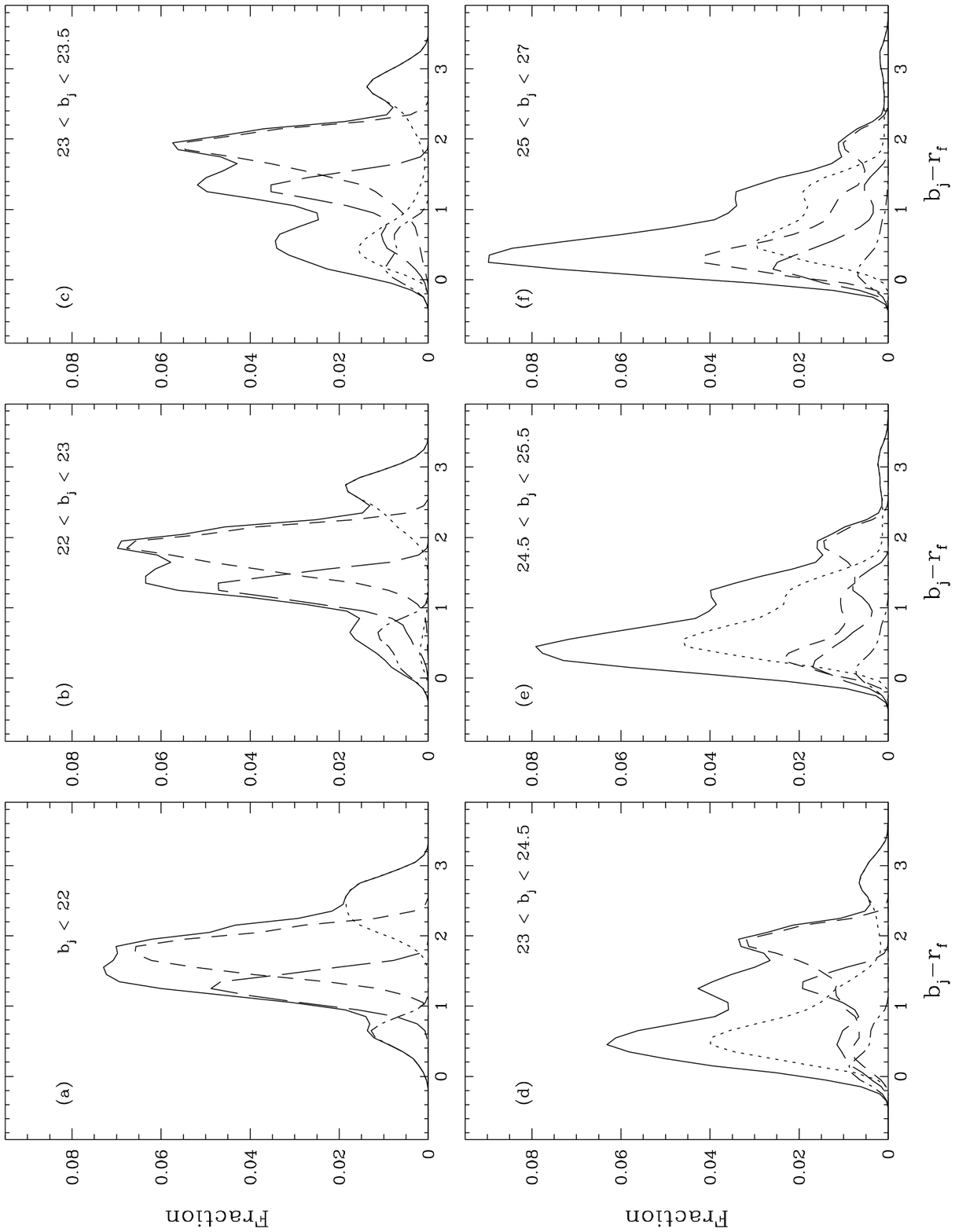}{12.0}{270}
\caption
{
{\bf Figure 7.}
$b_j-r_f$ predicted colour distribution for the various morphological types
in different $b_j$ bins for the $\Omega=0$ reference PLE model.
All galaxies: solid line;
E/S0: dotted line;
Sab-Sbc: short-dashed line;
Scd-Sdm: long-dashed line;
vB: dot-dashed line.
}
\endfigure

Fig 5 shows the contribution to the total $b_j$ and $K$ counts of each galaxy
type in the $\Omega = 0$ PLE model.
In $b_j$ and $U$ (not shown), early spirals are the dominant
population over most of the magnitude range.
Evolved E/S0s contribute only $\sim 20\%$ to the counts at $20 \le b_j \le 22$.
Young E/S0 galaxies at their maximum SFR period contribute $\sim 50\%$
and produce the hump in the total counts in the range $23.5 \le b_j \le 25.5$.
This hump has been noticed by Metcalfe et al. (1995a) in their counts.
E/S0 galaxies are also dominant for $K \la 22$.
There is no $K$ hump because even at $z=z_f=4.5$ the $K$ filter
samples the blue spectral region, not reaching the UV rest frame
and missing the signature of the high star formation episode.
The hump shifts to brighter magnitude for higher $\Omega$ and to fainter
magnitude for higher $z_f$.
The onset of the $b_j$ hump may seem particularly steep in our models because we
use a coarse grid of galaxy types.
With a finer grid of galaxy spectra the bright galaxy phases appear more
gradually and the onset of the hump is less evident (GK95).
The hump in the data (Fig 3b) suggests some degree of discreteness in the
distribution of galaxy types, intermediate between ours and GK95's.

\subsection{Colour distributions}

%
\fign\beginfigure*{\fignumber}
\putfigl{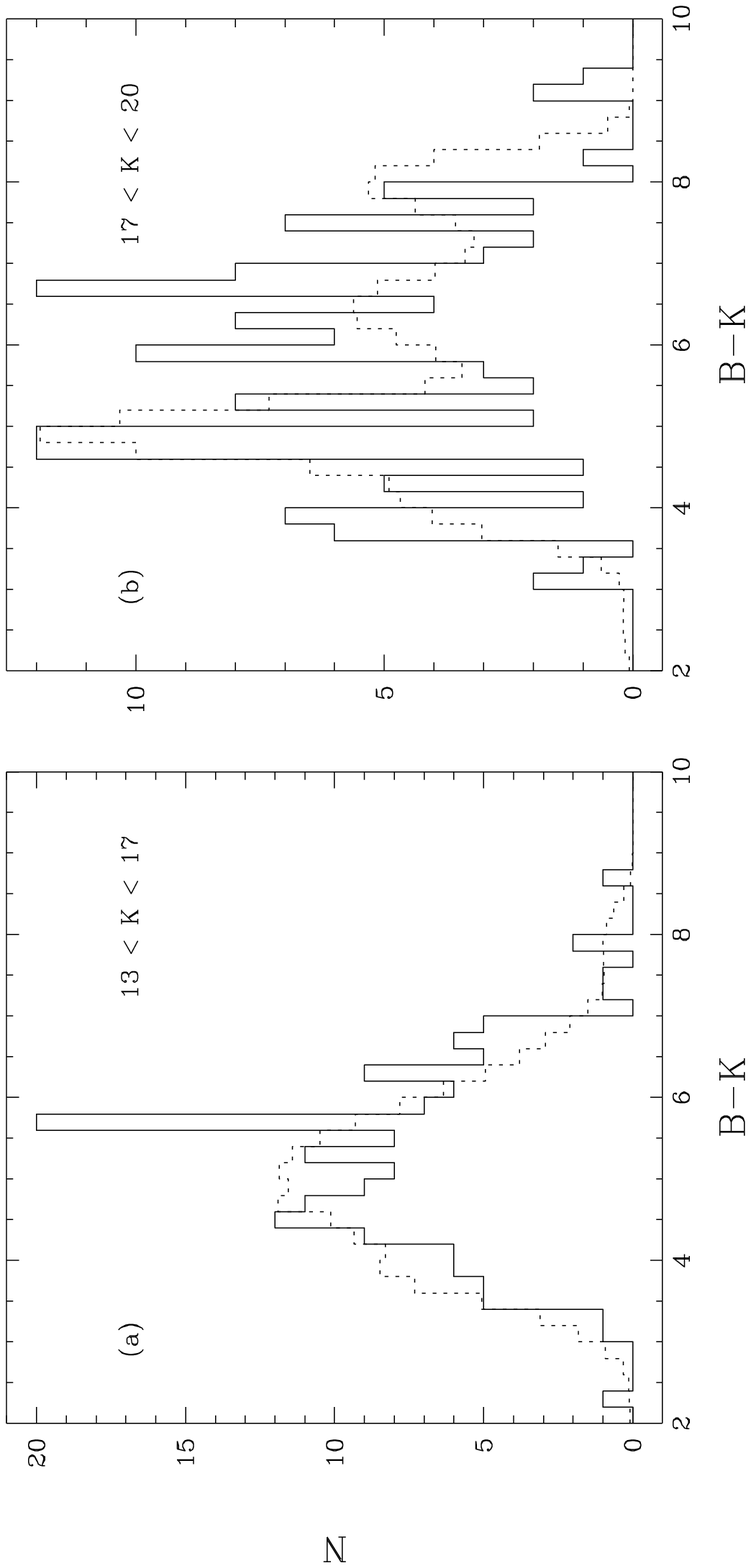}{8.5}{270}
\caption
{
{\bf Figure 8.}
$B-K$ colour distribution for different $K$ bins.
The solid histograms show the observed distributions from
the Hawaii $K$ band survey (Songaila et al. 1994).
The dotted histograms show the predictions of the reference $\Omega=0$
PLE model.
}
\endfigure

The colour distributions derived from faint galaxy surveys show a gradual
trend towards bluer mean colours at fainter magnitude.
The result that $b_j-r_f$ (photographic) becomes significantly bluer
beyond $b_j\sim22$ (Kron 1980), has been confirmed by various groups
(Shanks et al. 1984; Tyson 1984; Infante, Pritchet \& Quintana 1986).
With CCD detectors colour distributions have been extended up to
$b_j\sim27$ (Koo \& Kron 1992 and references therein; Metcalfe et al. 1995a).
The median $B-K$ colour of $K$-selected field galaxies also becomes bluer
beyond $K\sim 17.5$ (Gardner et al. 1993).
To make the comparison of the data and models more realistic, we have
applied a gaussian error function with $\sigma=0.15$ mag to the
colour vs. $z$ relations before deriving the colour distributions.
This procedure is expected to take into account, at least to first order, both
the observational errors in the colours and the intrinsic dispersion
in the colours of galaxies of the same morphological type.

Fig 6 shows the $b_j-r_f$ colour distribution for different bins of $b_j$.
At the brightest and faintest bins, panels (a), (d)-(f), the agreement between
the data and model distribution is quite good.
At intermediate magnitude, panels (b) and (c), the model distribution
is redder than the data by $\sim \Delta (b_j-r_f) = 0.4$.
Despite this discrepancy, the model predicts that
galaxies with $b_j-r_f < 0.8$ appear at $b_j \sim 23$,
and are essentially absent at brighter magnitude.


\fign\beginfigure{\fignumber}
\putfigl{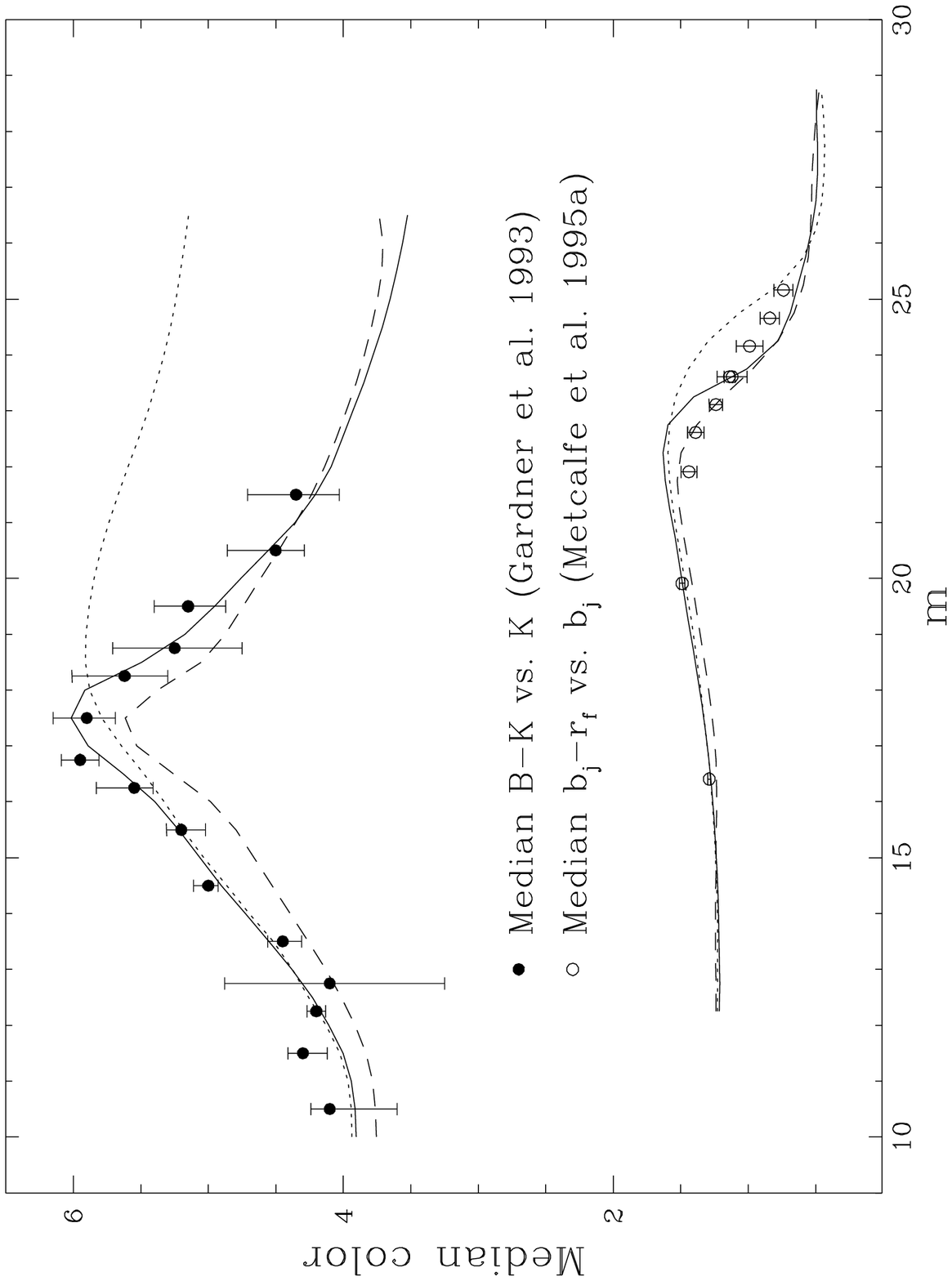}{6.0}{270}
\caption
{
{\bf Figure 9.}
Median colours as a function of apparent magnitude.
The bottom part of the figure shows $(b_j-r_f)_{med}$ vs. $b_j$.
The top part shows $(B-K)_{med}$ vs. $K$.
The data points are from the compilations by Metcalfe et al. (1995a) and
Gardner et al. (1993). The lines show the predictions for the
$\Omega=0$ reference PLE model (solid line),
$\Omega=1$ reference PLE model (dashed line),
$\Omega=0$ nE model (dotted line).
}
\endfigure

Fig 7 shows the $b_j-r_f$ fractional colour distribution predicted by the 
model for the same $b_j$ bins
of Fig 6 discriminated by galaxy morphological type.
While at bright magnitude, panels (a) and (b), the various morphological types
are reasonably well separated in colour, at the faintest bins, panels (e) and
(f), galaxies of all types share the same colours.
In particular, galaxies bluer than $b_j-r_f = 0.8$ are not just vB galaxies or
late type spirals, but $\sim 50\%$ of them in panel (c) and about 60\%
in panel (d), are high $z$ young E/S0 galaxies with $z_{med} \sim 1.5$.
These are the same galaxies responsible for the hump discussed in \S3.3.4.
The E/S0 galaxies are responsible for both the blue (at high $z$) and the red
(at low $z$) tails of the distribution in panels (c) and (d).

%
\fign\beginfigure*{\fignumber}
\putfigl{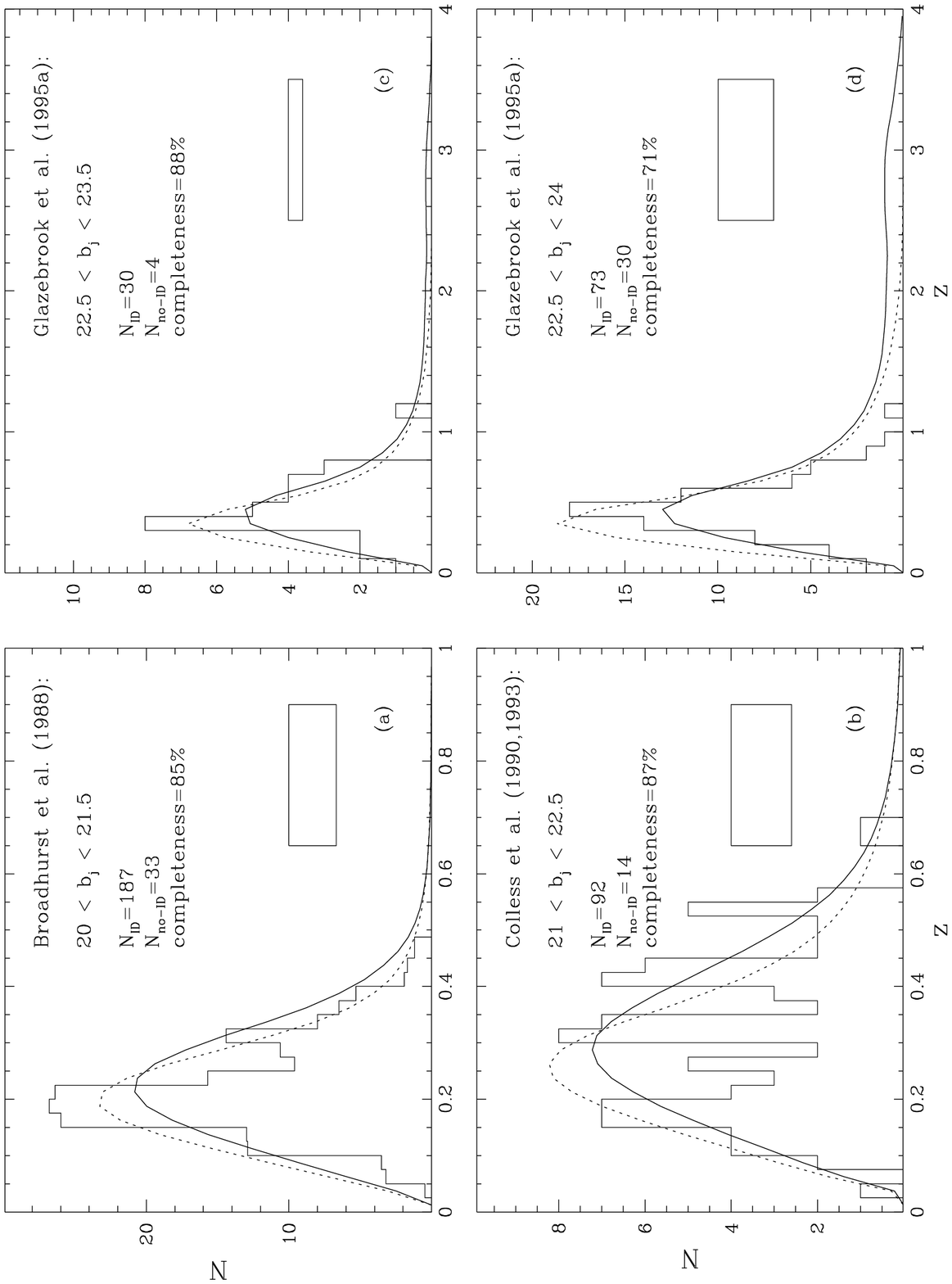}{11.0}{270}
\caption
{
{\bf Figure 10.}
$z$ distributions for $b_j$ selected samples.
The source for the observed distributions (solid histogram) is indicated in
each panel together with $N_{ID}$ = number of galaxies with measured $z$,
$N_{no-ID}$ = number of galaxies for which no $z$ could be measured, and
the completeness limit of the survey.
The area of the rectangle in each panel $= N_{no-ID}$.
Dotted line: nE model; solid line: PLE model; both for
$\Omega = 0,\ z_f = 4.5$.
The model predictions have been scaled to the total number of objects,
$N_{ID} + N_{no-ID}$.
}
\endfigure

Fig 8 compares the Hawaii $K$-band survey (Songaila et al. 1994)
$B-K$ colour distribution with our model.
This sample has been obtained as a combination of a number of different
$K$ magnitude limited surveys. The size of the surveyed area was diminished as
a function of limiting $K$ to provide a roughly constant number of
galaxies in each magnitude bin from $K = 13$ to $K = 20$. The model has been
scaled to the number of galaxies observed in each bin. Also in this
case the agreement between model and data is rather good. 

Fig 9 shows the observed $(b_j-r_f)_{med}$ and $(B-K)_{med}$ median colours
as a function of $b_j$ and $K$.
The $\Omega = 0$ PLE model reproduces rather well the data.
High-$z$ star forming galaxies are responsible for blueing the median model
colour at faint $K$.
This is one possible reason why galaxies with colour as red as local E
galaxies are not detected at faint $K$ (Gardner 1995).
The nE mode fails to reproduce the data, especially in $(B-K)_{med}$ vs. $K$,
where the difference between the PLE and the nE models is large.
Even though in the $\Omega = 1$ PLE model we have used intrinsically redder
galaxy SEDs than in the $\Omega = 0$ model (Table 3), the median $B-K$
for the $\Omega = 1$ model is bluer than for the data.
The main reason for this shift towards bluer colours in the $\Omega = 1$ model
is the younger age of the galaxies in this cosmology, despite the higher $z_f$.

\subsection{Redshift distributions}

Multi-object spectrograph surveys of large samples of faint galaxies have
reached faint enough magnitude for evolution to be
detectable.
At $b_j \ga 21$ large numbers of $z>1$ galaxies are predicted by PLE models
in which galaxies undergo {\it strong} luminosity evolution (Broadhurst et al.
1988, 1992).
The failure of the $z$ surveys to reveal these galaxies sets strong
upper limits on the amount of luminosity evolution that can be invoked to
explain the excess in the number counts and the blueing of the median
colour discussed in the previous sections.
Rather than being evidence against all PLE models
and in favour of the nE model, we argue below that the observed
$z$ distributions of blue-selected faint galaxies
are consistent with PLE models in which galaxies evolve mildly in
luminosity and rule out only the large amounts of luminosity evolution assumed
in early PLE models.

\subsubsection{$b_j$ band}

Fig 10 shows the $z$ distributions derived from four surveys.
Panels (a)-(c) correspond to $b_j$ limited samples, while the data in
panel (d) are a combination of seven fields with different limiting magnitude,
ranging from 23 to 24 in $b_j$, and completeness limit $> 70\%$ (Glazebrook et
al. 1995a). In the latter case, our simulation includes the different
$b_j$ limiting magnitude for each field.
The data in panel (b) correspond to the LDSS survey (Colless et al. 1990),
supplemented with additional $z$ measurements in a subsample of the original
LDSS survey areas by Colless et al. (1993).
The $z$ distribution of the $\Omega = 0$ reference PLE model shown in panels
(a) and (b) is consistent with the data, and does not show a significant high
$z$ tail. The mean $z$ for the PLE model distribution in these two panels are
$\langle z \rangle = (0.23,~0.32)$, in excellent agreement with the measured
values ($0.22,~0.31$).
Colless et al. (1993) reduced the $z$ incompleteness in their LDSS subsample
to $4.5\%$. From an analysis of these data they conclude that at most
$\sim4\%$ of galaxies with $b_j < 22.5$ have $z > 0.7$.
The PLE model predicts $\sim 3\%$ of galaxies with $z>0.7$ in this magnitude
range.

%
\begintable{6}
\nofloat
\caption{{\bf Table 6.} Glazebrook et al. (1995) $z$ surveys }
\halign{%
~#\hfil~~~ & ~~~#\hfil~~~ & ~~~\hfil#\hfil~~~ & ~~~\hfil#\hfil~~~ \cr
\noalign{\vskip 3pt}\noalign{\hrule}\cr\noalign{\vskip 3pt}
$b_j$ range & source & $z_{med}$ & $f_{0.7}$ \cr
\noalign{\vskip 3pt}\noalign{\hrule}\cr\noalign{\vskip 3pt}
$22.5-23.5$ & data     & 0.46        & 0.13        \cr
            & UL$^a$   & $\le 0.48$  & $\le 0.24$  \cr
            & PLE      & 0.50        & 0.27        \cr
            & G95$^b$  & 0.76-1.31   & 0.55-0.82   \cr
            &          &             &             \cr
$22.5-24$   & data     & 0.46        & 0.12        \cr
            & UL       & $\le 0.56$  & $\le 0.38$  \cr
            & PLE      & 0.59        & 0.40        \cr
            & G95      & 0.83-1.39   & 0.61-0.84   \cr
\noalign{\vskip 3pt}\noalign{\hrule}\cr
}
\tabletext{\par
\noindent $^a$ Upper limits if $z$-unidentified galaxies are at $z>0.7$\par
\noindent $^b$ PLE models by Glazebrook et al. (1995) }
\endtable

At fainter magnitude, panels (c) and (d), the PLE model predicts a
high $z$ tail which is not seen in the data.
Following Glazebrook et al. (1995a), we compare in Table 6 the
values of $z_{med}$ (median $z$) and $f_{0.7}$ (fraction of galaxies
with $z>0.7$) for their survey and our model.
For each magnitude range in Table 6, the first line gives
$z_{med}$ and $f_{0.7}$ for the galaxies with measured $z$,
the second line gives upper limits computed by
assuming that all the $z$-unidentified galaxies are at $z > 0.7$, and the
third line lists the predictions of the $\Omega = 0$ reference PLE model.
The model can be reconciled with the data only if all,
or at least most of the galaxies without measured $z$ in the
indicated magnitude range are at high $z$.
This assumption will be discussed in \S3.6.
Our PLE model is in good agreement with the spectroscopic data of the two
$B$-selected SSA13 and SSA22 Hawaii Survey fields (Cowie, Hu \& Songaila 1995):
all 9 galaxies with $B \le 23$ have $z\le0.7$, and for $23<B\le24$, 8 galaxies
out of 21 (38\%) with measured $z$ have $z>0.7$.

The nE model distribution is in good agreement with the data in the
four panels of Fig 10.
The nE model requires the $z$ distribution of the $z$-unidentified
galaxies not to be very different from that of the galaxies with measured $z$.
However, it is important to recall here the failure of the nE model to reproduce
the counts and the colour distributions at faint magnitude (\S3.3.1).

Note that we have not included in our models the $(1+z)^4$ dependence of
the surface brightness with $z$, which decreases the fraction of
expected high $z$ galaxies (Yoshii \& Peterson 1995).
The magnitude of this effect is a function not only of the intrinsic
parameters of the galaxies, but also of the details of the data
reduction procedure adopted in the construction of the photometric catalogs
from which the galaxies to be observed spectroscopically are selected.
Other effects, such as dust extinction inside galaxies (GK95), which have not
been considered in our models can decrease the predicted number of high $z$
galaxies.

To compare different PLE models, we list in Table 6 the range of $z_{med}$
and $f_{0.7}$ computed by Glazebrook et al. (1995a) from their PLE models.
The Glazebrook et al. values are significantly higher than ours.
The same is true for the GRV90 and Metcalfe et al. (1995a) PLE models.
The basic reasons for this difference are the milder galaxy evolution
assumed in our PLE models, and/or the more crude treatment of luminosity and
spectral evolution of galaxies in the quoted papers.
For example, in the Glazebrook et al. (1995a) model, a single SFR history
is adopted for all morphological types.
On the other hand, the KGB93 and GK95 models produce values of $z_{med}$ and
$f_{0.7}$ slightly lower than ours and, therefore, in good agreement
with the data.
However, the KGB93 model also predicts a significant population of low
luminosity galaxies at $z < 0.2$, which is not seen in the Glazebrook et al.
data.

%
\fign\beginfigure{\fignumber}
\putfig{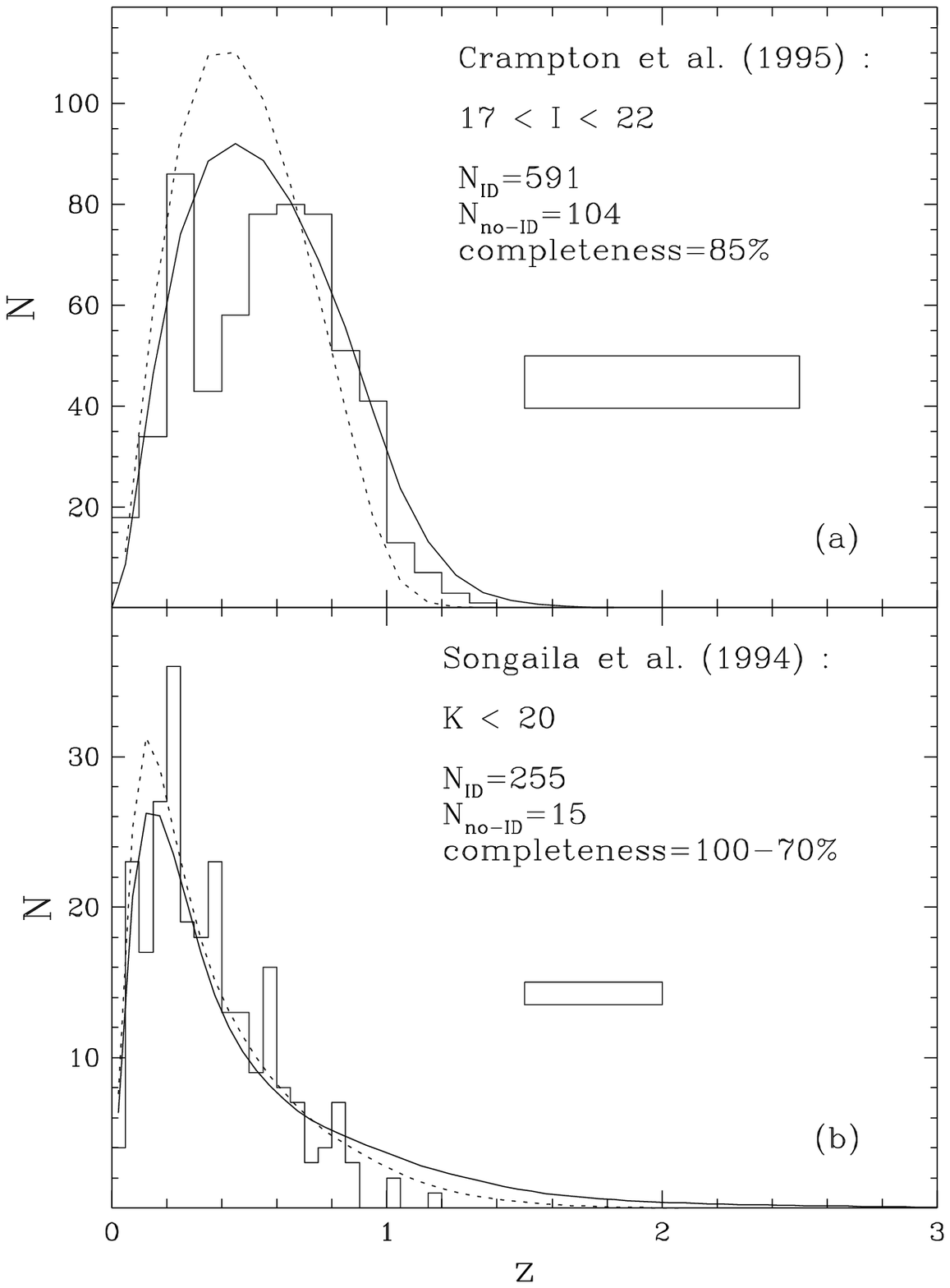}{11}{0}
\caption
{
{\bf Figure 11.}
Redshift distributions for $I$ and $K$ selected samples.
The source for the observed distributions (solid histogram) are indicated in
each panel.
The Crampton et al. (1995) data were kindly provided to us by L. Tresse.
The two values for completeness of the Songaila et al. sample refer
to $K<18$ and $K>18$, respectively.
Dotted line: nE model; solid line: PLE model; both for
$\Omega = 0,\ z_f = 4.5$.
}
\endfigure

%
\fign\beginfigure{\fignumber}
\putfigl{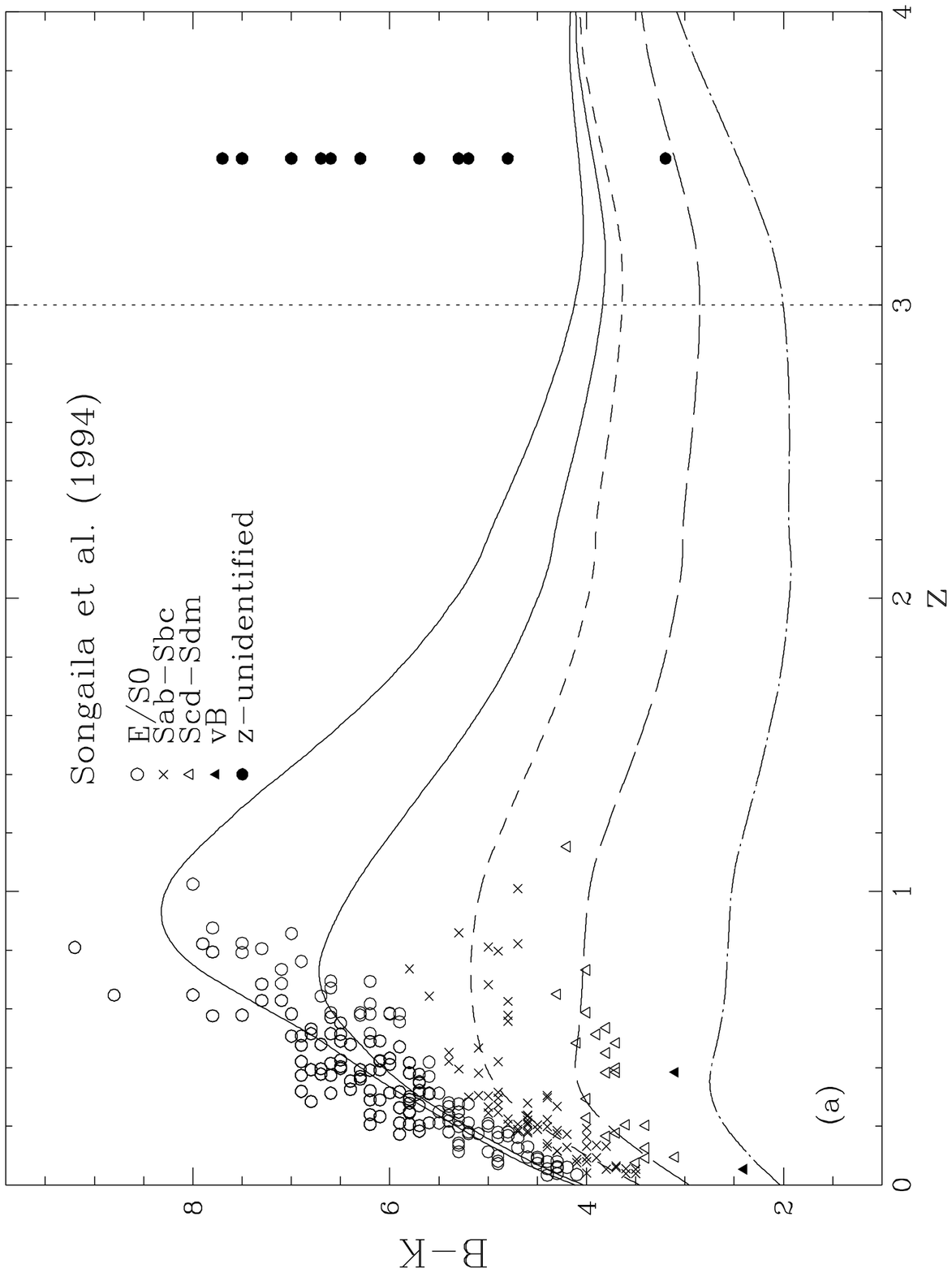}{6.0}{270}
\putfigl{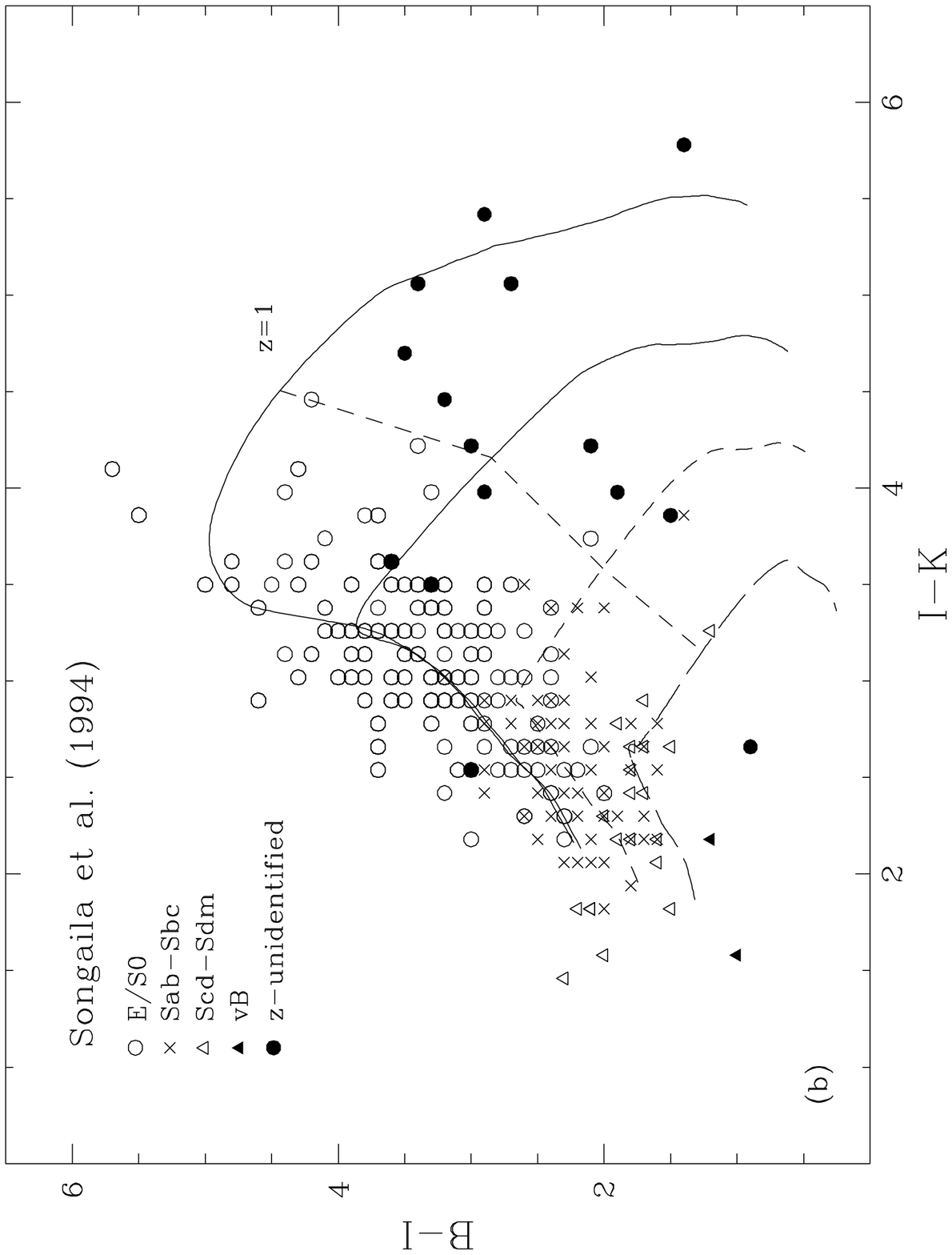}{6.0}{270}
\caption
{
{\bf Figure 12.}
{\it (a)}
$B-K$ colour as a function of $z$ for the Songaila et al. sample.
The solid dots correspond to the $z$-unidentified galaxies in the survey and
are plotted arbitrarily at $z=3.5$
The lines represent the colour predicted for each morphological type by the
BC93 models listed in Table 3 for the $\Omega = 0$ reference model.
E/S0: solid lines;
Sab-Sbc: short-dashed line;
Scd-Sdm: long-dashed line;
vB: dot-dashed line.
{\it (b)}
$B-I$ vs. $I-K$ for the same sample. The lines have the same meaning as in
{\it (a)}. The $z = 1$ locus is indicated. See text for details.
}
\endfigure

\subsubsection{$I$ and $K$ bands}

In Fig 11 we compare the $z$ distribution derived from the large and deep $z$
surveys of Crampton et al. (1995, $I$-selected) and Songaila et al. (1994,
$K$-selected) with the models.
The $z$ distribution in panel (a) is characterized by
$\langle z \rangle = 0.56$, $z_{med} = 0.57$, and $f_{0.7} = 0.34$.
The $\Omega = 0$ reference PLE model predicts
$\langle z \rangle = 0.58$, $z_{med} = 0.54$, and $f_{0.7} = 0.30$,
in excellent agreement with the data.
The high $z$ E/S0 galaxies predicted at $b_j \ga 23.5$ have a mean
$\langle b_j-I \rangle <1$, corresponding to $\langle I \rangle \ga 22.5$,
and are, therefore, not expected in significant numbers in this survey
limited at $I = 22.0$.
The nE model $z$ distribution,
$\langle z \rangle = 0.48$, $z_{med} = 0.47$, and $f_{0.7} = 0.18$,
does not reproduce well the tail of high $z$ galaxies observed
in this survey.
The $I$ band LF derived from the Crampton et al. (1995) data set (see
Fig 6 of Lilly et al. 1995b), as well as the E galaxy LF constructed from the
HST Medium Deep Survey (estimating $z$ photometrically, Im et al. 1995),
show clear evidence of evolution for $z\le 1$.
The amount of evolution is consistent with the prediction of our
PLE model: $\Delta I_z \sim -1.2$ up to $z = 1$.

To compare the models with the Songaila et al. (1994) $z$ distribution,
we have taken into account the decreasing area vs.
increasing limiting $K$ relationship of the survey as discussed in \S3.4.
The completeness of this spectroscopic survey is $\sim$ 100\% for
$K < 18$ and $\sim$ 70\% for $18 < K < 20$.
At variance with our $z$ distribution for $b_j$-selected samples (Table 6),
the $f_{0.7}$ upper limits derived from the $K$-selected data
are significantly lower than the model values.
For the observed $z$ distribution in Fig 11, panel (b), $f_{0.7} = 0.08$.
The PLE model predicts $f_{0.7} = 0.23$, well above the
upper limit $f_{0.7} < 0.13$ derived assuming that all galaxies
with no measured $z$ have $z > 0.7$.
This result indicates that there may be a problem in one or more
of the ingredients or assumptions that go into our PLE models.
The spectroscopic incompleteness of the sample is important at the faint
level, but it cannot be the only cause of the problem since the discrepancy
is already present in the bin $17 < K < 18$, where the spectroscopic survey
is complete. In this bin, the upper limit from the observed distribution is
$f_{0.7} < 0.16$, and the model predicts $f_{0.7} = 0.37$.
We think that the evolutionary rate in the $K$ band has to be revised in
order to explain the observations. We are currently exploring galaxy
evolution models based on different sets of evolutionary tracks and stellar
spectral libraries (Bruzual 1995) to evaluate their effect on PLE models.

\subsection{On the nature of $z$-unidentified galaxies}

Since most $z$ surveys are based on spectra which cover the range from
$\sim$3700 to $\sim$7500 \AA, the [OII] $\lambda3727$\ line (most prominent
feature in low S/N galaxy spectra) can be used as a $z$ indicator only up
to $z\sim 1$.
Higher $z$ galaxies are difficult to identify because of the lack of features
in the UV rest frame spectra of typical galaxies.
This fact, by itself, suggests the possibility that a large fraction
of the galaxies with no measured $z$ is at high $z$.
This assumption is also supported by an analysis of the colour distributions
of these galaxies.
We discuss here in detail the behavior of the $B-I$ and $I-K$ colours
in the Songaila et al. (1994) sample.
Fig 12a shows $B-K$ vs. $z$ for all galaxies in the sample, discriminated
by morphological type accordingly to the closest proximity of the galaxy
colour to the BC93 model colour for the class at the same $z$.
Most of the $z$-unidentified galaxies (black dots in the plot)
are consistent with the model colours of E/S0s at $z > 0.2$.
In Fig 12b the same galaxies are plotted in the $B-I$ vs. $I-K$ plane.
The lines show the expected location of galaxies of different morphological
type at increasing $z$.
Two things are interesting to note: first, galaxies of all types fall
reasonably well in the appropriate region in this plane.
Second, a large fraction of the $z$-unidentified galaxies has colours
consistent with being early-type galaxies at $z > 1.0$.
We have verified {\it a posteriori} the reliability of the photometric
$z$ estimates
by examining the measured $z$ of galaxies with colours close to the expected
colour at $z = 1$: 7 out of 8 such galaxies have $0.64 < z < 1.16$.
Thus, it seems reasonable to assume that most of the $z$-unidentified galaxies
in this sample are actually at high $z$.

This conclusion is consistent with the Keck telescope spectroscopic observations
of faint galaxies ($K<20, I<22.5, B<24.5$) by Cowie et al. (1995b).
Their data show strong evidence of a significant fraction of high-$z$ luminous
star forming galaxies.
Of 333 galaxies that have been observed, $z$ could be measured for 281 galaxies.
91 galaxies ($\sim 32\%$) have $z>0.7$ and 40 galaxies ($\sim 14\%$) have $z>1$.
Cowie et al. argue that inspection of the $B-I$ vs. $I-K$ colour-colour plane
suggests that most of the remaining unidentified objects are
luminous high $z$ star forming galaxies.

The same type of analysis cannot be applied to the $b_j$-selected survey of
Glazebrook et al. (1995a) because only $B-R$ is available for these galaxies.
The photometric $z$ estimates using a single colour are less accurate.
The $B-R$ distribution suggests a higher percentage of late type galaxies in
this sample than in the $K$-selected sample, in agreement with our
model for the number counts (\S3.3.4).

\section {Conclusions}
\def\ititem#1{\item{\it (#1)}}
The detailed comparison of a large amount of faint galaxy survey data and the
predictions of new models has provided the following results.

\beginlist

\ititem{i} The $\Omega=0$ nE model fits reasonably well the observed
$(U,b_j,r_f,I)$ counts up to $\sim$ (21,22,22,23), respectively.
At fainter magnitude the counts show an excess with respect to the nE model
in all bands, except in $K$. The excess in the $b_j$ counts
amounts to $N_{obs}/N_{nE}\sim 3$ at $b_j\sim 26$, and is lower
than in previous studies by a factor of $\sim$ 2.
This difference is due to our choice of normalizing the model counts
to the observed number in the range $19.0 < b_j < 19.5$,
and to the higher UV flux in the BC93 models of local E/S0's than in the SEDs
used in previous nE models (GRV90, Maddox et al. 1990).

\ititem{ii}
A PLE model, in which galaxies are characterized by the LF and SEDs
listed in Table 1 and Table 3 for the ($\Omega = 0,\ z_f = 4.5,\ t_g = 16$ Gyr)
Friedmann cosmology, provides good fits to most of the existing data
from faint galaxy surveys. In particular, it reproduces well the
faint galaxy counts in the $U,~b_j,~r_f,~I$ and $K$ bands, as well as
the colour distributions of blue and $K$-selected samples up to the faintest
observational limits, and the $z$ distributions of $b_j$-, $I$- and
$K$-selected samples up to $b_j<23, I<22$, and $K<17$. In the magnitude
range $23 \le b_j \le 24$, the predictions from the model can still
be reconciled with the data if most of the galaxies without a $z$
determination are at relatively high $z$ ($z>0.7)$.

\ititem{iii}
The steep slope of the APM counts at bright $b_j$ requires
strong evolution even at relatively small $z$ (Maddox et al. 1990).
This is in conflict with the mild amount of evolution required to fit
the $z$ and colour distributions at fainter magnitude.
It has been suggested that the low counts at bright magnitude may be produced
by either photometric errors (Metcalfe et al. 1995b) and/or by the fact that
photographic surveys may have missed a significant population of low surface
brightness galaxies because of the relatively high isophotal limits in these
surveys (McGaugh 1994; Ferguson \& McGaugh 1995).

\ititem{iv}
PLE models in a flat $\Omega = 1$ universe cannot reproduce several aspects
of the data. In particular, the faint counts are significantly below the
observed ones in all bands, including $K$.
A NLE model in an $\Omega = 1$ universe,
in which galaxy density increases as $(1+z)^2$ due to merger events
is found to reproduce satisfactorily the counts (see also RVG90).
However, we consider that a more realistic model, which takes into
account the change of the galaxy photometric properties when the merging
occurs, must be explored before this result can be taken at face value.

\ititem{v}
The Scalo IMF, being relatively poor in massive stars,
produces models for early-type galaxies which are considerably less luminous in
the UV at early epochs than the Salpeter IMF models.
This lower flux translates into a milder spectral evolution, as required
by the observed $z$ distributions of faint galaxies.
Scalo IMF models are thus preferred for early-type galaxies over Salpeter
IMF models. 

\ititem{vi}
The observed $b_j-r_f$ and $B-K$ colour distributions are
reproduced more closely by our PLE models if we assume that star formation
in E/S0 galaxies extends over a longer period of time ($\tau_1$ and $\tau_2$
models) than in a short box-type burst models ($B_1$ model).

\ititem{vii}
At faint magnitude ($b_j \ga 23, K \ga 18$) our PLE model predicts a significant
fraction of high-$z$ galaxies, mostly blue young early-type E/S0 and Sab-Sbc
galaxies. The predictions of this model are in excellent agreement
with the $B-K$ colours of the Hawaii $K$-selected survey which failed
to reveal galaxies at high $z$ with colours as red as those of
local E/S0 galaxies (Gardner 1995). On the other hand, recent results on the
LF of red and blue galaxies in the CFRS sample suggest evolution of the blue
galaxies and not of the red galaxies (Lilly et al. 1995b).
This appears to be in contradiction with our model and to favor models in
which a significant amount of the luminosity evolution needed to fit the
faint counts is due to spiral rather than early-type galaxies.
A discussion of these models can be found in Campos \& Shanks (1995).

\ititem{viii}
The most significant remaining discrepancy between our PLE model and the
data is in the $z$ distribution of the Songaila et al. (1994) $K$-selected
sample at $K>17$.
The fraction of galaxies at high $z$ predicted by the model is significantly
higher, by a factor of $\sim 2-3$, than observed.
The discrepancy persists even if, as suggested by their colours, most of the
galaxies which have been observed spectroscopically and for which no $z$ could
be measured, are at high $z$.
Thus, on the one hand, the $z$ distribution of these galaxies seems to rule
out evolution.
On the other hand, the $K$ band LF (Glazebrook et al. 1995b) indicates
that galaxies at $z=1$ are $\sim 0.75$ magnitude brighter than local ones,
consistent with PLE models.
Only additional spectroscopy of faint $K$-selected galaxies can resolve this
apparent contradiction between different data sets.

\endlist

In summary, simple PLE models, together with an appropriate selection of galaxy
evolution models (SFR and IMF) which provide mild luminosity evolution
at least up to $z=1$ in an $\Omega = 0$ universe, can still be considered
as baseline models.
Additional improvements should be considered in this framework, for example:
{\it (a)} Introduction of different $z_f$ for galaxies of different
morphological type and/or different luminosity. This could help in
explaining the colour-luminosity relation, observed in cluster
galaxies (see Bower, Lucey \& Ellis 1992 for ellipticals and
Gavazzi 1993 for both early and late types).
{\it (b)} Spectrophotometric models which include chemical and dynamical
evolution (Bressan, Chiosi \& Fagotto 1994) predict that E/S0 galaxies
approach their local colour at younger ages than in chemically homogeneous
models. This faster rate of evolution might bring the
predictions of the $\Omega = 1$ PLE model in better agreement with the
observations and should be explored in detail.
{\it (c)} Introduction in the models of at least the most important
observational selection effects, e.g. the
surface brightness selection (McGaugh 1994, Yoshii \& Peterson 1995).
{\it (d)} A detailed analysis of the effects of extinction by dust
(GK95, Campos \& Shanks 1995),
taking into account the time evolution of the amount of dust inside galaxies,
which can be particularly important at high $z$.

The interested reader may request the $k-$ and $(e+k)-$corrections required
to reproduce these models via e-mail from G.B.A. or L.P.

\section*{Acknowledgments}

G.B.A. was supported by SFB 328 during his visit to the Landessternwarte
Heidelberg K\"onigstuhl.
L.P. acknowledges the hospitality of the Landessternwarte Heidelberg
K\"onigstuhl and the Centro de Investigaciones de Astronom{\'\i}a during
the realization of this project, as well as the support of
the EEC program No. CHRX-CT92-0033.
We thank an anonymous referee for his/her careful reading of the first
version of this paper, and for challenging us to make this paper
shorter and more interesting. We expect to have succeeded.

\bye

\end